%% file: main.tex
\documentclass[12pt]{article}

\usepackage[a4paper,margin=2.5cm]{geometry}
\usepackage{setspace}
\usepackage[T1]{fontenc}
\usepackage[utf8]{inputenc}
\usepackage{float}
\usepackage[version=4]{mhchem}
\usepackage{newtxtext}
\usepackage[scale=0.95]{sourcesanspro}
\usepackage{xcolor}
\usepackage{titlesec}
\usepackage{caption}
\usepackage{fancyhdr}
\usepackage{amsmath,amssymb}
\usepackage{newtxmath}
\usepackage{graphicx}
\usepackage{booktabs}
\usepackage{tabularx}
\usepackage{microtype,pdflscape}
\usepackage{tikz}
\usetikzlibrary{calc}
\usepackage[superscript]{cite}
\usepackage{bibunits}
\usepackage{hyperref}
\usepackage{listings}
\usepackage[most]{tcolorbox}

\definecolor{skillbefore}{HTML}{747E89}
\definecolor{skillafter}{HTML}{32647D}
\definecolor{skillink}{HTML}{25323D}

\tcbset{
  skillpanel/.style={
    enhanced jigsaw,
    sharp corners,
    boxrule=0pt,
    frame hidden,
    borderline west={1.5pt}{0pt}{#1},
    colback=#1!4!white,
    colbacktitle=#1!9!white,
    coltitle=skillink,
    coltext=black,
    fonttitle=\sffamily\small\setstretch{1},
    fontupper=\small\setstretch{1.15},
    titlerule=0pt,
    left=10pt,
    right=10pt,
    top=6pt,
    bottom=6pt,
    toptitle=5pt,
    bottomtitle=5pt,
    before skip=8pt,
    after skip=8pt,
    pad at break*=2mm,
    before upper={
      \setlength{\parindent}{0pt}
      \setlength{\parskip}{5pt}
    }
  }
}

\newtcolorbox{skilltext}[3]{
  skillpanel={#1},
  breakable,
  title={\textbf{#2}\enspace\textcolor{#1}{\textbar}\enspace #3}
}

\newtcblisting{skillcode}[3]{
  skillpanel={#1},
  listing only,
  title={\textbf{#2}\enspace\textcolor{#1}{\textbar}\enspace #3},
  listing options={
    basicstyle=\ttfamily\footnotesize,
    columns=fullflexible,
    keepspaces=true,
    breaklines=true,
    breakatwhitespace=false,
    showstringspaces=false,
    tabsize=2,
    aboveskip=0pt,
    belowskip=0pt
  }
}

\defaultbibliographystyle{naturemag}
\defaultbibliography{references}

\definecolor{heading}{HTML}{202020}
\definecolor{rulegray}{HTML}{767676}
\definecolor{linkblue}{HTML}{1D5B87}
\definecolor{citegreen}{HTML}{2F6F5E}
\definecolor{urlpurple}{HTML}{68468A}
\definecolor{coverpaper}{HTML}{F4F7F6}
\definecolor{covernavy}{HTML}{071B2B}
\definecolor{coverblue}{HTML}{087EA4}
\definecolor{covercyan}{HTML}{18B7D4}
\definecolor{covergold}{HTML}{D7A83E}
\definecolor{covermist}{HTML}{DCE7E8}
\hypersetup{
  hidelinks,
  bookmarksopen=false
}
\titleformat{\section}
  {\normalfont\sffamily\fontsize{14}{17}\selectfont\bfseries\color{heading}}
  {\thesection}{0.7em}{}
\titleformat{\subsection}
  {\normalfont\sffamily\fontsize{12}{15}\selectfont\bfseries\color{heading}}
  {\thesubsection}{0.7em}{}
\titleformat{\subsubsection}
  {\normalfont\sffamily\fontsize{11}{14}\selectfont\bfseries\color{heading}}
  {\thesubsubsection}{0.7em}{}
\titleformat{\paragraph}[block]
  {\normalfont\normalsize\bfseries}
  {\theparagraph}{0.7em}{}
\titlespacing*{\section}{0pt}{18pt plus 3pt minus 2pt}{8pt}
\titlespacing*{\subsection}{0pt}{14pt plus 2pt minus 2pt}{6pt}
\titlespacing*{\subsubsection}{0pt}{12pt plus 2pt minus 2pt}{5pt}
\titlespacing*{\paragraph}{0pt}{10pt plus 2pt minus 1pt}{4pt}
\allowdisplaybreaks[1]
\renewcommand{\headrulewidth}{0.3pt}

\fancypagestyle{plain}{%
  \fancyhf{}%
  \fancyhead[L]{\sffamily\fontsize{9}{11}\selectfont\color{heading}\textsc{Gan Jiang}}%
  \fancyhead[R]{\sffamily\fontsize{9}{11}\selectfont\color{heading} Research manuscript}%
  \fancyfoot[C]{\sffamily\fontsize{10}{12}\selectfont\thepage}%
  \renewcommand{\headrulewidth}{0.3pt}%
}
\makeatletter
\renewcommand{\headrule}{\hbox to\headwidth{\color{rulegray}\leaders\hrule height \headrulewidth\hfill}}

\renewcommand{\@maketitle}{%
  \thispagestyle{empty}%
  \begin{tikzpicture}[remember picture,overlay]
    \fill[coverpaper] (current page.south west) rectangle (current page.north east);

    \draw[covermist,line width=0.55pt]
      ([xshift=18mm,yshift=-43mm]current page.north west) --
      ([xshift=-18mm,yshift=-43mm]current page.north east);
    \draw[covercyan,line width=2.2pt]
      ([xshift=18mm,yshift=-43mm]current page.north west) -- ++(26mm,0);
    \node[anchor=north east,align=right,text=coverblue]
      at ([xshift=-18mm,yshift=-16mm]current page.north east)
      {\sffamily\fontsize{8.2}{10}\selectfont\bfseries Research manuscript\\[-0.1em]
       \normalfont\sffamily\fontsize{8.2}{10}\selectfont October 2026};

    \node[anchor=north west,inner sep=0pt]
      at ([xshift=18mm,yshift=-15mm]current page.north west)
      {\includegraphics[width=10mm]{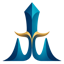}};
    \node[anchor=west,align=left,text=covernavy]
      at ([xshift=31mm,yshift=-20.5mm]current page.north west)
      {\sffamily\fontsize{10.5}{11}\selectfont\bfseries\textls[35]{GAN JIANG}\\[-0.05em]
       \normalfont\sffamily\fontsize{6.6}{8}\selectfont\color{coverblue}\textls[45]{Diffraction Intelligence}};

    \node[anchor=north west,align=left,text width=174mm,inner sep=0pt]
      at ([xshift=18mm,yshift=-59mm]current page.north west)
      {\sffamily\color{covernavy}
       \fontsize{35}{37}\selectfont\bfseries Gan Jiang\par
       \vspace{3.5mm}
       \fontsize{23}{28}\selectfont\mdseries
       A self-learning scientific agent for X-ray diffraction\par};

    \node[anchor=north west,align=left,text width=174mm,inner sep=0pt]
      at ([xshift=18mm,yshift=-125mm]current page.north west)
      {\color{covernavy}\@author};

    \draw[covermist,line width=0.55pt]
      ([xshift=18mm,yshift=42mm]current page.south west) --
      ([xshift=-18mm,yshift=42mm]current page.south east);
    \draw[covercyan,line width=2.2pt]
      ([xshift=18mm,yshift=42mm]current page.south west) -- ++(26mm,0);
    \node[anchor=north west,align=left,text width=174mm,inner sep=0pt]
      at ([xshift=18mm,yshift=36mm]current page.south west)
      {\sffamily\setstretch{1}\fontsize{10}{12}\selectfont
       \color{coverblue}\bfseries Correspondence\par
       \vspace{1.2mm}
       \color{covernavy}\mdseries
       \href{mailto:bincao@hkust-gz.edu.cn}{bincao@hkust-gz.edu.cn} (Bin Cao)\par
       \href{mailto:mezhangt@hkust-gz.edu.cn}{mezhangt@hkust-gz.edu.cn} (Tong-Yi Zhang)\par
       \href{mailto:jun.wang@ucl.ac.uk}{jun.wang@ucl.ac.uk} (Jun Wang)\par
       \vspace{1.5mm}
       \color{coverblue}\bfseries HomePage\quad
       \color{covernavy}\mdseries
       \href{https://ganjiang.asia}{ganjiang.asia}};

  \end{tikzpicture}%
  \null\vfill\clearpage
}
\makeatother

\newcommand{\SupplementCoverPage}{%
  \thispagestyle{empty}%
  \begin{tikzpicture}[remember picture,overlay]
    \fill[coverpaper] (current page.south west) rectangle (current page.north east);
    \node[anchor=north west,inner sep=0pt]
      at ([xshift=18mm,yshift=-15mm]current page.north west)
      {\includegraphics[width=10mm]{favicon.png}};
    \node[anchor=west,align=left,text=covernavy]
      at ([xshift=31mm,yshift=-20.5mm]current page.north west)
      {\sffamily\fontsize{10.5}{11}\selectfont\bfseries\textls[35]{GAN JIANG}\\[-0.05em]
       \normalfont\sffamily\fontsize{6.6}{8}\selectfont\color{coverblue}\textls[45]{Diffraction Intelligence}};
    \node[anchor=north east,align=right,text=coverblue]
      at ([xshift=-18mm,yshift=-16mm]current page.north east)
      {\sffamily\fontsize{8.2}{10}\selectfont\bfseries Supplementary\\[-0.1em]
       \normalfont\sffamily\fontsize{8.2}{10}\selectfont Information};
    \draw[covercyan,line width=2.2pt]
      ([xshift=18mm,yshift=-43mm]current page.north west) -- ++(26mm,0);
    \draw[covermist,line width=0.55pt]
      ([xshift=44mm,yshift=-43mm]current page.north west) --
      ([xshift=-18mm,yshift=-43mm]current page.north east);

    \node[anchor=north west,align=left,text width=174mm,inner sep=0pt]
      at ([xshift=18mm,yshift=-66mm]current page.north west)
      {\sffamily\color{coverblue}\fontsize{9}{11}\selectfont\bfseries
       Supplementary Information\par
       \vspace{5mm}
       \color{covernavy}\fontsize{29}{34}\selectfont\bfseries
       Gan Jiang\par
       \vspace{2mm}
       \fontsize{20}{25}\selectfont\mdseries
       A self-learning scientific agent for X-ray diffraction\par};

    \node[anchor=north west,align=left,text width=174mm,inner sep=0pt,text=covernavy]
      at ([xshift=18mm,yshift=-135mm]current page.north west)
      {\AuthorNames\par\vspace{3mm}\AuthorAffiliations};

    \draw[covermist,line width=0.55pt]
      ([xshift=18mm,yshift=42mm]current page.south west) --
      ([xshift=-18mm,yshift=42mm]current page.south east);
    \draw[covercyan,line width=2.2pt]
      ([xshift=18mm,yshift=42mm]current page.south west) -- ++(26mm,0);
    \node[anchor=north west,align=left,text width=174mm,inner sep=0pt]
      at ([xshift=18mm,yshift=36mm]current page.south west)
      {\sffamily\setstretch{1}\fontsize{10}{12}\selectfont
       \color{coverblue}\bfseries Correspondence\par
       \vspace{1.2mm}
       \color{covernavy}\mdseries
       \href{mailto:bincao@hkust-gz.edu.cn}{bincao@hkust-gz.edu.cn} (Bin Cao)\par
       \href{mailto:mezhangt@hkust-gz.edu.cn}{mezhangt@hkust-gz.edu.cn} (Tong-Yi Zhang)\par
       \href{mailto:jun.wang@ucl.ac.uk}{jun.wang@ucl.ac.uk} (Jun Wang)\par
       \vspace{1.5mm}
       \color{coverblue}\bfseries HomePage\quad
       \color{covernavy}\mdseries
       \href{https://ganjiang.asia}{ganjiang.asia}};

    \node[anchor=south east,text=coverblue]
      at ([xshift=-18mm,yshift=18mm]current page.south east)
      {\sffamily\fontsize{7}{9}\selectfont\textls[35]{October 2026}};
  \end{tikzpicture}%
  \null\vfill\clearpage
}

\renewenvironment{abstract}
  {\par\begingroup\normalfont\normalsize\doublespacing
   \noindent{\sffamily\bfseries\color{heading} ABSTRACT}\par\nobreak
   \vspace{6pt}\noindent\rmfamily\bfseries\color{black}\ignorespaces}
  {\par\endgroup\vspace{12pt}}

\newcommand{\AuthorNames}{%
  \sffamily\fontsize{11.4}{14.5}\selectfont\bfseries
  Bin Cao$^{1,\dagger,\ast}$, Huichi Zhou$^{2,\dagger}$, Runyu Yang$^{3}$,
  Jingsong Li$^{3}$, Shuchen Sun$^{1}$, Yan Song$^{2}$, Hanyu Gao$^{4}$,  Zhongwei Yu$^{1}$, 
  Tong-Yi Zhang$^{1,\ast}$ \& Jun Wang$^{2,\ast}$%
}
\newcommand{\AuthorAffiliations}{%
  \rmfamily\mdseries\fontsize{8.8}{11.2}\selectfont
  $^{1}$The Hong Kong University of Science and Technology (Guangzhou), China\\
  $^{2}$University College London, United Kingdom\\
  $^{3}$AI Lab, The Yangtze River Delta, China\\
  $^{4}$Institute of Automation, Chinese Academy of Sciences, China\\[0.25em]
  $^{\dagger}$These authors contributed equally to this work.\\
  $^{\ast}$Corresponding authors.%
}

\title{Gan Jiang: A self-learning scientific agent for X-ray diffraction}
\author{%
  \AuthorNames\\[0.65em]
  \AuthorAffiliations
}
\date{}

\begin{document}

\begin{bibunit}

\maketitle
\begin{abstract}
A central challenge for scientific agents is to turn analytical experience into reusable expertise grounded in physical evidence. Here we introduce Gan Jiang, a self-learning agent for powder X-ray diffraction built on a diffraction-analysis ecosystem we developed: XMatcher, XQueryer, XDecomposer and WPEM. Together, these engines span phase identification, multiphase decomposition and physics-constrained whole-pattern modelling. Gan Jiang converts analytical experience into executable skills by diagnosing failures, revising skill instructions and code, and validating revisions before reuse, without retraining the language model or changing the underlying physical models. Skills selected using development data and frozen before held-out evaluation achieve higher refinement scores than the original expert-designed skills across FullProf, GSAS-II and PyWPEM. The agent resolves strongly overlapping reflections, quantifies a five-phase ancient Egyptian cosmetic, tracks lattice evolution in an operating battery and compares atomic configurations in a disordered oxide catalyst. On DeltaXRDbench, it leads the evaluated methods in single- and multiphase identification across simulated and experimental data. Without supplied composition, single-phase top-1 accuracies reach 96.30\%, 81.78\% and 40.83\% on MP500, RRUFF and opXRD, respectively, compared with 58.00\%, 58.47\% and 26.45\% for the strongest comparator. These results demonstrate how an integrated scientific tool ecosystem can support agents that extract structural knowledge from measurements while accumulating validated analytical expertise that transfers to new samples.
\end{abstract}

\newpage

\section*{Main}

Every diffraction experiment poses two questions: what structure explains the measurement, and how should that explanation be tested? X-ray diffraction has connected synthesized matter to atomic structure for more than a century, underpinning crystallography and materials discovery~\cite{thomas2012birth,oganov2019structure}. In powder X-ray diffraction (PXRD), the scattering from many crystallites is compressed into a one-dimensional pattern. Peak positions constrain lattice spacings, intensities reflect atomic arrangements and phase abundances, and peak shapes carry information about microstructure and the instrument~\cite{cao2026physics,warren1990x}. Recovering a structural explanation is nevertheless an ill-posed inverse problem: overlapping reflections, disorder, preferred orientation and incomplete measurements can make different models difficult to distinguish~\cite{cao2026physics,yu2026equivariant}. Reliable interpretation therefore requires a sequence of decisions about candidate structures, refinement parameters and unexplained residuals~\cite{cao2026ai}.

High-throughput measurements and autonomous laboratories have increased the demand for such decisions~\cite{ma2025understanding,szymanski2023autonomous,chen2026agentic}. Machine-learning methods now classify phases, retrieve structures and generate structural candidates directly from diffraction patterns~\cite{cao2025xqueryer,maffettone2021crystallography,salgado2023automated,lee2020deep,chen2024crystal,lai2025end}. Automated refinement programs and emerging agents further connect these predictions to numerical verification~\cite{tian2013srrietveld,cui2015autofp,wu2026autoxrd,fei2026dara,yadav2026automated,li2026rongzai,campbell2026hybrid}. Yet closing the analysis loop for one sample does not by itself improve the procedure used for the next. Failed refinements, expert corrections and successful recovery strategies must be converted into reusable knowledge, and that knowledge must be tested beyond the episode that produced it.

Here we introduce Gan Jiang, a self-improving scientific agent that makes this conversion explicit. Gan Jiang combines a diffraction-analysis toolchain with a learning architecture. Its persistent memory consists of executable skills: procedures specifying what to do, when to do it and how to assess the result. Diffraction evidence guides both structural decisions within an analysis and revisions to the skills used in later analyses. For example, a GSAS-II training episode showed that changing several parameter groups together worsened the fit; the revised skill therefore emphasizes starting from defaults and adjusting one group at a time. This rule is retained for subsequent analyses without retraining the underlying language model. Candidate revisions are evaluated before adoption, and held-out comparisons test whether the resulting improvements transfer to unseen samples. We demonstrate this approach through challenging materials applications, controlled skill-learning experiments and single- and multiphase identification benchmarks. The resulting system couples the ability to interpret measurements with the ability to improve how those measurements are interpreted.
\section*{Results}

\begin{figure}[H]
  \centering
  \includegraphics[width=0.95\linewidth]{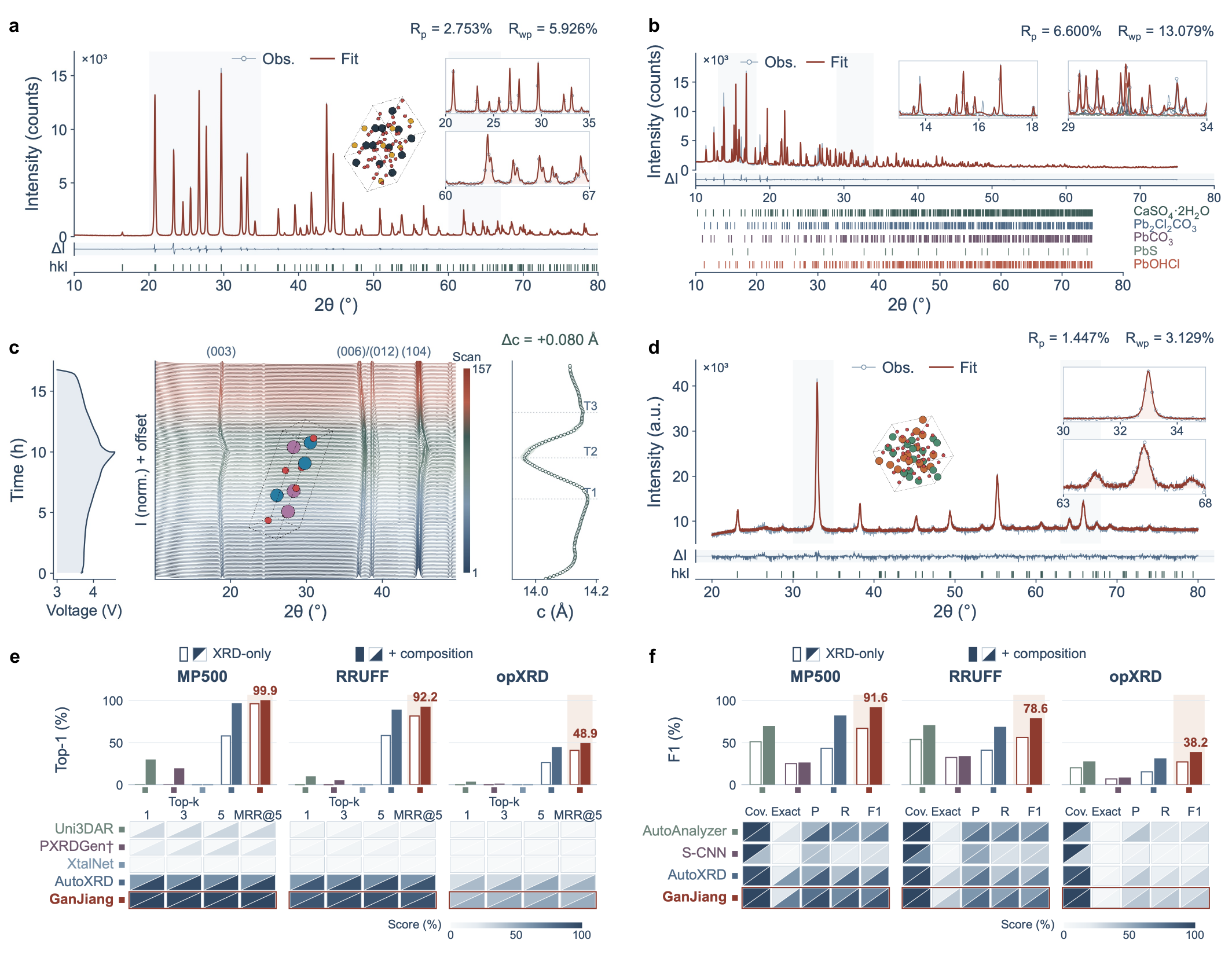}
  \caption{\textbf{Diverse materials applications and leading diffraction-identification performance.}
  \textbf{a}, Whole-pattern refinement of orthorhombic \ce{PbSO4} with strongly overlapping reflections ($R_\mathrm{p}=2.753\%$, $R_\mathrm{wp}=5.926\%$).
  \textbf{b}, Five-phase analysis of an ancient Egyptian cosmetic powder containing gypsum, phosgenite, cerussite, galena and laurionite ($R_\mathrm{p}=6.600\%$, $R_\mathrm{wp}=13.079\%$).
  \textbf{c}, Operando analysis of an O3-type \ce{Li(Ni_{0.8}Co_{0.1}Mn_{0.1})O2} cathode. The voltage--time trace (left), diffraction waterfall (centre; colour indicates scan order) and refined $c$ axis (right) connect electrochemical cycling to lattice evolution.
  \textbf{d}, Refinement of the selected Ru--Mn oxide configuration ($R_\mathrm{p}=1.447\%$, $R_\mathrm{wp}=3.129\%$). In \textbf{a,b,d}, pale blue symbols and red curves denote observed and calculated intensities; $\Delta I=I_\mathrm{obs}-I_\mathrm{calc}$ is the difference profile, and ticks mark calculated Bragg positions.
  \textbf{e}, Single-phase identification on MP500, RRUFF and opXRD: bars show top-1 accuracy and heat maps show top-1, top-3, top-5 and mean reciprocal rank (MRR@5).
  \textbf{f}, Multiphase identification on the same sources: bars show sample-wise macro-F1 and heat maps show prediction coverage (Cov.), exact-set recovery (Exact), precision (P), recall (R) and F1. Open and filled bars denote XRD-only and composition-assisted conditions. Gan Jiang is highlighted in red. Higher scores are better. Full values and composition-input protocols are given in Tables~\ref{tab:xrdbench-single} and \ref{tab:xrdbench-multi}.}
  \label{fig:cases}
\end{figure}

\subsection*{Broad materials applications and leading performance}

Gan Jiang applies a unified analytical framework to questions spanning mineral composition, battery operation and atomic disorder (Fig.~\ref{fig:cases}). Four applications demonstrate its ability to resolve overlapping reflections, quantify a complex mixture, track structural changes and distinguish between competing atomic configurations. Alongside these applications, DeltaXRDbench establishes the breadth of its identification performance. Without supplied composition, Gan Jiang achieves single-phase top-1 accuracies of 96.30\%, 81.78\% and 40.83\% on MP500, RRUFF and opXRD, exceeding the strongest comparator by 38.30, 23.31 and 14.38 percentage points, respectively (Fig.~\ref{fig:cases}\textbf{e}). It also leads the evaluated methods in multiphase precision, recall, F1 and exact-set recovery across all three sources, with and without composition assistance (Fig.~\ref{fig:cases}\textbf{f}). These comparisons establish state-of-the-art performance among the methods evaluated under the declared benchmark conditions.

In orthorhombic \ce{PbSO4} ($Pnma$), extensive reflection overlap makes isolated peak assignments unreliable. The pattern contains 383 pairs of overlapping Bragg reflections under Cu $K\alpha$ radiation ($\lambda_{\alpha1}=1.540560$~\AA\ and $\lambda_{\alpha2}=1.544330$~\AA). Gan Jiang uses whole-pattern modelling to retain the constraints supplied by the complete profile~\cite{cao2026ai}. The resulting refinement reaches $R_\mathrm{p}=2.753\%$ and $R_\mathrm{wp}=5.926\%$, with lattice parameters $a=8.4851(0)$~\AA, $b=5.4016(1)$~\AA\ and $c=6.9643(3)$~\AA\ (Fig.~\ref{fig:cases}\textbf{a}). This case illustrates how  Gan Jiang links candidate identification to an analysis appropriate for strongly overlapping signals.

The same approach quantifies a five-phase ancient Egyptian cosmetic powder. Previous studies identified gypsum (\ce{CaSO4.2H2O}), phosgenite (\ce{Pb2Cl2CO3}), cerussite (\ce{PbCO3}), galena (\ce{PbS}) and laurionite (\ce{PbOHCl}) in this material~\cite{forbes1965studies,walter1999making}. Gan Jiang resolves their contributions to the synchrotron profile and estimates respective mass fractions of 12.53\%, 18.53\%, 32.02\%, 9.69\% and 27.23\%, with $R_\mathrm{wp}=13.079\%$ (Fig.~\ref{fig:cases}\textbf{b}; Supplementary Table~\ref{tab:egyptian-lattice}). The analysis recovers a quantitative mineral assemblage from densely overlapping signatures. Its substantial lead-chloride component is consistent with earlier evidence for deliberate chemical processing~\cite{walter1999making}, connecting the recovered composition to a historical materials-processing question.

For an operating O3-type \ce{Li(Ni_{0.8}Co_{0.1}Mn_{0.1})O2} cathode cycled between 3.0 and 4.6~V~\cite{cao2026ai}, Gan Jiang treats successive scans as a connected trajectory. Each accepted structural state initializes the next refinement, while background and diffraction parameters are reassessed for the new measurement. The resulting $c$-axis trajectory captures expansion and high-voltage contraction consistent with the peak motion in the measured profiles (Fig.~\ref{fig:cases}\textbf{c}). Here, retaining experimental context allows the analysis to follow structural evolution across the series.

In a Ru--Mn oxide electrocatalyst obtained by cation exchange from orthorhombic \ce{Mn2O3}~\cite{qin2023orthorhombic}, the challenge is to distinguish chemically plausible Ru-site occupations. Gan Jiang screens candidate configurations within a $3\times3$ supercell and compares their whole-pattern refinements. The selected model reaches $R_\mathrm{p}=1.447\%$ and $R_\mathrm{wp}=3.129\%$ (Fig.~\ref{fig:cases}\textbf{d}) and is deposited as CCDC 2530452~\cite{CCDC2530452}. This comparison identifies a diffraction-supported model within the tested configuration space, providing a structural hypothesis for subsequent investigation of catalytic behaviour.

\subsection*{From diffraction fingerprints to structural explanations}

\begin{figure}[H]
  \centering
  \includegraphics[width=\linewidth,clip]{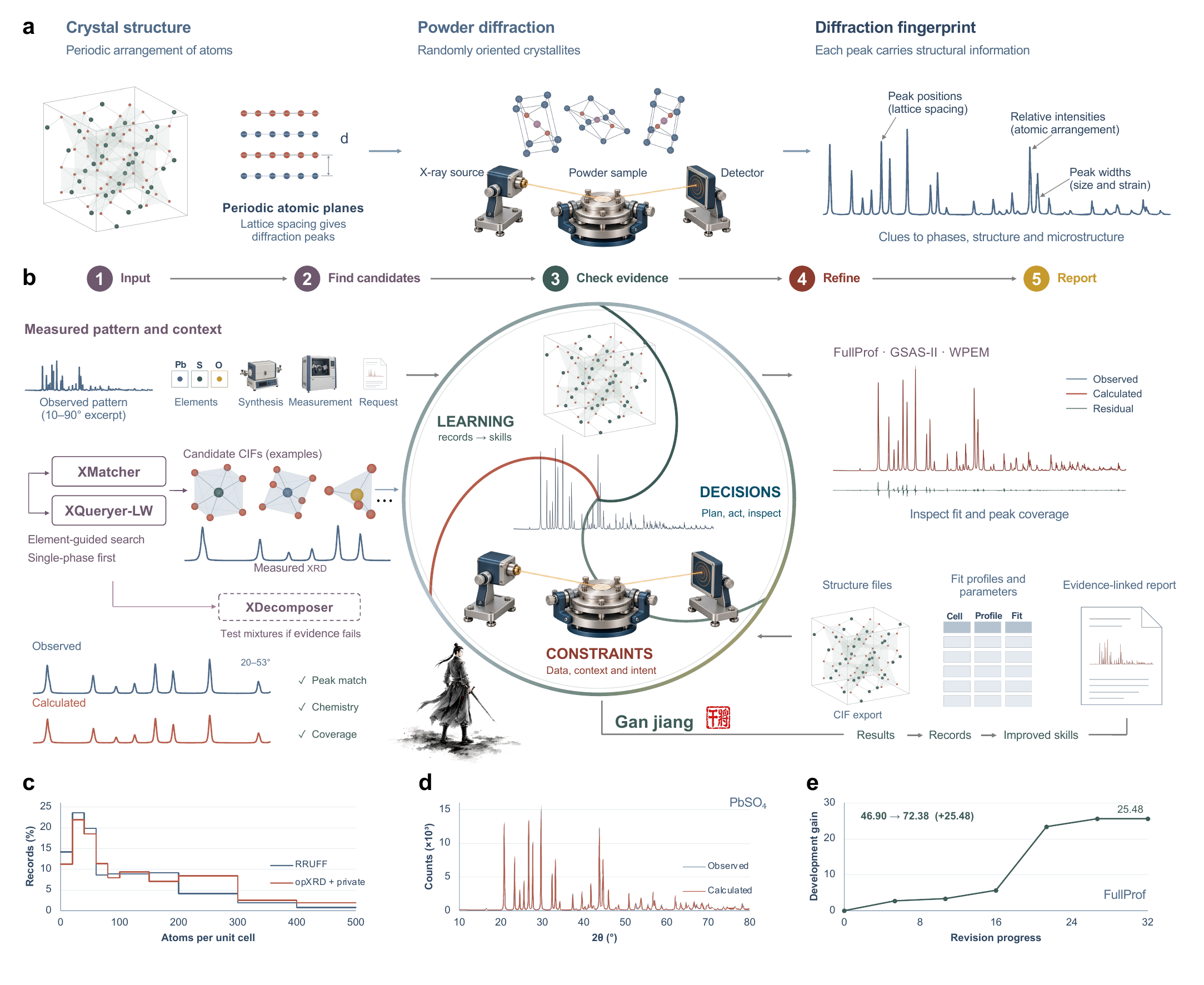}
  \caption{\textbf{Diffraction physics, analytical workflow and reusable learning in Gan Jiang.}
  \textbf{a}, A crystal's periodic atomic arrangement produces diffraction at angles determined by lattice spacings. Randomly oriented crystallites generate a powder pattern whose peak positions, relative intensities and widths provide information about the lattice, atomic arrangement and microstructure, subject to instrumental effects.
  \textbf{b}, Gan Jiang links five analysis stages: input, candidate retrieval, evidence checking, refinement and reporting. The measured pattern is combined with available chemical and experimental context. XMatcher and XQueryer-LW propose single-phase candidates; XDecomposer and XMatcher-AutoMix support mixture analysis when single-phase hypotheses fail the evidence checks. Candidate structures are assessed against peak agreement, chemistry and coverage, then refined with eligible engines. Outputs include CIFs, observed and calculated profiles, refined parameters and an evidence-linked report. The central loop connects scientific constraints, analytical decisions and learning from completed episodes.
  \textbf{c}, Distribution of atoms per unit cell in the RRUFF and combined opXRD/private branches of the experimental structure collection, expressed as percentages of records within each branch.
  \textbf{d}, Representative \ce{PbSO4} whole-pattern fit.
  \textbf{e}, Summary of FullProf workflow revision, illustrating the accumulation of improvements. Development trajectories and frozen-version held-out comparisons are distinguished in Fig.~\ref{fig:self-evolving}.}
  \label{fig:ganjiang-overview}
\end{figure}

The physical basis of these analyses is accessible through a simple relationship: a periodic arrangement of atoms produces a characteristic pattern of scattered X-rays (Fig.~\ref{fig:ganjiang-overview}\textbf{a}). In a powder, many crystallite orientations are measured together, yielding intensity as a function of scattering angle, $2\theta$. Bragg's law, $2d\sin\theta=n\lambda$, links each reflection to a lattice-plane spacing $d$ at wavelength $\lambda$. Peak positions therefore constrain the unit cell; relative intensities constrain the atomic arrangement; and widths reflect crystallite size, strain and instrumental effects~\cite{warren1990x}. When several crystalline phases coexist, their reflections overlap. The scientific task is to find a set of structures and parameters that jointly accounts for this measured fingerprint.

Gan Jiang organizes that task into five connected stages: receive the pattern and context, find candidate structures, check their evidence, refine supported models and report the result (Fig.~\ref{fig:ganjiang-overview}\textbf{b}). The user supplies a scientific objective and may add elemental composition, synthesis history or instrument information. Candidate structures are retrieved as crystallographic information files (CIFs), which describe the lattice and atomic positions. Identification asks which structures could be present; refinement adjusts a proposed model to reproduce the measured profile. The difference between observed and calculated intensities then exposes features that the current explanation leaves unresolved.

Four complementary engines provide the analytical foundation. XMatcher performs interpretable search-and-match analysis~\cite{cao2026xmatcheropensourceframeworkxray}; XQueryer Lightweight (XQueryer-LW) provides learned structure retrieval when elemental information is available~\cite{cao2025xqueryer}; XDecomposer proposes phase-resolved descriptions of mixtures~\cite{gao2026xdecomposer}; and PyWPEM performs physics-constrained whole-pattern modelling~\cite{cao2026ai}. FullProf~\cite{rodriguez2001fullprof} and GSAS-II~\cite{toby2013gsas} provide established refinement branches. Their outputs are brought back to the same experimental question: do the proposed phases explain the observed peaks, and are the refined parameters physically plausible?

The reference collection connects this workflow to a broad structural search space: 100,315 MP500 structures and 2,007 experimentally derived structures (Fig.~\ref{fig:ganjiang-overview}\textbf{c}; Methods). The \ce{PbSO4} profile illustrates how a retrieved model becomes a whole-pattern explanation (Fig.~\ref{fig:ganjiang-overview}\textbf{d}). Completed analyses also retain the actions, diagnostics and corrections that produced that explanation. These records can be distilled into revised skills, allowing experience from one analysis to change the procedure used in another (Fig.~\ref{fig:ganjiang-overview}\textbf{e}). Gan Jiang thus operates at two levels: it tests structural hypotheses for the current material and tests procedural improvements for future materials.

\subsection*{An executable architecture for evidence-guided analysis}

\begin{figure}[H]
  \centering
  \includegraphics[width=\linewidth,clip]{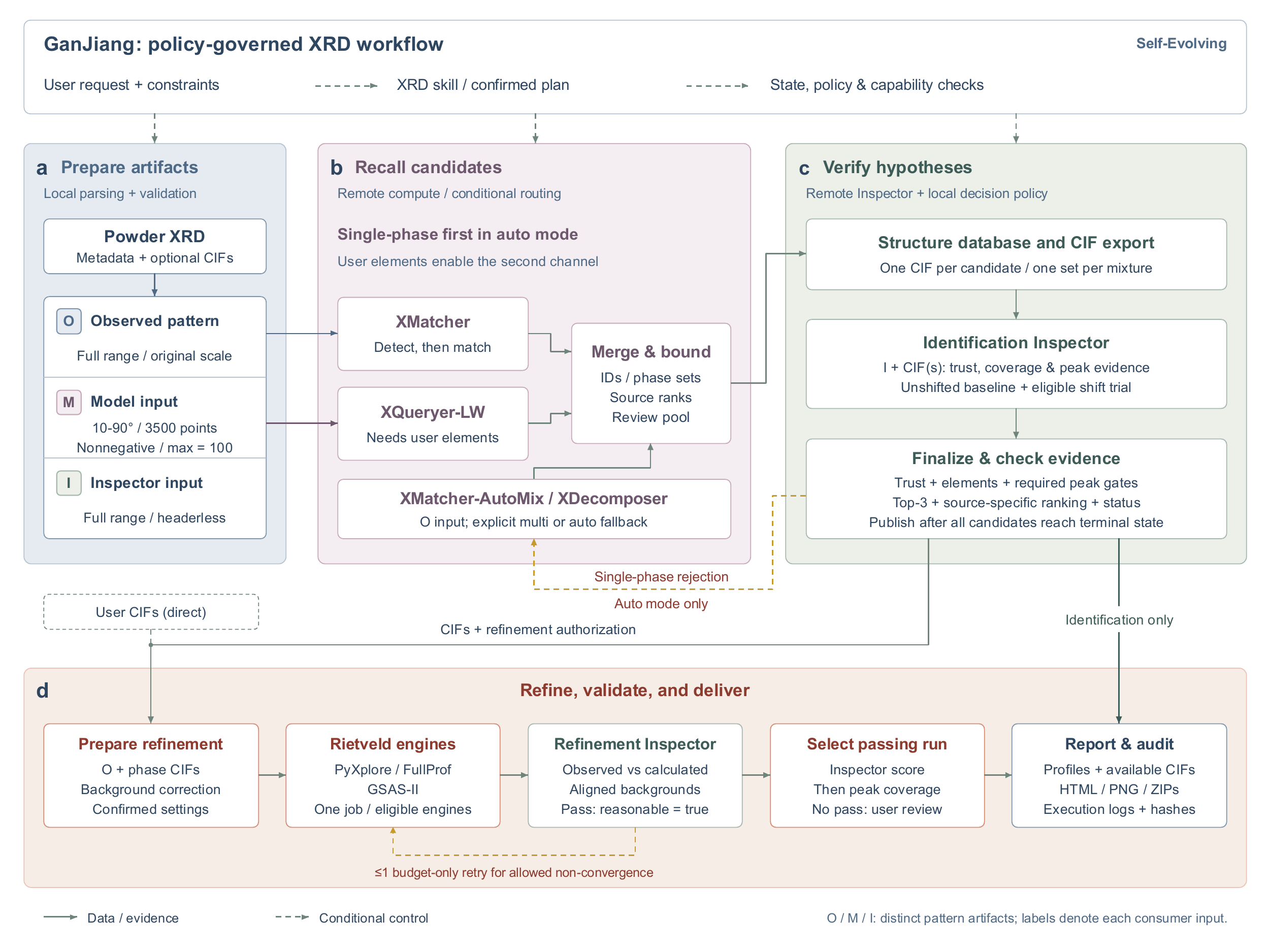}
  \caption{\textbf{Executable architecture for evidence-governed diffraction analysis.}
  The user objective and constraints determine the skill and analysis plan; state, policy and capability checks govern execution.
  \textbf{a}, Local parsing and validation produce an observed pattern (O; full range and original scale), a model input (M; 3,500 normalized points over $10^\circ$--$90^\circ$) and an inspector input (I; full range in the required format).
  \textbf{b}, XMatcher retrieves single-phase candidates; user-supplied elements enable XQueryer-LW. Candidates are merged while retaining identities and source ranks. XMatcher-AutoMix and XDecomposer support requested mixture analysis or automatic fallback after rejection of single-phase hypotheses.
  \textbf{c}, Candidate CIFs are exported and inspected against the measurement. Finalization requires the inspector verdict, elemental consistency where applicable and mandatory peak-evidence checks. Eligible angular-shift trials supplement the unshifted comparison.
  \textbf{d}, Supported phase sets or user-provided CIFs enter refinement through PyXplore (the PyWPEM service), FullProf or GSAS-II. An inspector checks the resulting profiles; passing runs are selected by inspection score and then peak coverage. At most one eligible budget-related retry is allowed. Solid arrows indicate data or evidence flow; dashed arrows indicate conditional control, including rejection and retry paths.}
  \label{fig:ganjiangframework}
\end{figure}

To implement this reasoning, Gan Jiang separates language-model planning, numerical calculation and scientific acceptance (Fig.~\ref{fig:ganjiangframework}). The language model interprets the objective and selects skills; specialized engines perform retrieval, decomposition and refinement; and explicit decision policies govern acceptance, workflow transitions and retries. The system supports identification alone, identification followed by refinement, and direct refinement of a supplied structure. Each analysis retains its input data, tool responses and policy version, so that a conclusion can be traced to the procedure that produced it.

The input layer preserves three representations with distinct purposes (Fig.~\ref{fig:ganjiangframework}\textbf{a}). The observed pattern retains its full measured angular range and original intensity scale for peak analysis and refinement. A separate model input is resampled to 3,500 points over $10^\circ$--$90^\circ$ and normalized for XQueryer-LW; unmeasured regions are treated as padding. The inspector receives the full-range measurement in its required format. Keeping these representations separate prevents a retrieval-specific transformation from replacing the experimental evidence used to judge a candidate.

XMatcher retrieves single-phase candidates from the observed peaks, and supplied elemental constraints enable the complementary XQueryer-LW channel (Fig.~\ref{fig:ganjiangframework}\textbf{b}). The system merges candidates by material identity while preserving source scores and provenance. The identification inspector then compares each candidate CIF, or jointly proposed phase set, with the measurement (Fig.~\ref{fig:ganjiangframework}\textbf{c}). Acceptance requires a favourable inspector verdict and all applicable composition and peak-verification checks. Agreement between retrieval channels helps prioritize inspection but cannot override a failed evidence check. Bounded angular-offset trials may assist matching; these trials do not constitute instrument calibration.

In automatic mode, completed rejection of the available single-phase hypotheses triggers mixture analysis through XMatcher-AutoMix or XDecomposer. A failed tool call alone does not establish that a sample is multiphase. This distinction separates computational failure from evidence that the structural model is incomplete. Explicitly requested mixture analyses can enter the multiphase branch directly, and all resulting phase sets undergo joint inspection.

Accepted phase models, or user-supplied structures, proceed to eligible refinement engines with recorded background preparation and instrument settings (Fig.~\ref{fig:ganjiangframework}\textbf{d}). PyWPEM is exposed through the PyXplore service, alongside FullProf and GSAS-II. The refinement inspector compares observed and calculated profiles on aligned angular coordinates and compatible intensity and background definitions. Passing runs are ranked by inspection score and peak coverage, with fit indices retained as diagnostics. At most one automatic retry is permitted for an eligible convergence or iteration-budget limitation. Unexplained peaks and phase-model mismatch instead prompt reconsideration of the hypothesis. Reports retain structures, profiles and diagnostics, while the associated audit record preserves execution parameters, failures and retry decisions. This record supports both scientific review and subsequent skill revision.

\subsection*{Self-improvement transfers analytical skills to unseen samples}

The defining test of self-improvement is whether experience from one analysis improves the procedure used for subsequent samples. Gan Jiang implements this through our proposed read--write reflective learning mechanism~\cite{zhou2026memento}. In that framework, skills form external memory: each combines instructions, prompts and executable support code. A router retrieves a skill for the current task, the agent executes it, and reflection on the resulting trace proposes changes to the skill library. Failure attribution directs revision towards the responsible procedure, while validation checks guard against regressions. The underlying language-model parameters remain fixed. Gan Jiang specializes this mechanism by tying revision to diffraction residuals, execution outcomes and the physical plausibility of refined models.

\begin{figure}[H]
  \centering

  \includegraphics[width=\linewidth]{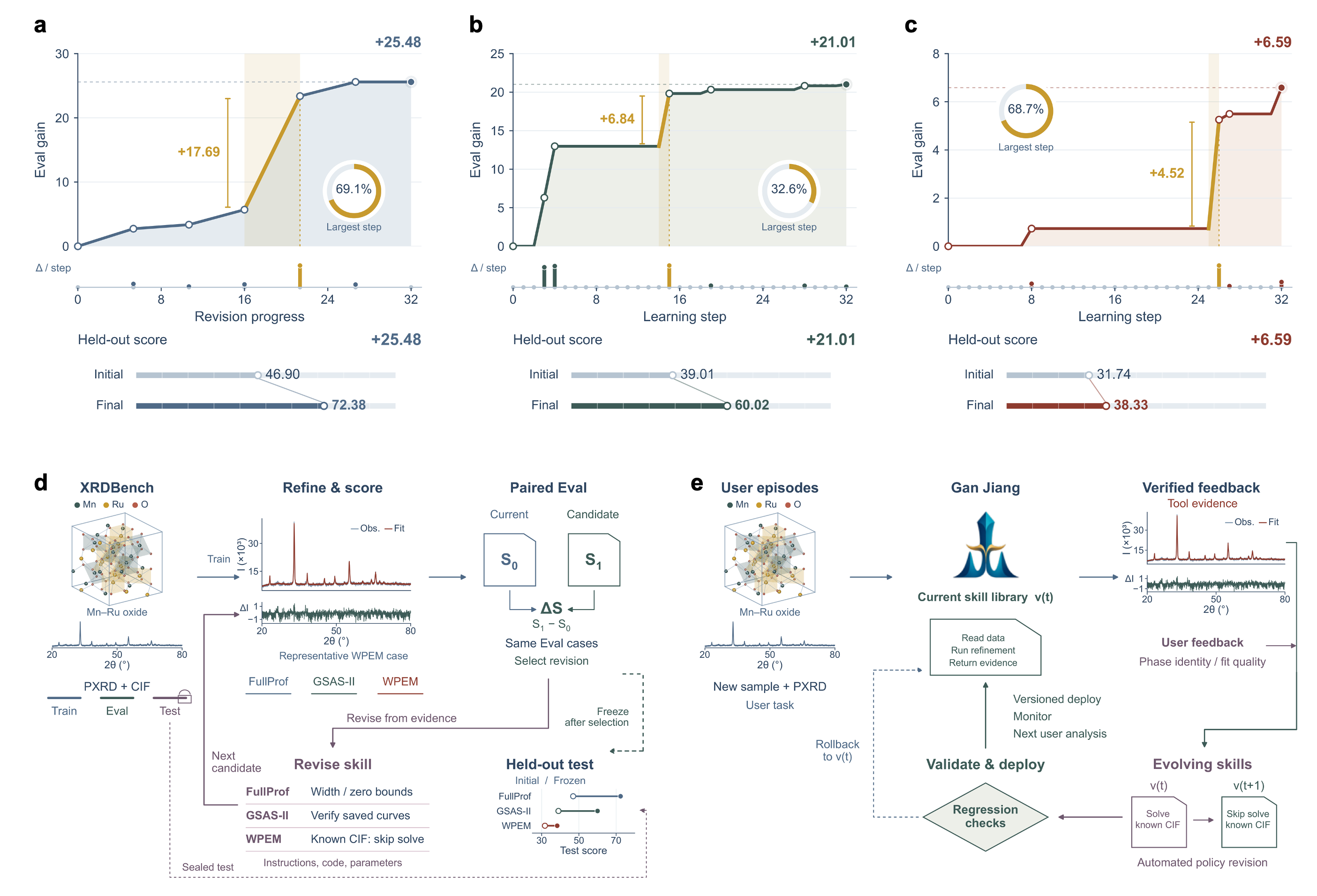}
  \caption{\textbf{Self-improvement of refinement workflows and transfer to held-out samples.}
  \textbf{a--c}, Learning trajectories for FullProf workflow revision, GSAS-II joint optimization and PyWPEM skill optimization (labelled WPEM), respectively. Upper plots show cumulative development-score gains; lower step plots show the increments associated with revisions. Highlighted intervals identify the large increments marked in each panel. Comparisons beneath the trajectories report mean held-out scores for the frozen initial and final workflows: 46.90 to 72.38 for FullProf, 39.01 to 60.02 for GSAS-II and 31.74 to 38.33 for PyWPEM. 
  \textbf{d}, Controlled learning protocol. Training episodes provide execution and diffraction feedback for revising instructions, code and parameters. Paired development checks select candidate revisions. The selected version is frozen before held-out evaluation, and test results do not select the learning trajectory.
  \textbf{e}, Deployment-time learning architecture. New user episodes are analysed with the current skill library; verified tool evidence and user feedback inform candidate revisions. Regression checks precede versioned deployment, with monitoring and rollback. }
  \label{fig:self-evolving}
\end{figure}

The distinction between sample-level optimization and skill learning is operational. Refining a lattice parameter changes the model for one sample. Revising when lattice parameters should be released, which diagnostics must be checked or how a recurrent failure should be handled changes the procedure for subsequent samples. After each learning episode, Gan Jiang uses the recorded trajectory to propose modifications to skill instructions, supporting code or refinement settings while preserving the input and output interfaces. Candidate versions can therefore be compared directly with the preceding version under the same evaluation conditions (Fig.~\ref{fig:self-evolving}\textbf{d}).

We evaluated workflow improvement across FullProf, GSAS-II and PyWPEM, starting from expert-optimized baseline procedures. Training episodes supplied the evidence for proposing changes, and paired development evaluations determined which revisions were retained. The selected version was then frozen before evaluation against the initial workflow on held-out samples. This design separates the data used to learn a procedure from the data used to assess its transfer. The FullProf trajectory summarizes workflow revisions, the GSAS-II experiment jointly optimizes skill and refinement settings, and the PyWPEM experiment evaluates skill optimization; their gains therefore quantify improvement under the stated adaptation regimes rather than a single common intervention.

All three final workflows improved their mean held-out refinement score (Fig.~\ref{fig:self-evolving}\textbf{a--c}). FullProf increased from 46.90 to 72.38, GSAS-II from 39.01 to 60.02, and PyWPEM from 31.74 to 38.33. These scores summarize profile agreement and include failed or timed-out evaluations as zero; they are not percentages of correctly identified structures. The development trajectories show when useful revisions were retained, whereas the frozen-version comparisons show that the final procedures improved performance beyond the samples used to select them. The smaller PyWPEM gain also indicates that the benefit depends on the initial workflow, engine and permitted revisions.

The learned changes make this mechanism concrete (Supplementary Section~\ref{sec:skill}). FullProf revisions replace a generic release sequence with a bounded, task-specific search over profile width and angular offset. GSAS-II skill revisions add explicit parameter-passing and saved-output checks, reducing the risk that a requested change is silently ignored or that reported diagnostics do not match the stored result. PyWPEM revisions distinguish refinement of a known structure from structure solving and introduce targeted responses to divergence or timeouts. These examples show how execution experience can become an actionable rule with a defined scope. 

The same architecture supports learning during deployment: verified user episodes can generate candidate updates that undergo regression checks before versioned release, with monitoring and rollback (Fig.~\ref{fig:self-evolving}\textbf{e}). The present experiments establish transfer after controlled revision and freezing. They provide a basis for, rather than a longitudinal demonstration of, open-ended improvement during routine use. Self-improvement in Gan Jiang is thus a testable property of revised analytical procedures, with each adopted skill linking its corrective action to the evidence that motivated it.

\subsection*{Benchmark gains across single- and multiphase identification}
\label{sec:xrdbench-identification}

To test whether the application results extend across a broader range of structures and measurements, we developed DeltaXRDbench. Its 62,044 patterns combine simulated MP500 data~\cite{cao2025simxrd}, experimental mineral patterns from RRUFF~\cite{lafuente2015rruff} and experimental data from opXRD~\cite{hollarek2026opxrd}. The collection contains 12,044 single-phase patterns and 50,000 two- or three-phase mixtures constructed from the source patterns. Evaluation compares submitted CIFs with reference structures, rather than requiring database identifiers, and separates ranked single-phase prediction from set-valued multiphase prediction (Methods). Mixtures generated from experimental profiles retain measured peak complexity but do not reproduce every effect present in a physically mixed specimen.

For single-phase identification, we evaluated Gan Jiang against Uni3DAR~\cite{lu2025uni3dar}, PXRDGen-Flow~\cite{li2025pxrdgen}, XtalNet-HMOF100~\cite{lai2025end} and AutoXRD~\cite{wu2026autoxrd}. Generative baselines use their available pretrained checkpoints. Under XRD-only inference, Uni3DAR predicts the elements required by its downstream model; its element predictor also supplies the inputs for PXRDGen-Flow and XtalNet-HMOF100. The composition-assisted condition instead supplies reference composition. AutoXRD uses the same candidate-library construction procedure as Gan Jiang, and we score its initial phase-identification output. PXRDGen-Flow is evaluated on eligible subsets satisfying its 100-atom limit.

Gan Jiang achieves the highest top-1, top-3, top-5 and MRR@5 scores on all three datasets in both input conditions (Table~\ref{tab:xrdbench-single}; Fig.~\ref{fig:cases}\textbf{e}). Its XRD-only top-1 accuracies of 96.30\%, 81.78\% and 40.83\% compare with 58.00\%, 58.47\% and 26.45\% for AutoXRD. The ranking advantage persists beyond the first candidate: MRR@5 reaches 96.94\%, 86.01\% and 42.22\%, versus 60.17\%, 60.89\% and 27.48\%. Supplied composition raises Gan Jiang's top-1 accuracies to 99.90\%, 92.16\% and 48.93\%. Thus the identification gains extend from simulated patterns to experimental measurements and remain substantial without reference composition.

\begin{table}[h]
\centering
\caption{\textbf{Single-phase structure identification across simulated and
experimental diffraction patterns.} Results are reported for end-to-end
XRD-only inference and XRD + oracle composition. All metrics are higher-is-better;
the best and second-best values within each dataset and condition are shown
in \textbf{bold} and \underline{underlined}, respectively. 
AutoXRD uses a fixed
candidate library (GanJiang dataset). PXRDGen-Flow's reduced RRUFF and opXRD subsets satisfy
its $\leq 100$-atom limit. PXRDGen and XtalNet abbreviate PXRDGen-Flow and
XtalNet-HMOF100 (944 RRUFF and 654 opXRD).}
\label{tab:xrdbench-single}
\scriptsize
\setstretch{1}
\setlength{\tabcolsep}{1.5pt}
\renewcommand{\arraystretch}{1.3}
\resizebox{\textwidth}{!}{%
\begin{tabular}{@{}lrrrr@{\hspace{0.55em}}rrrr@{}}
\toprule
& \multicolumn{4}{c}{XRD-only}
& \multicolumn{4}{c}{XRD + composition} \\
\cmidrule(lr){2-5}\cmidrule(l){6-9}
Model
& Top-1 $\uparrow$ (\%) & Top-3 $\uparrow$ (\%) & Top-5 $\uparrow$ (\%) & MRR@5 $\uparrow$ (\%)
& Top-1 $\uparrow$ (\%) & Top-3 $\uparrow$ (\%) & Top-5 $\uparrow$ (\%) & MRR@5 $\uparrow$ (\%) \\
\midrule
\multicolumn{9}{@{}l}{\textbf{MP500} ($N=10{,}000$)} \\
Uni3DAR
& 0.02 & 0.02 & 0.02 & 0.02
& 29.24 & 30.89 & 31.16 & 30.07 \\
PXRDGen
& 0.01 & 0.01 & 0.01 & 0.01
& 18.79 & 24.58 & 26.53 & 21.82 \\
XtalNet
& 0.00 & 0.00 & 0.00 & 0.00
& 0.00 & 0.00 & 0.20 & 0.04 \\
AutoXRD
& \underline{58.00} & \underline{61.90} & \underline{63.60} & \underline{60.17}
& \underline{96.20} & \underline{97.90} & \underline{98.30} & \underline{97.04} \\
Gan Jiang
& \textbf{96.30} & \textbf{97.60} & \textbf{97.70} & \textbf{96.94}
& \textbf{99.90} & \textbf{99.90} & \textbf{99.90} & \textbf{99.90} \\
\midrule
\multicolumn{9}{@{}l}{\textbf{RRUFF} ($N=1164$)} \\
Uni3DAR
& 0.00 & 0.00 & 0.00 & 0.00
& 9.22 & 9.85 & 10.06 & 9.55 \\
PXRDGen
& 0.00 & 0.00 & 0.00 & 0.00
& 4.58 & 7.67 & 8.58 & 6.13 \\
XtalNet
& 0.00 & 0.00 & 0.00 & 0.00
& 0.00 & 0.00 & 0.00 & 0.00 \\
AutoXRD
& \underline{58.47} & \underline{63.24} & \underline{64.41} & \underline{60.89}
& \underline{88.67} & \underline{93.86} & \underline{95.13} & \underline{91.23} \\
Gan Jiang
& \textbf{81.78} & \textbf{90.47} & \textbf{91.95} & \textbf{86.01}
& \textbf{92.16} & \textbf{95.02} & \textbf{95.23} & \textbf{93.64} \\
\midrule
\multicolumn{9}{@{}l}{\textbf{opXRD} ($N=880$)} \\
Uni3DAR
& 0.00 & 0.00 & 0.00 & 0.00
& 2.91 & 3.21 & 3.21 & 3.03 \\
PXRDGen
& 0.00 & 0.00 & 0.00 & 0.00
& 0.66 & 0.99 & 1.32 & 0.91 \\
XtalNet
& 0.00 & 0.00 & 0.00 & 0.00
& 0.00 & 0.00 & 0.00 & 0.00 \\
AutoXRD
& \underline{26.45} & \underline{28.13} & \underline{29.66} & \underline{27.48}
& \underline{44.04} & \underline{49.54} & \underline{52.60} & \underline{47.08} \\
Gan Jiang
& \textbf{40.83} & \textbf{43.43} & \textbf{44.19} & \textbf{42.22}
& \textbf{48.93} & \textbf{51.68} & \textbf{54.28} & \textbf{50.72} \\
\bottomrule
\end{tabular}%
}
\end{table}

For multiphase identification, we additionally compared AutoAnalyzer~\cite{yadav2026automated} and SimonnetCNN (S-CNN)~\cite{simonnet2024phase}, whose pipelines can be retrained on the Gan Jiang crystal dataset. Each evaluation samples 1,000 mixtures per source, and reported metrics are means across three fixed-seed runs. Exact-set recovery requires every reference phase to be found with no extra phase; sample-wise macro precision, recall and F1 measure partial recovery. Prediction coverage records only whether a nonempty valid output was returned. In the composition-assisted condition, Gan Jiang and AutoXRD receive chemical elements at inference, whereas oracle per-phase compositions post-filter the frozen XRD-only predictions of AutoAnalyzer and S-CNN.

\begin{table}[H]
\centering
\caption{\textbf{Multiphase identification across simulated and experimental
diffraction patterns.} Each dataset contains 1,000 hidden mixtures.
Results are reported under XRD-only and composition-assisted conditions.
Higher values are better for all metrics; the best and second-best values
within each dataset and condition are shown in \textbf{bold} and
\underline{underlined}, respectively. Cov. denotes the percentage of
mixtures with at least one valid predicted phase after parsing,
deduplication and any declared composition filtering. Exact denotes complete phase-set
recovery. Exact, precision (P), recall (R), and F1 are evaluated over all
1,000 mixtures in each run. P, R, and F1 are sample-wise macro
averages.}
\label{tab:xrdbench-multi}
\scriptsize
\setlength{\tabcolsep}{3pt}
\renewcommand{\arraystretch}{1.25}
\resizebox{\linewidth}{!}{%
\begin{tabular}{@{}lrrrrr@{\hspace{0.8em}}rrrrr@{}}
\toprule
& \multicolumn{5}{c}{XRD-only}
& \multicolumn{5}{c}{XRD + composition} \\
\cmidrule(lr){2-6}\cmidrule(l){7-11}
Model
& Cov. $\uparrow$ (\%) & Exact $\uparrow$ (\%) & P $\uparrow$ (\%) & R $\uparrow$ (\%) & F1 $\uparrow$ (\%)
& Cov. $\uparrow$ (\%) & Exact $\uparrow$ (\%) & P $\uparrow$ (\%) & R $\uparrow$ (\%) & F1 $\uparrow$ (\%) \\
\midrule
\multicolumn{11}{@{}l}{\textbf{MP500} ($N=1{,}000$)} \\
AutoAnalyzer
& \textbf{100.00} & 1.70 & 46.49 & \underline{58.08} & \underline{51.12}
& 95.63 & 23.37 & 85.08 & 58.08 & 69.25 \\
S-CNN
& \textbf{100.00} & 1.27 & 40.11 & 18.99 & 25.06
& 43.17 & 1.60 & 42.46 & 18.99 & 25.72 \\
AutoXRD
& \underline{92.40} & \underline{15.80} & \underline{50.22} & 40.48 & 43.24
& \underline{98.50} & \underline{59.00} & \underline{85.57} & \underline{80.12} & \underline{81.75} \\
Gan Jiang
& \textbf{100.00} & \textbf{22.80} & \textbf{76.85} & \textbf{64.77} & \textbf{66.98}
& \textbf{100.00} & \textbf{69.70} & \textbf{94.97} & \textbf{89.68} & \textbf{91.61} \\
\midrule
\multicolumn{11}{@{}l}{\textbf{RRUFF} ($N=1{,}000$)} \\
AutoAnalyzer
& \textbf{100.00} & 6.40 & \underline{56.55} & \underline{51.62} & \underline{53.78}
& \underline{98.97} & 20.03 & \underline{75.85} & \underline{69.62} & \underline{70.13} \\
S-CNN
& \textbf{100.00} & 2.13 & 49.74 & 25.17 & 32.34
& 64.57 & 2.47 & 53.12 & 25.17 & 33.33 \\
AutoXRD
& \underline{92.40} & \underline{13.80} & 48.70 & 37.83 & 41.07
& 98.40 & \underline{35.10} & 74.45 & 65.37 & 68.17 \\
Gan Jiang
& \textbf{100.00} & \textbf{15.90} & \textbf{59.90} & \textbf{54.55} & \textbf{56.23}
& \textbf{100.00} & \textbf{38.20} & \textbf{81.23} & \textbf{78.15} & \textbf{78.56} \\
\midrule
\multicolumn{11}{@{}l}{\textbf{opXRD} ($N=1{,}000$)} \\
AutoAnalyzer
& \textbf{100.00} & 0.00 & 18.42 & \underline{22.73} & \underline{20.14}
& 51.83 & 1.70 & \underline{39.24} & 22.54 & 27.02 \\
S-CNN
& \textbf{100.00} & 0.07 & 9.68 & 5.82 & 6.91
& 31.93 & 0.13 & 12.73 & 5.82 & 7.79 \\
AutoXRD
& \underline{90.20} & \underline{1.30} & \underline{20.77} & 13.12 & 15.31
& \underline{97.70} & \underline{5.20} & 37.73 & \underline{27.77} & \underline{30.55} \\
Gan Jiang
& \textbf{100.00} & \textbf{3.10} & \textbf{28.63} & \textbf{26.47} & \textbf{27.04}
& \textbf{100.00} & \textbf{8.20} & \textbf{39.83} & \textbf{37.90} & \textbf{38.21} \\
\bottomrule
\end{tabular}%
}
\end{table}

Gan Jiang leads exact-set recovery, precision, recall and macro-F1 under both conditions on all three sources (Table~\ref{tab:xrdbench-multi}; Fig.~\ref{fig:cases}\textbf{f}). Without supplied composition, its exact-set recovery is 22.80\%, 15.90\% and 3.10\% on MP500, RRUFF and opXRD, compared with 15.80\%, 13.80\% and 1.30\% for the next-best method, AutoXRD. Its corresponding macro-F1 scores are 66.98\%, 56.23\% and 27.04\%; the strongest alternatives are 51.12\%, 53.78\% and 20.14\%, all from AutoAnalyzer. With composition assistance, Gan Jiang reaches exact-set recovery of 69.70\%, 38.20\% and 8.20\%, and macro-F1 of 91.61\%, 78.56\% and 38.21\%. It returns a nonempty prediction for every evaluated mixture, although coverage alone does not establish correctness.

The consistent ranking across conditions supports the breadth of Gan Jiang's identification capability. The absolute scores also delimit it: many opXRD measurements cover incomplete angular ranges, and exact recovery of all mixture constituents remains difficult even with composition information (Supplementary Fig.~\ref{fig:xrdbench-overview}). 

\section*{Discussion}

Gan Jiang makes the procedure connecting a measurement to a structural conclusion an explicit object of learning. It integrates diffraction-analysis tools with a persistent library of executable skills, allowing analytical experience to inform subsequent analyses. The applications and identification benchmarks establish the system's analytical capabilities, while held-out comparisons of frozen workflow versions demonstrate that learned skills improve refinement beyond the original expert-designed procedures.

The applications illustrate the value of this approach beyond routine phase identification. Archaeological materials, operating batteries and disordered catalysts require decisions about which structural hypotheses the available evidence can support. Recording these decisions and their supporting evidence provides a basis for improving analytical procedures and, ultimately, choosing what to measure next. An unresolved reflection, for example, may warrant a targeted scan to distinguish competing hypotheses.

The incomplete patterns common in opXRD highlight the need to extend learning from interpretation to data acquisition: structural ambiguity may persist because the distinguishing information was never measured. A next step for Gan Jiang is therefore to close the loop between analysis and experiment. By identifying the missing or uncertain diffraction features that distinguish competing hypotheses, the agent could direct a diffractometer to acquire the necessary data, reassess its interpretation and incorporate the resulting experience into its analytical skills. Prospective experiments will test whether this feedback improves experimental efficiency and structural inference. The longer-term goal is a scientific agent that learns both how to interpret measurements and which measurements to make.

\section*{Data availability}

The DeltaXRDbench benchmark, including data resources and evaluation tools, is
available at \url{https://github.com/DeltaGanjiang/deltaxrdbench}.

\section*{Code availability}
The publicly available portions of the code are hosted on GitHub at \url{https://github.com/DeltaGanjiang}.

\clearpage
\section*{Methods}
\subsection*{Gan Jiang agent}

Gan Jiang is a self-learning diffraction-analysis agent that connects a language
model to explicit, versioned scientific skills and executable tools. It accepts
an observed PXRD pattern, a user objective and optional compositional,
structural or instrumental information. The orchestration layer validates these
inputs, maintains the evolving task state and routes the analysis through
candidate retrieval, phase-set construction, evidence inspection and, when
requested, structural refinement. XMatcher and XQueryer provide complementary
single-phase retrieval channels, XDecomposer and  XMatcher-AutoMix support multiphase analysis,
PyWPEM, FullProf and GSAS-II provide independent whole-pattern or refinement
branches, and XRDinspector evaluates the resulting evidence.

Reusable procedures are stored outside the underlying language model as skill
specifications, prompts and supporting code. Adaptation changes these artefacts
while the underlying language-model parameters remain fixed. Execution traces, diffraction
agreement, structural plausibility and expert feedback are used to attribute
failures and propose targeted skill revisions; candidate revisions are
versioned and can be tested before adoption. This design separates scientific
decision policies from model parameters and preserves the provenance of both
successful and failed analyses. The agent architecture, task-state definition,
skill routing and adaptation procedure are described in Supplementary Methods,
Section~\ref{sec:ganjiang}.

\subsection*{Evaluation of self-improvement}
\label{method:self-improvement}

Workflow improvement was evaluated separately for FullProf, GSAS-II and PyWPEM.
Each branch began with an expert-optimized baseline. Training episodes provided
execution traces and refinement diagnostics for proposing revisions to skill
instructions, supporting code or permitted settings. Candidate revisions were
compared with the current version through paired development evaluation; held-out
samples were not used to propose or select revisions. After selection, the skill
version was frozen and compared with the initial workflow on the held-out set.
Failed or timed-out evaluations contributed zero to the mean score. Development
trajectories and held-out initial/final comparisons are reported separately in
Fig.~\ref{fig:self-evolving}. The FullProf workflow revisions, GSAS-II joint
optimization and PyWPEM skill optimization have different adaptation scopes;
their score changes should therefore be interpreted within each branch.
The reported comparisons assess transfer of the final workflow, rather than
isolating the contribution of any individual edit. Selected before-and-after
skill excerpts are provided in Supplementary Section~\ref{sec:skill}.

\subsection*{Gan Jiang crystal dataset}

The Gan Jiang candidate library combines MP500 with experimentally derived
structures. MP500 is an ASE-compatible SQLite database containing 100,315
periodic crystal structures across 84 elements. Each record stores the atomic
structure and a unique Materials Project-style identifier. Stored cells contain
1--360 atoms (median 20). 

The experimental collection contains 2,007 complete, parseable periodic
structures: 1,159 from RRUFF and 848 from a combined opXRD and private-data
branch. It spans 80 elements and 141 space groups, with stored cells containing
2--496 atoms (median 54). Source identifiers, original CIF filenames and
SHA-256 checksums preserve provenance and support exact duplicate detection. Gan Jiang also supports
external ICSD~\cite{hellenbrandt2004inorganic} and CCDC~\cite{groom2014cambridge} databases through plugins. Dataset composition and structural
statistics are reported in Supplementary Results,
Section~\ref{sec:mp500}.

\subsection*{XRD Benchmark}
\label{method:delatxrdbench}

We developed DeltaXRDbench, a model-agnostic benchmark for phase identification
and refinement assessment across simulated and experimental powder-XRD data.
The benchmark contains 62,044 patterns: 40,000 derived from simulated MP500 data
\cite{cao2025simxrd}, 11,164 derived from experimental RRUFF data
\cite{lafuente2015rruff} and 10,880 derived from experimental opXRD data
\cite{hollarek2026opxrd}. These comprise 12,044 single-phase patterns and
50,000 deterministic two- or three-phase mixtures. All patterns are represented
on a common $10^{\circ}$--$80^{\circ}$ $2\theta$ grid at $0.01^{\circ}$
intervals and processed using baseline subtraction, non-negative clipping
and integrated-intensity normalization.

DeltaXRDbench evaluates ranked single-phase identification, set-valued
multiphase identification and numerical agreement between observed and
calculated or refined profiles. Submitted CIFs are parsed, structurally
deduplicated and compared with the reference structures using the benchmark
\texttt{StructureMatcher}. Multiphase predictions are evaluated by
maximum-cardinality one-to-one structural matching, so neither a predicted
structure nor a reference structure can be counted twice.

For both single-phase and multiphase identification, let $\mathcal I$ denote
the complete eligible evaluation set for a stated method, dataset and input
condition, with $N=|\mathcal I|$. This set is fixed independently of whether
a method returns a valid prediction. Let
$\mathcal V\subseteq\mathcal I$ contain samples with at least one valid
prediction after parsing, structural deduplication and any declared
post-processing. Prediction coverage is
\begin{equation}
\mathrm{Coverage}=100\frac{|\mathcal V|}{N}.
\label{eq:xrdbench-coverage}
\end{equation}
All identification metrics in both tracks are evaluated over $\mathcal I$.
Missing, empty, or invalid candidate lists or phase sets contribute zero and
remain in the denominator $N$. The subset $\mathcal V$ is used only as the
numerator of prediction coverage, never as the scoring denominator.

For single-phase sample $i$, let $S_i$ be the reference structure and
$\widehat{\mathcal S}_i=(\hat S_{i,1},\ldots,\hat S_{i,K_i})$ the ordered,
structurally deduplicated candidate list, with $0\leq K_i\leq5$. Define
$h_i^{(k)}=1$ if at least one of the first $\min(k,K_i)$ candidates
structurally matches $S_i$, and zero otherwise. Let $r_i$ be the rank of the
first matching candidate, with $r_i=\infty$ and $1/\infty=0$ when no match
exists. Missing, empty or invalid candidate lists have $K_i=0$ and contribute
zero to every single-phase metric. The aggregate scores are
\begin{equation}
\mathrm{Top\text{-}}k
=\frac{100}{N}\sum_{i\in\mathcal I}h_i^{(k)},
\quad k\in\{1,3,5\},
\qquad
\mathrm{MRR@5}
=\frac{100}{N}\sum_{i\in\mathcal I}\frac{1}{r_i}.
\label{eq:xrdbench-single-metrics}
\end{equation}

For multiphase sample $i$, let $p_i$ and $t_i$ be the sizes of the
structurally deduplicated predicted and reference phase sets, respectively,
with $t_i>0$, and let $m_i$ be the cardinality of their maximum one-to-one
structural matching. The per-sample scores are
\begin{equation}
\label{eq:xrdbench-multiphase-metrics}
P_i=
\begin{cases}
m_i/p_i,&p_i>0,\\
0,&p_i=0,
\end{cases}
\qquad
R_i=\frac{m_i}{t_i},\qquad
F1_i=\frac{2m_i}{p_i+t_i},\qquad
E_i=\mathbf{1}[m_i=p_i=t_i].
\end{equation}
Missing, empty or invalid phase sets have $p_i=m_i=0$ and therefore receive
zero for precision, recall, F1 and exact-set recovery. These metrics are
macro-averaged over all eligible samples:
\begin{equation}
(P,R,F1,\mathrm{Exact})
=\frac{100}{N}\sum_{i\in\mathcal I}(P_i,R_i,F1_i,E_i).
\label{eq:xrdbench-multiphase-aggregate}
\end{equation}
Oracle composition filtering does not change $\mathcal I$. 

For multiphase evaluation, we sample 1,000 mixtures per dataset using each
of three fixed seeds. Metrics are computed separately over each seed's
complete eligible evaluation set and reported as the unweighted mean of
the three seed-level values; predictions are not pooled across seeds.
Coverage is averaged in the same way.

Refinement assessment reports $R_p$, $R_{wp}$, whole-pattern correlation
and a composite diagnostic score. Dataset construction, structural-matching
tolerances, empty-output handling, worked metric examples and reproducibility
artifacts are detailed in Supplementary Methods,
Section~\ref{sec:xrdbench}.

\subsection*{XMatcher}

XMatcher follows a search–match strategy. It constructs a theoretical peak library from Gan Jiang crystal structures, validates and preprocesses experimental profiles, detects peaks, and retrieves candidate phases subject to explicit chemical constraints. For each candidate, XMatcher searches over a bounded global peak shift and determines a one-to-one assignment between experimental and calculated reflections using a cost function that combines peak-position and relative-intensity discrepancies. Candidates are ranked according to matched-peak precision, theoretical-peak recall, intensity coverage, and mean assignment quality. The corresponding peak assignments, unmatched theoretical reflections, and residual experimental peaks are retained for expert inspection.

For multiphase analysis, AutoMix forms small combinations of the leading
single-phase candidates and estimates their non-negative contributions to the
retained experimental peaks, with an explicit penalty for unnecessary phases.
Element sets or structure identifiers can be imposed as search constraints.
The fitted contributions quantify shares of retained diffraction evidence and
are not interpreted as Rietveld-refined mass or volume fractions. Library
construction, peak detection, matching, ranking and AutoMix fitting are
detailed in Supplementary Methods, Section~\ref{sec:xmatcher}.

\subsection*{XQueryer lightweight}

XQueryer Lightweight is a compact neural retrieval model for rapid
single-phase identification against the 100,315-structure MP500 label space
\cite{cao2025xqueryer}. Each pattern is interpolated to a fixed angular grid
and represented by the original trace together with three low-pass Fourier
reconstructions. A depthwise-separable one-dimensional convolutional encoder
extracts multiscale peak features, and compact element-guided queries attend to
the resulting tokens. The fused representation is passed to a normalized cosine classifier, which produces a ranked list of candidate structures. Unlike the original XQueryer, this model requires elemental information at inference because it is trained for the element-aware setting used in the Gan Jiang system. Details of the simulations, architecture, training, and inference are provided in Supplementary Methods, Section~\ref{sec:xqueryer}.

\subsection*{XDecomposer}

XDecomposer separates a multiphase pattern into a set of component diffraction
profiles before identification \cite{gao2026xdecomposer}. A hierarchical
convolutional encoder and a frozen Transformer pretrained by masked
single-phase reconstruction provide local and global pattern features.
Learnable phase queries predict slot activity and modulate a symmetric decoder,
which generates input-masked component profiles and supports an unknown phase
count up to four. Training uses permutation-invariant component assignment,
slot-activity supervision and mixture-reconstruction consistency, so no
predefined candidate list or phase ordering is required. Data generation,
network design, losses and inference are described in Supplementary Methods,
Section~\ref{sec:XDecomposer}.

\subsection*{WPEM}

PyWPEM, implemented in PyXplore, performs statistical whole-pattern
decomposition for strongly overlapping diffraction signals \cite{cao2026ai}.
It represents a background-subtracted profile as a non-negative mixture of
pseudo-Voigt peak-shape functions and estimates soft component responsibilities
with an expectation--maximization procedure. Bragg-position and profile
constraints couple the statistical mixture to the crystallographic model,
while soft assignments allow coincident reflections to share observed
intensity without introducing negative peak contributions. The fitted peak
positions, shapes and integrated intensities support subsequent refinement and
phase-fraction analysis. The probabilistic formulation, constrained update
rules and fraction calculation are provided in Supplementary Methods,
Section~\ref{sec:PyXplore}.

\subsection*{XRDinspector}

XRDinspector is an evidence-screening layer for phase-identification and
refinement outputs. In identification mode, it compares observed peaks with
reflections calculated from one or more submitted CIFs and reports
intensity-weighted observed coverage, predicted-peak precision, per-phase
support and, for mixtures, non-negative intensity agreement. In refinement
mode, it aligns observed and calculated profiles and reports $R_p$, $R_{wp}$,
whole-pattern correlation and local-residual diagnostics. A conservative
four-level policy converts these diagnostics into delivery decisions while
retaining the underlying matches, warnings and settings; acceptance indicates
profile consistency, not structural uniqueness. Scoring rules and default
thresholds are specified in Supplementary Methods,
Section~\ref{sec:XRDinspector}.

\section*{Acknowledgements}

We thank all members of the Gan Jiang team at AI Lab, Yangtze River Delta, for their contributions to the development of the Gan Jiang system and to this work (see the team page at \url{https://ganjiang.asia/team.html}). Funding information will be provided upon acceptance of the manuscript.

\section*{Author contributions}

B.C. led and supervised the project, developed the physics-based analysis tools and designed the agent workflow and benchmark.
H.Z., R.Y. and J.L. developed the agent learning mechanisms and the analysis workflows.
S.S. performed the benchmark evaluations.
Y.S. developed the API infrastructure.
H.G. developed XDecomposer for multiphase decomposition.
Z.Y. conducted system testing and contributed to benchmark evaluation.
T.-Y.Z. and J.W. supervised the research and contributed to refining the study.

\section*{Competing interests}

The authors declare no competing interests.

\section*{Additional information}

Supplementary information is available for this paper. Correspondence and
requests for materials should be addressed to Corresponding Author.

\newpage
\putbib
\end{bibunit}

\clearpage
\begin{bibunit}
\setcounter{section}{0}
\setcounter{figure}{0}
\setcounter{table}{0}
\setcounter{equation}{0}
\renewcommand{\thesection}{Supplementary Note \arabic{section}}
\renewcommand{\thesubsection}{\arabic{section}.\arabic{subsection}}
\renewcommand{\thefigure}{S\arabic{figure}}
\renewcommand{\thetable}{S\arabic{table}}
\renewcommand{\theequation}{S\arabic{equation}}
\renewcommand{\theHsection}{supp.\arabic{section}}
\renewcommand{\theHfigure}{supp.\arabic{figure}}
\renewcommand{\theHtable}{supp.\arabic{table}}
\renewcommand{\theHequation}{supp.\arabic{equation}}

\SupplementCoverPage

\begingroup
\setcounter{tocdepth}{2}
\renewcommand{\contentsname}{Supplementary contents}
\makeatletter
\renewcommand*{\l@section}{\@dottedtocline{1}{0em}{11em}}
\makeatother
\begin{spacing}{1.15}
\tableofcontents
\end{spacing}
\endgroup

\clearpage
\section{Supplementary methods}

\subsection{Gan Jiang agent}
\label{sec:ganjiang}

\paragraph{Architecture and skill representation.}
Gan Jiang backbone stores reusable skills as external memory.
Each skill combines a declarative specification
(\texttt{SKILL.md}), prompts and supporting executable code.
An orchestration layer coordinates the LLM, context management,
tools and skill library. During deployment, adaptation modifies
these external artefacts while keeping the underlying LLM
parameters fixed.

In Gan Jiang, domain-specific skills connect the analysis
objective to the available diffraction tools. The toolchain
includes XTranscoder for data conversion, XQueryer and XMatcher
for phase identification, XDecomposer for multiphase decomposition,
and PyWPEM, GSAS-II and FullProf for refinement. A skill specifies
how a tool is used within an analysis procedure, including the
required inputs, relevant parameters and interpretation of its
outputs. This organisation allows the same tool to participate
in different workflows according to the sample, available
information and analysis objective.

\paragraph{Task state and sequential analysis.}
An XRD analysis episode begins with diffraction data and a user
objective, such as identifying constituent phases or refining
a candidate structure. For the present description, the evolving
analysis state is represented as
\begin{equation}
s_t =
\left(
q,\mathcal{D},\mathcal{C},\mathcal{H}_t,\mathcal{O}_t
\right),
\label{eq:ganjiang-state}
\end{equation}
where $q$ denotes the objective, $\mathcal{D}$ the supplied data,
$\mathcal{C}$ the available experimental or compositional
constraints, $\mathcal{H}_t$ the analysis history, and
$\mathcal{O}_t$ the intermediate outputs at step $t$.
Experimental constraints may include the radiation wavelength,
measurement range and known elements, when supplied.
Intermediate outputs can include candidate phases, calculated
patterns, refined structures and residual profiles.

This state representation separates the requested scientific
conclusion from the information currently available to support
it. For example, identifying a plausible candidate does not
complete a task that also requires structural refinement.
Similarly, a successful refinement run does not by itself
establish that all phases in a mixture have been identified.
Subsequent actions therefore depend on both the requested
endpoint and the evidence accumulated during analysis.

\paragraph{Skill retrieval and routing.}
The backbone combines sparse BM25 and dense retrieval, merges
candidates through score-aware reciprocal rank fusion, and
optionally applies a cross-encoder reranker. Its trainable router
uses synthetic positive queries and hard negatives with a
multi-positive contrastive objective to favour behavioural
compatibility.

For diffraction analysis, behavioural compatibility concerns
whether a procedure can address the current analytical question
with the available inputs. Phase identification, quantitative
phase analysis and structural refinement may involve similar
terminology but require different input artefacts and execution
steps. Routing must therefore account for the analysis stage.
For instance, a refinement procedure requires an initial
structural model, whereas an identification procedure may be
needed to obtain that model from the measured pattern.

\paragraph{Tool execution and diffraction evidence.}
Selected skills guide the preparation of tool inputs, execution
of calculations and interpretation of returned results.
The relevant outputs depend on the task: phase identification
produces candidate assignments, decomposition estimates
phase-resolved contributions, and refinement updates the
parameters of a structural or profile model.

For describing the the evidence available to the agent, a
multiphase calculated pattern can be written schematically as
\begin{equation}
I_{\mathrm{calc}}(x)
=
B(x;\boldsymbol{\beta})
+
\sum_{k=1}^{K}
s_k I_k(x;\boldsymbol{\phi}_k,\boldsymbol{\eta}),
\label{eq:ganjiang-pattern}
\end{equation}
where $x=2\theta$, $B$ is the background, $K$ is the number of
modelled phases, $s_k$ is a phase scale factor,
$\boldsymbol{\phi}_k$ contains phase-specific parameters, and
$\boldsymbol{\eta}$ denotes shared profile or instrumental
parameters. This expression provides a common description of
the calculated contributions; the detailed parameterisation
depends on the refinement engine.

The residual
\begin{equation}
\Delta I(x)
=
I_{\mathrm{obs}}(x)-I_{\mathrm{calc}}(x)
\label{eq:ganjiang-residual}
\end{equation}
provides information beyond an aggregate agreement factor.
Unexplained peaks may motivate reconsideration of the phase
assignment, whereas systematic peak-position discrepancies
may motivate examination of calibration or lattice parameters.
These are diagnostic possibilities rather than unique
assignments of an error source. The interpretation must also
consider the experimental conditions and the current model.

\paragraph{Failure attribution and skill revision.}
In the backbone, unsuccessful execution triggers trace-based
failure attribution, followed by targeted skill rewriting.
Persistent low utility can instead trigger skill restructuring
or discovery. Utility is estimated from observed successes and
failures. An optional test gate evaluates proposed mutations
and rolls them back when validation fails.

Within the diffraction workflow, failure attribution should
distinguish an execution problem from an inadequate scientific
model. A malformed input file, an unsuccessful optimisation
and an incomplete phase assignment require different remedies.
Likewise, changing a parameter for one sample is distinct from
revising a reusable analysis procedure. A transferable skill
update should capture the conditions under which a corrective
action is appropriate, rather than preserve a sample-specific
answer as a general rule.

\paragraph{Feedback and scope of adaptation.}
For Gan Jiang, feedback can be organised into execution status,
diffraction agreement, structural plausibility and expert
assessment. These components answer different questions:
whether a calculation completed, whether it reproduces the
observations, whether its parameters remain physically
credible, and whether the interpretation is supported by
additional knowledge. Expert corrections can further identify
limitations that are not captured by a numerical residual.

Evaluation of skill adaptation should distinguish improvement
on a revisited sample from transfer to previously unseen
samples. The skill-library version used for each evaluation
should therefore be specified, together with whether updates
are enabled during testing. This distinction is necessary to
separate the benefit of additional attempts on an individual
pattern from a reusable improvement in the analysis procedure.

\subsection{DeltaXRDbench: a reproducible benchmark}
\label{sec:xrdbench}

XRD encodes information about crystalline phases and is widely used to identify
materials and validate structural models. Automated analysis must nevertheless
contend with peak overlap, instrumental effects, background, noise and
multiphase mixtures. Large materials databases
\cite{hollarek2026opxrd,lafuente2015power,jain2020materials} and software
libraries \cite{cao2025simxrd,ong2013python} have accelerated structure-based
analysis, but evaluation protocols for XRD phase identification remain
fragmented. In particular, database-specific identifiers can reward
memorization and hinder fair assessment of methods that predict crystal
structures directly.

We present DeltaXRDbench, a reproducible benchmark that unifies simulated and
experimental XRD data under a common representation and submission contract.
DeltaXRDbench is designed to evaluate single- and multiphase identification, accept
predicted structures independently of any specific phase database, and make
data construction and scoring fully inspectable. A complementary refinement
track quantifies agreement between calculated or refined patterns and
experimental measurements. The benchmark comprises 62,044 standardized XRD
patterns, including 50,000 multiphase examples; CIF-based structural matching
with Top-1, Top-3, Top-5 and MRR@5 for ranked single-phase predictions and
exact-set, sample-wise macro precision, recall, F1 and coverage for multiphase
predictions; a refinement-quality protocol
reporting $R_p$, $R_{wp}$, correlation and a continuous score; and reproducible
dataset builders, JSONL manifests and a model-agnostic submission interface. Code and benchmark resources are
available at \url{https://github.com/DeltaGanjiang/deltaxrdbench}.

\paragraph{Tasks}
Each sample contains an input XRD pattern on a common grid. The single-phase
task accepts an ordered list of at most five predicted CIFs. The multiphase
task accepts an unordered predicted phase set; the reference mixtures contain
two or three phases, although composition filtering may retain fewer predicted
phases. In a hosted evaluation, the
reference CIFs for the test split are withheld, and the evaluator parses each
submission and performs one-to-one structural matching against the reference
phases. The refinement task receives an experimental pattern together with a
calculated or fitted pattern and evaluates their numerical agreement.

\begin{table}[h]
\centering
\caption{DeltaXRDbench task contract. A phase set is evaluated after one-to-one matching between submitted and reference structures.}
\label{tab:tasks}
\scriptsize
\begin{tabularx}{\linewidth}{@{}
  >{\centering\arraybackslash}p{0.26\linewidth}
  >{\centering\arraybackslash}p{0.26\linewidth}
  >{\centering\arraybackslash}X@{}}
\toprule
Track & Required prediction & Primary metrics \\
\midrule
Single-phase identification & Ranked list of at most five CIFs & Top-1, Top-3, Top-5; MRR@5 \\
Multi-phase identification & Unordered CIF set & coverage; exact set; macro precision, recall and F1 \\
Refinement & Calculated intensity pattern & $R_p$, $R_{wp}$, correlation, score \\
\bottomrule
\end{tabularx}
\end{table}

\begin{figure}[htbp]
  \centering
  \includegraphics[width=0.8\linewidth,height=0.72\textheight,keepaspectratio]{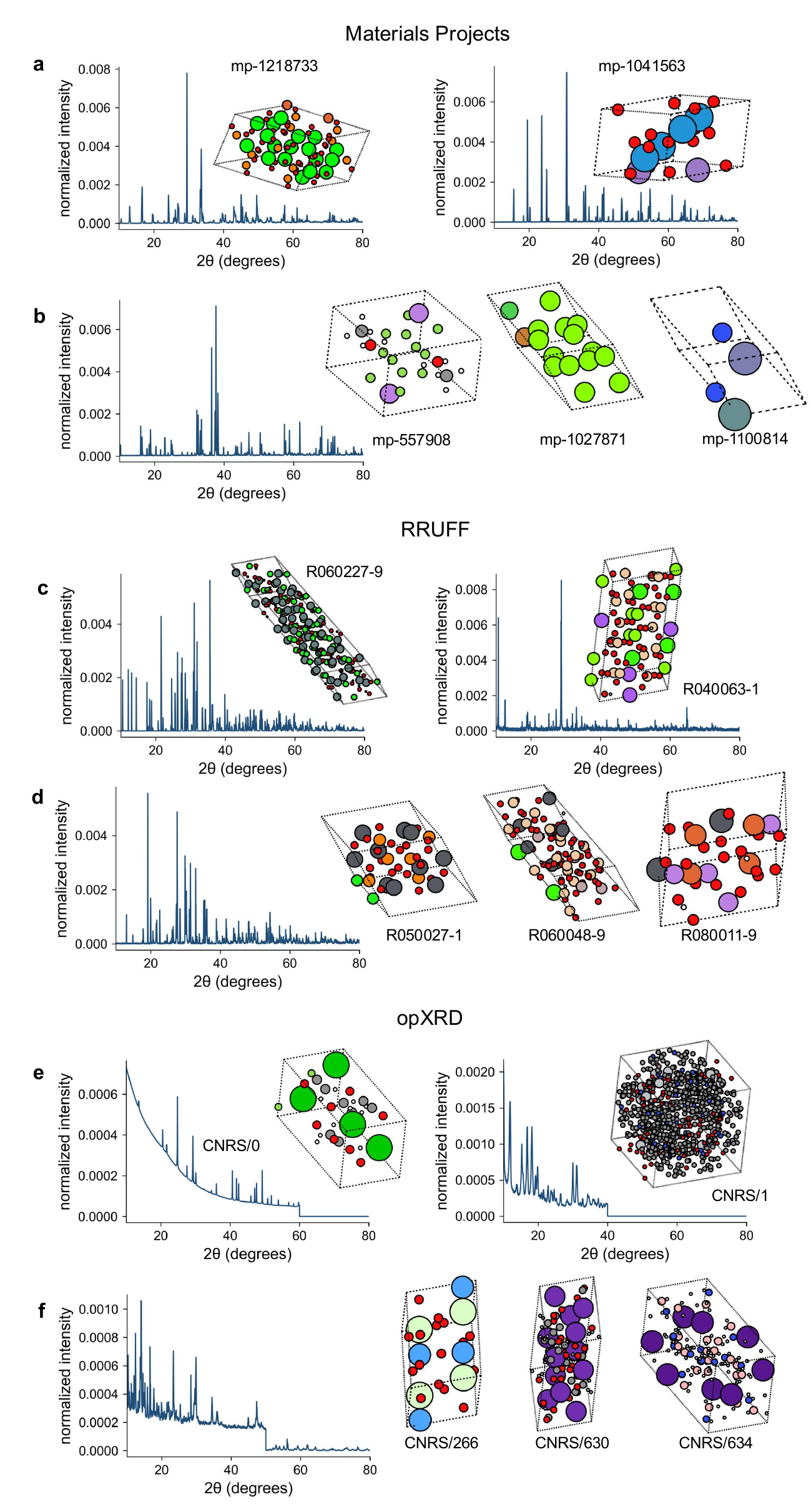}
  \caption{\textbf{Representative single- and multiphase samples in DeltaXRDbench.} \textbf{a,b}, Simulated MP500 patterns; \textbf{c,d}, experimental RRUFF patterns; and \textbf{e,f}, experimental opXRD patterns. \textbf{a,c,e} show representative single-phase examples, pairing each normalized powder-XRD profile with its crystal structure. \textbf{b,d,f} show representative multiphase mixtures, with the combined profile at left and the constituent phase structures at right. The horizontal axis is $2\theta$ (degrees) and the vertical axis is normalized intensity; labels identify the corresponding database records. Sphere colours distinguish atomic elements and dashed polyhedra indicate unit-cell boundaries.}
  \label{fig:xrdbench-overview}
\end{figure}

\paragraph{Data sources and standardization}
Table~\ref{tab:data} summarizes the current release. MP500 is a simulated
collection generated from crystal structures. Its Cu K$\alpha$ patterns
incorporate randomized peak broadening, small zero shifts, background signals
and intensity-dependent noise. The experimental RRUFF collection comprises
usable structure--pattern pairs from the RRUFF mineral database
\cite{lafuente2015rruff}, while opXRD provides additional experimental
structure--pattern pairs. A notebook for inspecting and validating sampled
data is available at
\url{https://github.com/DeltaGanjiang/deltaxrdbench/blob/main/explore_datasets.ipynb}.

All patterns are interpolated onto a common grid spanning
$10^\circ \leq 2\theta \leq 80^\circ$ at $0.01^\circ$ intervals (7,001
points). The first intensity percentile is subtracted as a baseline, after
which negative values are clipped to zero and each pattern is normalized to
unit integrated intensity. This standardized representation ensures consistent
input dimensionality and transparent preprocessing while preserving the
provenance of each raw pattern in the release metadata.

\begin{table}[t]
\centering
\caption{Dataset composition in the current DeltaXRDbench release. ``Single''
denotes single-phase patterns, whereas ``Multi'' denotes multiphase patterns
generated by mixing single-phase patterns.}
\label{tab:data}
\small
\begin{tabular}{@{}lrrr@{}}
\toprule
Source & Single & Multi & Total \\
\midrule
MP500 (simulated) & 10,000 & 30,000 & 40,000 \\
RRUFF (experimental) & 1,164 & 10,000 & 11,164 \\
opXRD (experimental) & 880 & 10,000 & 10,880 \\
\midrule
All sources & 12,044 & 50,000 & 62,044 \\
\bottomrule
\end{tabular}
\end{table}

\paragraph{Multiphase XRD generation}
For each multi-phase XRD, the builder samples either two or three distinct single-phase patterns without replacement. Fractions are sampled from a Dirichlet distribution after reserving a minimum fraction of 0.10 for every constituent phase. For $k \in \{2,3\}$ phases, the mixture is
\begin{equation}
I_{\mathrm{mix}}(2\theta) = \sum_{i=1}^{k} w_i I_i(2\theta), \qquad w_i \geq 0.10, \quad \sum_{i=1}^{k}w_i = 1.
\end{equation}
The random seed, grid, preprocessing settings, and mixture fractions are recorded in manifests. Mixtures probe peak overlap and set-valued prediction, but do not model every experimental nonlinearity, such as absorption, or microstructural interaction.

\paragraph{Task 1: Phase identification}
The evaluator reads a UTF-8 JSONL submission containing one record per sample.
Each record specifies a \texttt{sample\_id} and an ordered list of predicted
CIF files or canonical phase identifiers for the single-phase track, or an
unordered phase list with optional fractions for the multiphase track.
Predictions are resolved to structures before scoring; database identifiers
themselves are never used as the correctness criterion. A CIF that cannot be
parsed is invalid. Structurally equivalent duplicates are collapsed, retaining
the earliest occurrence in a ranked single-phase list. Fractions affect only
whether a multiphase candidate is retained during the specified filtering step;
they do not weight the identification scores.

The evaluator uses the benchmark \texttt{StructureMatcher} implementation
\cite{ong2013python}, with lattice-length, site and angular tolerances of 0.2,
0.3 and $5^\circ$, respectively. Define
\begin{equation}
\mathrm{SM}(A,B)=
\begin{cases}
1,&\text{if structures $A$ and $B$ are equivalent under these settings},\\
0,&\text{otherwise}.
\end{cases}
\label{eq:xrdbench-supp-sm}
\end{equation}
This criterion permits crystallographically equivalent cell descriptions while
requiring the benchmark's site and lattice agreement. For a multiphase sample,
all predicted--reference pairs satisfying $\mathrm{SM}=1$ form the edges of a
bipartite graph. A maximum-cardinality matching on this graph supplies the
number of correct phases, so neither one predicted CIF nor one reference CIF
can be counted twice.

For both single-phase and multiphase identification, let $\mathcal I$ denote
the complete eligible evaluation set for a stated method, dataset and input
condition, with $N=|\mathcal I|$. This set is fixed independently of whether
a method returns a valid prediction. Let $\mathcal V\subseteq\mathcal I$
contain the samples for which at least one valid prediction remains after
parsing, deduplication and any declared composition filter. Prediction
coverage is
\begin{equation}
\mathrm{Coverage}=100\frac{|\mathcal V|}{N}.
\label{eq:xrdbench-supp-coverage}
\end{equation}
All samples in $\mathcal I$ contribute to the denominators of single-phase
Top-1, Top-3, Top-5 and MRR@5, and multiphase exact-set recovery and
macro-averaged precision, recall and F1. Missing, empty, or invalid candidate
lists or phase sets contribute zero to every identification metric in the
corresponding track and remain in $\mathcal I$. This rule also applies when
no valid candidate remains after parsing, deduplication and any declared
composition filter. The subset $\mathcal V$ is used only as the numerator
of prediction coverage, never as the scoring denominator. For the multiphase
results in Table~\ref{tab:xrdbench-multi}, $N=1{,}000$ for each dataset,
input condition and run.

\paragraph{Single-phase identification metrics}
Each eligible single-phase sample $i\in\mathcal I$ has one reference structure $S_i$ and an
ordered, structurally deduplicated candidate list
\begin{equation}
\widehat{\mathcal S}_i=(\hat S_{i,1},\ldots,\hat S_{i,K_i}),
\qquad 0\leq K_i\leq5.
\end{equation}
Missing, empty, or invalid candidate lists are represented by $K_i=0$.
For cutoff $k\in\{1,3,5\}$, the sample is a hit if a correct structure occurs
within the first $k$ available candidates:
\begin{equation}
h_i^{(k)}=
\mathbf 1\!\left[\exists j\leq\min(k,K_i):
\mathrm{SM}(\hat S_{i,j},S_i)=1\right].
\label{eq:xrdbench-supp-hit}
\end{equation}
The first correct rank and truncated reciprocal-rank contribution are
\begin{align}
r_i&=\min\{j:\mathrm{SM}(\hat S_{i,j},S_i)=1\},
\quad r_i=\infty\ \text{if the set is empty},\\
q_i&=\begin{cases}
1/r_i,&r_i\leq5,\\
0,&r_i>5.
\end{cases}
\label{eq:xrdbench-supp-rank}
\end{align}
For $K_i=0$, $h_i^{(k)}=0$, $r_i=\infty$ and $q_i=0$; the sample remains
in the evaluation set. The dataset-level percentages over all $N$ eligible
samples are
\begin{align}
\mathrm{Top\text{-}}k
&=\frac{100}{N}
\sum_{i\in\mathcal I}h_i^{(k)},
\qquad k\in\{1,3,5\},\\
\mathrm{MRR@5}
&=\frac{100}{N}
\sum_{i\in\mathcal I}q_i.
\label{eq:xrdbench-supp-single-aggregate}
\end{align}
Thus Top-$k$ measures whether a correct structure is retrieved by rank $k$,
whereas MRR@5 also rewards placing it earlier. If the first match is at rank
two, that sample contributes 0 to Top-1, 1 to Top-3 and Top-5, and $1/2$ to
MRR@5. As a complete numerical example, suppose there are four eligible
samples: three return valid candidate lists with first-match ranks 1, 2 and
$\infty$, and the fourth returns no valid candidates. All four samples
remain in the denominator. Then Top-1 is $100(1/4)=25.00\%$, Top-3 and
Top-5 are each $100(2/4)=50.00\%$, MRR@5 is
$100(1+1/2+0+0)/4=37.50\%$, and coverage is $100(3/4)=75.00\%$.
Because there is exactly one
reference per sample, applying set-based Recall@1 or F1@1 to the top-ranked
candidate produces the same value as Top-1 and adds no independent information.

\paragraph{Multiphase identification metrics}
For each eligible mixture $i\in\mathcal I$, let
$\widehat{\mathcal T}_i$ be the predicted phase set after parsing,
structural deduplication and any declared composition filtering, and let
$\mathcal T_i$ be the non-empty reference phase set. If the submission is
missing or no valid candidate remains, set
$\widehat{\mathcal T}_i=\varnothing$. Denote their sizes by
\begin{equation}
p_i=|\widehat{\mathcal T}_i|,\qquad
t_i=|\mathcal T_i|>0,
\end{equation}
and let $m_i$ be the size of the maximum one-to-one structural matching defined
above. Hence $0\leq m_i\leq\min(p_i,t_i)$, with $m_i=0$ when $p_i=0$.
The per-sample precision, recall and F1 score are
\begin{equation}
\begin{aligned}
P_i&=\begin{cases}
\dfrac{m_i}{p_i}, & p_i>0,\\[3pt]
0, & p_i=0,
\end{cases}
& R_i&=\frac{m_i}{t_i},
& F1_i&=\frac{2m_i}{p_i+t_i}.
\end{aligned}
\label{eq:xrdbench-supp-multi-prf}
\end{equation}
Thus a missing, empty, or invalid prediction has $P_i=R_i=F1_i=0$;
precision is explicitly defined at $p_i=0$ to avoid division by zero.

Precision penalizes extra predicted phases, whereas recall penalizes missed
reference phases. Exact-set recovery is stricter:
\begin{equation}
E_i=\mathbf 1[m_i=p_i=t_i].
\label{eq:xrdbench-supp-multi-exact}
\end{equation}
It equals one only when every predicted phase is matched, every reference phase
is recovered and the two set sizes are equal. Merely matching the correct
number of phases is insufficient if any structural assignment is wrong.

The reported multiphase metrics are sample-wise macro averages over
all $N$ eligible mixtures, including failures:
\begin{align}
P_{\mathrm{macro}}&=\frac{100}{N}
\sum_{i\in\mathcal I}P_i,&
R_{\mathrm{macro}}&=\frac{100}{N}
\sum_{i\in\mathcal I}R_i,\\
F1_{\mathrm{macro}}&=\frac{100}{N}
\sum_{i\in\mathcal I}F1_i,&
\mathrm{Exact}&=\frac{100}{N}
\sum_{i\in\mathcal I}E_i.
\label{eq:xrdbench-supp-multi-aggregate}
\end{align}
Because $t_i>0$, an empty prediction also has $E_i=0$. For example,
if 800 of 1,000 eligible mixtures receive predictions and every returned
phase set is exactly correct, coverage, macro precision, macro recall,
macro F1 and exact-set recovery are all 80\%, not 100\%.

These are unweighted means of sample scores, not micro-averages formed by
pooling phase counts across samples. For example, consider a three-phase
reference $\mathcal T_i=\{A,B,C\}$. If the prediction is
$\widehat{\mathcal T}_i=\{A,B,D\}$ and $A$ and $B$ match one-to-one, then
$p_i=t_i=3$, $m_i=2$, $P_i=R_i=F1_i=2/3$, and $E_i=0$. If the prediction is
instead $\{A,B\}$, then $p_i=2$, $t_i=3$, $m_i=2$, giving $P_i=1$,
$R_i=2/3$, $F1_i=4/5$, and still $E_i=0$. The second prediction has perfect
precision but is not an exact identification because phase $C$ is missing.

\paragraph{Sampling Strategy for Multiphase Evaluation}

For the benchmark results reported in this paper, we randomly sample 1,000 patterns from the multiphase patterns to reduce the computational cost. All multiphase experiments are repeated using three fixed random seeds. For any metric $M$, the reported value is the arithmetic mean of the three
seed-level values computed independently:
\begin{equation}
\overline{M}=\frac{1}{3}\sum_{s=1}^{3}M^{(s)},
\label{eq:xrdbench-supp-seed-average}
\end{equation}
Each seed-level identification metric uses the complete set of
$N=1{,}000$ eligible mixtures as its denominator, with the zero-for-failure
convention defined above. Coverage is computed over the same set and
averaged across seeds separately. The predictions from the three runs are
not pooled before metric calculation.

\paragraph{Task 2: Refinement assessment}
The calculated pattern is linearly interpolated onto the experimental grid and
scaled using an optional non-negative global factor. Profile agreement is
quantified using
\begin{equation}
R_p = \frac{\sum_i |y_i^{\mathrm{obs}}-y_i^{\mathrm{calc}}|}{\sum_i |y_i^{\mathrm{obs}}|}
\end{equation}
alongside the weighted residual $R_{wp}$, for which approximate Poisson weights
are used when point-wise uncertainties are unavailable. Pearson correlation
provides a complementary measure of whole-pattern agreement. 

\paragraph{Reproducibility}
Each source directory contains \texttt{patterns.h5}, \texttt{manifest.jsonl}, and reference structures. HDF5 stores the common $2\theta$ grid and dense intensity arrays; manifests map a sample to its array location, task, source, and audit metadata. The reference implementation exposes builders and an evaluator through a Python package and command-line interface. A portable evaluation report contains aggregate metrics and sample-level evidence, including missing predictions and extra submission identifiers.

\subsection{XMatcher: search-and-match XRD phase identification}
\label{sec:xmatcher}
Search-and-match remains central to expert XRD analysis because it combines
chemical and experimental priors with reflection-level evidence
\cite{warren1990xSupplement}. This interpretability is particularly valuable
for samples affected by impurities, polymorphism, preferred orientation, peak
overlap or incomplete reference data. XMatcher therefore reports which
observations support or contradict each candidate and identifies residual
features requiring inspection.

XMatcher is a component of Gan Jiang. While its full implementation remains under closed-source development, a basic version is available to the community for matching experimental powder X-ray diffraction patterns against theoretical peak libraries. In contrast to established commercial packages, such as JADE~\cite{JADE2026}, HighScore~\cite{HighScorePlus2026}, Match!~\cite{Match2026}, DIFFRAC.EVA~\cite{DIFFRACEVA2026}, and PDXL~\cite{rigaku2010integrated}, the open version provides access to its reference data, candidate-generation procedure, and matching evidence, enabling inspection, reproducibility, and further development.

\paragraph*{Overview and notation}
The XMatcher workflow consists of theoretical library construction,
experimental profile preprocessing and peak detection, single-phase
candidate retrieval, AutoMix fitting of multiphase hypotheses, and
whole-pattern validation against the supplied structures.

We denote an experimental profile by
$D=\{(x_i,I_i)\}_{i=1}^{N}$, where $x_i=2\theta_i$ and $I_i\geq 0$.
The detected peaks retained for matching are
$E=\{(e_i,a_i)\}_{i=1}^{m}$, where $e_i$ and $a_i$ are the position
and intensity of peak $i$, respectively. For each candidate phase $q$,
the theoretical library stores
$T_q=\{(t_{qj},b_{qj},h_{qj},d_{qj})\}_{j=1}^{n_q}$, where
$t_{qj}$, $b_{qj}$, $h_{qj}$, and $d_{qj}$ denote the calculated
position, relative intensity, reflection family, and interplanar
spacing of reflection $j$, respectively.

\paragraph*{Theoretical pattern calculation}
For radiation of wavelength $\lambda$, a reflection from planes indexed by Miller indices $(hkl)$ and separated by $d_{hkl}$ occurs at Bragg angle $\theta_{hkl}$ according to Bragg's law,
\begin{equation}
2d_{hkl}\sin\theta_{hkl}=\lambda,
\qquad
2\theta_{hkl}=2\arcsin\left(\frac{\lambda}{2d_{hkl}}\right).
\end{equation}
XMatcher uses the pymatgen's XRD calculator, after converting each ASE
record to a pymatgen structure, to obtain ideal reflection intensities. Each
phase is normalized independently:
\begin{equation}
b_{qj}^{(\%)}=100\frac{b_{qj}}{\max_k b_{qk}},
\end{equation}
where $k$ spans the reflections of phase $q$. Measured intensities may differ
because of texture, absorption, extinction, crystallite size, fluorescence,
background, microstrain and instrumental response.

\paragraph*{Gan Jiang dataset indexing}
The offline builder converts \texttt{GanJiang.db} into
\texttt{GJ\_xrd.pkl}, storing identifiers, composition,
space-group metadata and calculated peaks. The released library uses Cu
K$\alpha$ radiation, $10^{\circ}\leq2\theta\leq90^{\circ}$, a minimum
$d$ spacing of $0.5$ \AA{} and at most 30 reflections per structure.

An exact index maps $\operatorname{sort}(\mathcal{E}_q)$ to phases, whereas an
inverted index maps each element to phases containing it. For query set $Q$,
exact and contains filtering return $\{q:\mathcal{E}_q=Q\}$ and
$\{q:Q\subseteq\mathcal{E}_q\}$, respectively. In unconstrained AutoMix, each
component may contain any subset of the permitted elements, allowing, for
example, TiO$_2$ and SiO$_2$ under the joint scope
$\{\mathrm{Ti,Si,O}\}$.

\paragraph*{Experimental peak detection}
Invalid angle--intensity pairs are removed, negative intensities are clipped
to zero and observations are sorted by $2\theta$. The raw profile is retained,
while peak detection uses configurable smoothing and baseline subtraction:
\begin{equation}
I_i^{\prime}=\max\left(0,\tilde I_i-B_i\right).
\end{equation}
Local maxima must satisfy minimum height, prominence and angular-separation
criteria and may be refined from neighbouring samples. Retrieval uses the
leading $m$ peaks, normalized as
\begin{equation}
a_i^{(\%)}=100\frac{a_i}{\max_{r\in\{1,\ldots,m\}}a_r}.
\end{equation}
Detector settings are recorded because retaining too few peaks can suppress
evidence from minor phases.

\paragraph*{Single-phase searching}
\label{Single-phase searching}
For each phase $q$, XMatcher evaluates a common shift
$\delta\in[-\Delta,\Delta]$ using a regular grid augmented by experimental--
theoretical peak differences. Pair $(i,j)$ is eligible when
\begin{equation}
|e_i-(t_{qj}+\delta)|\leq\tau,
\end{equation}
where $\tau$ is the positional tolerance. Eligible-pair cost combines angular
and relative-intensity disagreement:
\begin{equation}
C_{ij}(\delta)=w_x\frac{|e_i-(t_{qj}+\delta)|}{\tau}
+w_I\frac{|a_i^{(\%)}-b_{qj}^{(\%)}|}{100},
\qquad w_x+w_I=1.
\end{equation}
where $w_x,w_I\geq0$. Ineligible pairs have infinite cost. Linear-sum
assignment is restricted to rows and columns with finite candidates, so
unmatched peaks remain residual evidence without invalidating other matches.
The resulting $M_q(\delta)$ is one-to-one.

For each shift, precision $P=|M_q(\delta)|/m$ and theoretical recall
$R=|M_q(\delta)|/n_q$ measure matched peak counts. Intensity-weighted coverage
is
\begin{equation}
C_E=\frac{\sum_{(i,j)\in M_q(\delta)}a_i^{(\%)}}{\sum_i a_i^{(\%)}},
\qquad
C_T=\frac{\sum_{(i,j)\in M_q(\delta)}b_{qj}^{(\%)}}{\sum_j b_{qj}^{(\%)}}.
\end{equation}
The quality factor is
$Q=100\exp[-\operatorname{mean}_{(i,j)\in M_q(\delta)}C_{ij}]$.
The default ranking score combines assignment quality, peak-matching
precision and recall, and intensity coverage:
\begin{equation}
H=w_Q Q\left(\frac{P+R}{2}\right)
  +w_T(100C_T)
  +w_E(100C_E)
  +w_B[100\min(P,R)],
\end{equation}
where $w_Q,w_T,w_E,w_B$ are nonnegative weights that sum to one.
Here $C_E$ and $C_T$ are fractions, whereas $Q$ and $H$ range from
0 to 100. The score $H$ is used to rank hypotheses and does not, by
itself, establish phase identity. XMatcher retains the optimized shift,
peak assignments, and unmatched peaks for inspection.
\paragraph*{AutoMix searching}
\label{AutoMix searching}
AutoMix forms combinations of up to $p$ phases from the leading $n$
single-phase candidates, evaluating $\sum_{k=1}^{p}\binom{n}{k}$ hypotheses.
The default $n=8$ and $p=3$ produce 92 combinations. In known-phase mode,
exact element sets and structure IDs are resolved locally and combined with logical
AND; candidates satisfying each required family are retained even when their
single-phase scores are lower.

For an $r$-phase hypothesis, $A\in\mathbb{R}^{m\times r}$ contains the
normalized intensity assigned from phase $\ell$ to experimental peak $i$,
with unmatched entries set to zero and non-zero columns scaled to unity.
Given $y=(a_1,\ldots,a_m)^\mathsf{T}$, non-negative coefficients are fitted by
\begin{equation}
\hat w=\arg\min_{w\geq0}\left\|Aw-y\right\|_2^2.
\end{equation}
with $\hat y=A\hat w$, $\varepsilon=y-\hat y$ and
$\mathrm{SSE}=\sum_i\varepsilon_i^2$. The unpenalized fit score is
\begin{equation}
V=100\max\left(0,1-\frac{\mathrm{SSE}}{\sum_i y_i^2}\right).
\end{equation}
AutoMix ranks combinations by $S=V-0.75(r-1)$. The displayed contribution of
phase $\ell$ is
\begin{equation}
\rho_\ell=100\frac{\sum_i A_{i\ell}\hat w_\ell}{\sum_i\hat y_i}.
\end{equation}
Non-zero $\rho_\ell$ values sum to 100\% but represent only the fitted
contribution to retained peak evidence, not Rietveld-refined mass or volume
fractions. Peaks are assigned to the largest fitted component, marked as
overlapping when several phases contribute substantially, and retained as
residuals when $y_i>\hat y_i$. A required structure ID must exceed the configured
minimum contribution (3\% by default), preventing its inclusion with a
numerically negligible coefficient.

\subsection{XQueryer Lightweight}
\label{sec:xqueryer}

XQueryer Lightweight is a compact variant of XQueryer designed for efficient
training and inference and integration into automated diffraction workflows,
while retaining the original framework's core structure-identification
capabilities~\cite{cao2025xqueryer}. It combines multiscale, frequency-aware
peak encoding with compact element-guided cross-attention to reduce
computational cost while capturing the diffraction and compositional
information needed for large-scale phase identification. The model is trained
on the MP500 dataset (Section~\ref{sec:mp500}).

\paragraph{Problem formulation and data generation}

Let $s_i$ denote the $i$-th structure in the MP500 database, with label $y_i$. For each requested sample, the simulator draws a parameter vector $\phi$ and produces an angle--intensity pair $(2\boldsymbol{\theta},\mathbf{z})=\mathcal{S}(s_i;\phi)$. The intensity is linearly interpolated to
\begin{equation}
  \mathbf{x}\in\mathbb{R}^{3500},\qquad 2\theta_j\in[10,90]\ \text{degrees},
\end{equation}
then clipped to non-negative values and scaled so that its maximum is 100. The same interpolation is applied to experimental inputs.

Rather than splitting structures across folds, this repository uses an \emph{XRD-level split}: every split covers every structure but uses a disjoint range of deterministic simulation seeds. Thus validation and test patterns have different simulated conditions from their training counterparts, while the classification label space is fixed. Default counts are 20 training, one validation, and two test patterns per structure. The train/validation/test sample counts under this default are therefore 2,006,300, 100,315, and 200,630, respectively.

\begin{table}[t]
\centering
\caption{Default ranges sampled for dynamic Pysimxrd simulation.}
\label{tab:simulation}
\small
\begin{tabular}{@{}cc@{}}
\toprule
Quantity & Range / setting \\
\midrule
Grain size & 10--100 \\
Preferred orientation (each component) & $-0.3$--$0.3$ \\
Thermal vibration & 0--2 \\
Zero shift & 0--2 \\
Detector--sample distance & 300--600 \\
Detector half-height / sample half-height & 3--8 / 1--4 \\
Background order / ratio & 6 / 0.02--0.20 \\
Mixture noise ratio & 0.01--0.10 \\
XRD representation & reciprocal \\
\bottomrule
\end{tabular}
\end{table}

\paragraph{Frequency-aware peak encoding}

The normalized input is first transformed with a real FFT. In addition to the unmodified trace, three low-pass reconstructions are obtained by retaining 0.35, 0.60, and 0.85 fractions of the real-frequency spectrum and applying the inverse real FFT. Concatenating these views gives a four-channel signal. This parameter-free operation supplies the encoder with peak representations at multiple frequency scales.

The four channels are processed by a 1D CNN. A stem convolution is followed by five residual depthwise-separable blocks with kernel sizes 17, 15, 11, 9, and 7; each block has stride two. Depthwise filtering captures local line-shape changes efficiently, while pointwise convolutions mix the frequency views. For the default width of 64, the final feature map has 192 channels. Adaptive average pooling compresses it to 96 tokens, which are projected to an attention width of 192 and augmented by learned positional embeddings.

\paragraph{Compact element-guided cross-attention}

For a structure with atomic numbers no greater than 92, an element-presence vector $\mathbf{e}\in\{0,1\}^{92}$ records which elements occur. An MLP maps $\mathbf{e}$ to four vectors of dimension 192. These vectors are added to four learned query embeddings and read the XRD token sequence using multi-head cross-attention. Each of two attention blocks applies residual cross-attention, query self-attention, and a feed-forward network:
\begin{align}
  \mathbf{q}' &= \mathbf{q} + \mathrm{MHA}(\mathrm{LN}(\mathbf{q}),\mathrm{LN}(\mathbf{t}),\mathrm{LN}(\mathbf{t})),\\
  \mathbf{q}^{+} &= \mathbf{q}' + \mathrm{MHA}(\mathrm{LN}(\mathbf{q}'),\mathrm{LN}(\mathbf{q}'),\mathrm{LN}(\mathbf{q}')) + \mathrm{FFN}(\mathrm{LN}(\cdot)).
\end{align}
Here $\mathbf{t}$ denotes the XRD tokens. This design retains composition-guided matching while replacing a heavy element-to-length-3500 expansion with only four compact queries.

The flattened queries are concatenated with global CNN mean/max statistics and global FFT mean/standard-deviation statistics. A two-layer MLP produces a 256-dimensional descriptor. By default, a normalized cosine classifier assigns the descriptor to one of the 100,315 structures among MP500:
\begin{equation}
  \ell_k=\alpha\frac{\mathbf{h}^{\top}\mathbf{w}_k}{\lVert\mathbf{h}\rVert_2\lVert\mathbf{w}_k\rVert_2},
\end{equation}
where $\alpha$ is a learned, bounded temperature initialized to 30.

\begin{table}[t]
\centering
\caption{Default model configuration. The parameter count was computed from \texttt{count\_parameters(Xmodel())}.}
\label{tab:model}
\small
\begin{tabular}{@{}cc@{}}
\toprule
Component & Setting \\
\midrule
Input grid & 3,500 samples; 10--90$^\circ$ \\
FFT views & original + 3 low-pass views \\
CNN base channels & 64 \\
Attention dimension / heads & 192 / 6 \\
XRD tokens / element queries & 96 / 4 \\
Attention refinement blocks & 2 \\
Classifier & cosine, 100,315 classes \\
Dropout & 0.15 \\
Trainable parameters & 28,012,497 \\
\bottomrule
\end{tabular}
\end{table}

\paragraph{Training and inference}

Training uses AdamW with learning rate $3\times10^{-4}$ and weight decay $10^{-4}$ by default. The learning-rate schedule linearly warms up for five epochs and then follows cosine decay. Cross-entropy uses label smoothing 0.05; gradient norms are clipped to 1.0. Optional weak XRD mixup combines intensity traces and uses the element-wise maximum of their element vectors. An EMA of model weights with decay 0.999 is maintained and is used for validation checkpoints and inference when available. The trainer supports CPU, one GPU, and distributed \texttt{torchrun} execution using PyTorch \cite{paszke2019pytorch}.

At inference time, an XRD file containing \texttt{angle} and
\texttt{intensity} columns is interpolated onto the training grid and passed
to the classifier. The command-line interface reports the top-$k$ predicted
class IDs, the corresponding MP500 database entry IDs, associated MP identifiers
when available, and softmax scores. XQueryer-LW is trained for the element-aware setting used in Gan Jiang and
requires the constituent elements as input at inference time. It therefore
does not generate predictions when elemental information is unavailable.
This differs from the original XQueryer, which was designed for more general
structure-identification settings.

\subsection{XDecomposer: set-based decomposition of multiphase XRD}
\label{sec:XDecomposer}
XDecomposer~\cite{gao2026xdecomposer} decomposes the XRD pattern of a mixed powder into the diffraction contributions of individual components prior to phase identification. This method integrates a convolutional encoder, a pre-trained Transformer, and a learnable phase query mechanism to simultaneously estimate multiple components. The decomposition process requires neither a predefined list of candidate structures nor a specified number of phases. The predicted component diffraction patterns and their activity scores provide the inputs for subsequent phase identification.

\begin{figure}[h]
  \centering
  \includegraphics[width=0.95\linewidth]{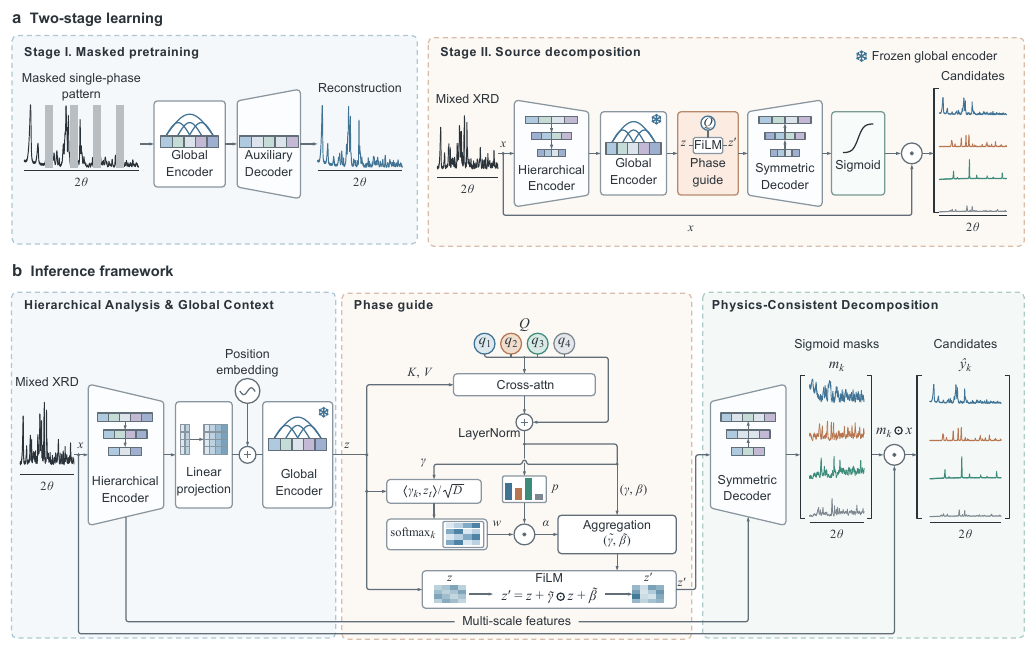}
  \caption{\textbf{XDecomposer learns to separate multiphase diffraction signals without a predefined phase list.} \textbf{a}, Two-stage learning. Stage I pretrains a global encoder by reconstructing masked regions of single-phase patterns. Stage II combines a hierarchical encoder, the frozen global encoder (snowflake), a feature-wise linear modulation (FiLM) phase guide and a symmetric decoder to produce a set of candidate component patterns from a mixed XRD input. \textbf{b}, Inference framework. Hierarchical multi-scale features and global context are conditioned by learned phase queries through cross-attention and aggregation; the decoder outputs sigmoid masks $m_k$, which are applied to the input profile $x$ to yield physics-consistent component candidates $\hat{y}_k$. Curves represent diffraction intensity as a function of $2\theta$; colours distinguish predicted components.}
  \label{fig:XDecomposer}
\end{figure}

\paragraph{Problem formulation and data preparation}
Input XRD profiles are interpolated onto a uniform grid of $L=3{,}500$
points spanning $10^{\circ}\leq2\theta\leq80^{\circ}$, with non-negative intensities normalized to a unit maximum. For a mixture containing $K$ phases, the profile $\mathbf{x}\in\mathbb{R}_{\geq0}^{L}$ is represented under the linear superposition approximation as
\begin{equation}
  \mathbf{x}=\sum_{k=1}^{K}\mathbf{c}_k+\boldsymbol{\varepsilon},
  \qquad \mathbf{c}_k=w_k\mathbf{s}_k,
  \qquad w_k\geq0,\quad\sum_{k=1}^{K}w_k=1,
\end{equation}
where $\mathbf{s}_k$ denotes the single-phase diffraction profile on a specified intensity scale, $w_k$ is the corresponding mixing coefficient, and $\boldsymbol{\varepsilon}$ accounts for noise and other residual contributions. XDecomposer accommodates an unknown phase count using $K_{\max}=4$ output slots, each predicting a component contribution $\hat{\mathbf{c}}_k$ and an associated activity probability $p_k$.

The simulated corpus comprises 2,006,300 patterns generated using
pysimxrd~\cite{cao2025simxrd} from 100,315 MP500 data, with 20 profiles per structure
incorporating variations in peak positions, widths, background and
instrumental conditions. Simulation parameters are summarized in
Table~\ref{tab:xdecomposer_simulation}. Crystal IDs are partitioned into training,
validation and test subsets in an 8:1:1 ratio before mixture construction, with all simulated variants of a given crystal assigned to the same subset. During data loading, mixtures are constructed by combining an anchor phase with $K-1$ additional distinct phases sampled from the same subset, where $K\in\{2,3,4\}$. Mixing coefficients are obtained through Dirichlet-based sampling subject to $w_k\geq0.15$. A common normalization factor is applied to each mixture and its weighted component targets to preserve their relative amplitudes; unused target slots are zero-padded.
\begin{table}[t]
\centering
\caption{Parameters used to generate simulated pretraining data for XDecomposer.}
\label{tab:xdecomposer_simulation}
\small
\begin{tabular}{@{}ccc@{}}
\toprule
Category & Parameter & Value or range \\
\midrule
Random perturbations
  & Crystallite size (nm) & 10--120 \\
  & Thermal vibration coefficient & 0.01--0.20 \\
  & Zero shift & 0--0.20 \\
  & Detector--sample distance (mm) & 300--600 \\
  & Detector slit half-height (mm) & 3--8 \\
  & Sample half-height (mm) & 1--4 \\
\midrule
Fixed parameters
  & Preferred-orientation factor & 0.15 \\
  & Background polynomial order & 6 \\
  & Mixture noise ratio & 0.02 \\
  & Lattice extinction/torsion ratio & 0.01/0.01 \\
\midrule
Data configuration
  & $2\theta$ range & $10^{\circ}$--$80^{\circ}$ \\
  & Step size & $0.02^{\circ}$ \\
  & Patterns per crystal ID & 20 \\
\bottomrule
\end{tabular}
\end{table}

\paragraph{Diffraction encoding and phase-query modulation}
A hierarchical one-dimensional convolutional encoder extracts multiscale features from the input diffraction profile. Successive layers capture local peak shapes and relationships among reflections over progressively larger angular intervals. Intermediate feature maps are retained for skip connections to the reconstruction decoder. The final convolutional features are projected into a latent sequence, augmented with positional encodings and processed by a pretrained Transformer encoder~\cite{vaswani2017attention} to model dependencies across the diffraction profile. The Transformer encoder is pretrained on single-phase XRD profiles using a masked reconstruction objective~\cite{he2022masked}, with an auxiliary decoder reconstructing masked regions from the visible input. The pretrained encoder is then incorporated into the decomposition network, and its parameters remain fixed during training on multiphase mixtures~\cite{gao2026xdecomposer}. Its output is denoted by $\mathbf{Z}\in\mathbb{R}^{T\times d}$, where $T$ is the latent sequence length and $d$ is the feature dimension.

Each learnable phase query $\mathbf{q}_k$ aggregates information from $\mathbf{Z}$ through multi-head cross-attention. The resulting query representation is mapped to a slot activity probability and parameters for feature-wise linear modulation (FiLM)~\cite{perez2018film}:
\begin{align}
  \mathbf{h}_k &=\mathrm{LN}\!\left[\mathbf{q}_k+\mathrm{MHA}(\mathbf{q}_k,\mathbf{Z},\mathbf{Z})\right],\\
  p_k&=\sigma\!\left(f_{\mathrm{act}}(\mathbf{h}_k)\right),\qquad
  (\boldsymbol{\gamma}_k,\boldsymbol{\beta}_k)=f_{\mathrm{mod}}(\mathbf{h}_k),
\end{align}
where $\mathrm{LN}$, $\mathrm{MHA}$ and $\sigma$ denote layer normalization, multi-head cross-attention and the sigmoid function, respectively. The prediction heads $f_{\mathrm{act}}$ and $f_{\mathrm{mod}}$ produce the activity probability $p_k$ and the scaling and offset vectors $(\boldsymbol{\gamma}_k,\boldsymbol{\beta}_k)$. At each latent position $t$, compatibility scores between the modulation vectors $\boldsymbol{\gamma}_k$ and the latent feature $\mathbf{Z}_t$ are
normalized across slots and scaled by the corresponding activity probabilities:
\begin{align}
  a_{kt}&=p_k\,\frac{\exp(\boldsymbol{\gamma}_k^{\mathsf T}\mathbf{Z}_t/\sqrt d)
  }{\sum_{j=1}^{K_{\max}}\exp(\boldsymbol{\gamma}_j^{\mathsf T}\mathbf{Z}_t/\sqrt d)
  },\\
\mathbf{Z}'_t&=\mathbf{Z}_t\odot\left(\mathbf{1}+\sum_{k=1}^{K_{\max}}a_{kt}\boldsymbol{\gamma}_k\right)+\sum_{k=1}^{K_{\max}}a_{kt}\boldsymbol{\beta}_k.
\end{align}
Here, $\odot$ denotes element-wise multiplication, and $a_{kt}$ determines the contribution of query $k$ to the affine transformation at position $t$. The modulated sequence $\mathbf{Z}'$ and the retained convolutional feature maps provide the inputs to the reconstruction decoder.

\paragraph{Component reconstruction}
Transposed convolutions and skip-feature fusion restore angular resolution and produce mask logits $\mathbf{u}_k$. The reconstructed components and their sum are
\begin{equation}
  \hat{\mathbf{c}}_k=\sigma(\mathbf{u}_k)\odot\mathbf{x},
  \qquad
  \hat{\mathbf{x}}=\sum_{k=1}^{K_{\max}}\hat{\mathbf{c}}_k.
\end{equation}
Independent sigmoid masks assign a continuous fraction of the observed intensity to each output component. This parameterization ensures $0\leq\hat c_{k\ell}\leq x_\ell$ at each angular sample $\ell$. Continuous masks accommodate contributions from multiple phases at overlapping reflections. A mixture-reconstruction loss encourages their summed intensity to reproduce the input profile.

\paragraph{Decomposition objective}

For predicted and reference profiles $\mathbf{u}$ and $\mathbf{v}$, the component loss combines intensity-weighted amplitude error, shape similarity and finite differences in the square-root domain:
\begin{align}
  \ell(\mathbf{u},\mathbf{v})={}&A\!\left[(\mathbf{u}-\mathbf{v})\odot
    (\mathbf{1}+\alpha\mathbf{v})\right] -\lambda_{\mathrm{shape}}\,
    \mathrm{SI\text{-}SDR}(\mathbf{u},\mathbf{v})\nonumber\\
  &+\lambda_{\mathrm{geo}}\left\{A\!\left(\Delta\sqrt{\mathbf{u}}-\Delta\sqrt{\mathbf{v}}\right)+\beta A\!\left(\Delta^2\sqrt{\mathbf{u}}-
                        \Delta^2\sqrt{\mathbf{v}}\right)\right\}.
\end{align}
Here, $A$ denotes the mean absolute value, $\Delta$ the first-order finite difference and SI-SDR the scale-invariant signal-to-distortion ratio~\cite{le2019sdr}. SI-SDR evaluates profile shape by comparing the prediction with its least-squares projection onto the target. Square roots use a positive numerical floor, and the SI-SDR term is omitted for zero targets. Amplitude weighting emphasizes stronger reflections, while the difference terms constrain peak slopes and curvature.

Permutation-invariant training~\cite{yu2017permutation} minimizes the total component error over all permutations of the zero-padded reference set $\{\bar{\mathbf{c}}_k\}_{k=1}^{K_{\max}}$:
\begin{equation}
  \mathcal{L}_{\mathrm{sep}}= \min_{\pi\in\mathfrak{S}_{K_{\max}}}
  \sum_{k=1}^{K_{\max}}\ell(\hat{\mathbf{c}}_{\pi(k)},\bar{\mathbf{c}}_k),
\end{equation}
where $\mathfrak{S}_{K_{\max}}$ contains slot permutations. The minimizing assignment defines binary activity targets $b_j$, with $b_j=1$ for slots matched to non-zero components. This assignment makes supervision independent of the ordering of reference phases. The complete objective is
\begin{equation}
  \mathcal{L}=\mathcal{L}_{\mathrm{sep}} +\frac{\lambda_{\mathrm{act}}}{K_{\max}}
    \sum_{j=1}^{K_{\max}}\mathrm{BCE}(p_j,b_j) +\lambda_{\mathrm{mix}}A(\hat{\mathbf{x}}-\mathbf{x}),
\end{equation}
where binary cross-entropy (BCE) supervises slot occupancy and the final term encourages agreement with the observed mixture.

\paragraph{Inference and evaluation}

A single forward pass predicts component profiles and activity probabilities. Thresholding at $\tau$ yields the retained components and phase count:
\begin{equation}
  \hat{\mathcal{C}}=\{\hat{\mathbf{c}}_k:p_k>\tau\},\qquad
  \hat K=\sum_{k=1}^{K_{\max}}\mathbf{1}[p_k>\tau].
\end{equation}
Retained profiles preserve their amplitudes on the mixture scale and serve as queries for phase identification.

\subsection{Statistical whole-pattern decomposition with PyXplore}
\label{sec:PyXplore}

Bragg peak positions, intensities and line shapes jointly encode geometric and chemical information, but extracting these quantities from a one-dimensional powder profile is intrinsically ill-posed because multiple reflections can coincide in $2\theta$ \cite{dinnebier2015powder,debye1916interferenzen}. Classical whole-pattern refinement methods (Rietveld, Pawley and Le Bail) enforce crystallographic consistency by coupling all reflections through a single structural model \cite{rietveld1967line,rietveld1969profile,pawley1981unit,le1988ab}. Under severe overlap, however, the resulting least-squares problem can be numerically fragile and may admit non-physical solutions (e.g., poorly determined or even negative intensities).

PyXplore addresses this bottleneck by recasting whole-pattern decomposition as probabilistic inference subject to hard diffraction constraints. Specifically, we treat the background-subtracted profile as a mixture of peak-shape functions (PSFs) and estimate component responsibilities via expectation maximization (EM), which naturally handles strong overlap through soft assignments. The observed intensity profile $y(2\theta)$ is approximated as
\begin{equation}
 y(2\theta) \approx \sum_{i=1}^{m} w_i\, p_i(2\theta;\,\boldsymbol{\phi}_i),
 \label{eq:mixture_main}
\end{equation}
where $p_i$ is a normalized PSF (we use a pseudo-Voigt family in practice), $w_i \ge 0$ is the component weight (proportional to integrated intensity), and $\boldsymbol{\phi}_i$ collects peak-shape parameters.

In PyXplore, this mixture view is made explicit by interpreting the background-subtracted diffraction profile as a continuous \emph{measure function} assembled from $m$ peak-shape functions. Each PSF is modelled using a thin-tailed pseudo-Voigt (PV) density, i.e., a convex combination of Lorentzian and Gaussian components,

\begin{equation}
p_{\mathrm{pv}}(x)=
\Delta\cdot\frac{1}{\pi}\frac{\gamma}{(x-\mu)^2+\gamma^2}
+ (1-\Delta)\cdot\frac{1}{\sqrt{2\pi}\sigma}\exp\!\left(-\frac{(x-\mu)^2}{2\sigma^2}\right),
\end{equation}

where $\mu$ is the Bragg-peak centre, $\gamma$ is the Lorentzian half width at half maximum (HWHM), $\sigma$ is the Gaussian standard deviation, and $\Delta\in(0,1)$ controls the Lorentzian--Gaussian mixing (with $\Delta\to1$ recovering a Lorentzian and $\Delta\to0$ recovering a Gaussian). Unlike Le Bail-type profile matching, which often constrains the Lorentzian and Gaussian widths through a shared full width at half maximum (FWHM) model (for example, $\Gamma^2=U\tan^2\theta+V\tan\theta+W$), we treat $\gamma$ and $\sigma$ as independent variables and optimize them directly within EM--Bragg to accommodate heterogeneous broadening and severe overlap.

The closed-form update rules in the $(t+1)$-th maximization step are given by
\begin{align}
\mu_i^{(t+1)}
&=
\frac{
\frac{2\pi}{\gamma_i^{(t)}}
\sum_{j=1}^{N}
y_j^{\mathrm{nb}}
\beta_{ji}^{L(t)} x_j
+
\frac{1}{\sigma_i^{2(t)}}
\sum_{j=1}^{N}
y_j^{\mathrm{nb}}
T_{ji}^{R(t)} x_j
}{
\frac{2\pi}{\gamma_i^{(t)}}
\sum_{j=1}^{N}
y_j^{\mathrm{nb}}
\beta_{ji}^{L(t)}
+
\frac{1}{\sigma_i^{2(t)}}
\sum_{j=1}^{N}
y_j^{\mathrm{nb}}
T_{ji}^{R(t)}
},
\\
\gamma_i^{(t+1)}
&=
\left[
\frac{
\sum_{j=1}^{N}
y_j^{\mathrm{nb}}
\beta_{ji}^{L(t)}
\big(x_j-\mu_i^{(t)}\big)^2
}{
\sum_{j=1}^{N}
y_j^{\mathrm{nb}}
\beta_{ji}^{L(t)}
}
\right]^{1/2},
\\
\sigma_i^{2(t+1)}
&=
\frac{
\sum_{j=1}^{N}
y_j^{\mathrm{nb}}
T_{ji}^{R(t)}
\big(x_j-\mu_i^{(t)}\big)^2
}{
\sum_{j=1}^{N}
y_j^{\mathrm{nb}}
T_{ji}^{R(t)}
},
\\
\Delta_i^{(t+1)}
&=
\frac{
\sum_{j=1}^{N}
y_j^{\mathrm{nb}}
T_{ji}^{L(t)}
}{
\sum_{j=1}^{N}
y_j^{\mathrm{nb}}
T_{ji}^{(t)}
}.
\label{core_em_bragg}
\end{align}

Here $y_j^{\mathrm{nb}}$ denotes the background-subtracted intensity at grid point $j$, $T_{ji}^{L}$ is the (normalized) responsibility/weight that assigns point $j$ to peak component $i$ in the E-step (with superscripts indicating the update used), and $\beta_{ji}^{L}$ and $T_{ji}$ are auxiliary quantities used in the closed-form M-step updates; their explicit definitions provided below.

Under the Bragg-law constraint on the peak centres, Eqs.~(\ref{8a})--(\ref{8d}) are iterated to update the mixture parameters, including the peak position $\mu_i$, the Lorentzian and Gaussian width parameters $\gamma_i$ and $\sigma_i^2$, and the mixing coefficient $\Delta_i$.

\paragraph{Mathematical framework}
We first define the notation used throughout the derivation. Let $Y^{\mathrm{ex}}(x_j)$ denote the experimentally observed XRD profile, $Y^{\mathrm{nb}}(x_j)$ the measured profile after background removal, and $Y^{\mathrm{WPEM}}(x_j)$ the calculated pattern obtained from an $m$-component mixture model. Here, $x_j$ $(j = 1, 2, \ldots, |\widetilde{2\theta}|)$ represents the sampling points corresponding to the diffraction angle $\widetilde{2\theta}$, and $|\widetilde{2\theta}|$ denotes the total number of sampled points. The objective of the model is to minimize the residual between the calculated pattern $Y^{\mathrm{WPEM}}(x_j)$ and the experimental observation $Y^{\mathrm{ex}}(x_j)$.

\textbf{M-component Mixture Model.}
The profile fitting task is formulated as a density estimation problem based on an $m$-component mixture model and is solved using a batch optimization strategy termed the Bragg-based expectation maximization process for whole-pattern decomposition. 
The background-free powder XRD profile is interpreted as a measure function composed of $m$ peak shape functions, each representing an underlying diffraction regime. 
A soft assignment associates each observation $\widetilde{2\theta}$ with all components, yielding
\begin{equation}
Y^{\mathrm{WPEM}}(x_j) = \sum_{i=1}^{m} W_i \, Y_i(x_j),
\end{equation}
where $Y_i(x_j)$ denotes the $i$-th PSF and $W_i$ is its corresponding weight, reflecting the effective number of observations explained by that component. 
The location parameter of each PSF determines the diffraction index of the associated Bragg peak and converges to the value prescribed by Bragg's law. 
Because PSFs are defined as continuous functions, smooth optimization schemes can be employed, enabling a continuous and physically consistent representation of diffraction intensities over the entire angular range.

In the framework, an XRD pattern is constructed as a measure function under the assumption that each component in the mixture model follows a specific probability density function (p.d.f.). 
Under this formulation, profile fitting is converted into a density estimation problem with latent variables, which is solved using the EM algorithm to perform maximum likelihood estimation subject to the constraint imposed by Bragg's law. 
The generative process can be described explicitly as follows. 
A latent variable $W$ is drawn from a discrete p.d.f.\ $w$, such that $P(W = W_i) = w_i$, and is assigned to component $Y_i$. 
Each component $Y_i$ follows a p.d.f.\ $y_i$ parameterized by $\boldsymbol{\theta}$, that is, $Y_i \sim y_i(\boldsymbol{\theta})$. 
Accordingly, the resulting $m$-component mixture model can be written as
\begin{equation}
y^{\mathrm{WPEM}}(x_j) = \sum_{i=1}^{m} w_i \, y_i(x_j).
\label{m-mixture-eq}
\end{equation}

From this perspective, the XRD pattern is a continuous measure function defined on the field of real numbers, satisfying
\begin{equation}
\Omega\!\left(y^{\mathrm{WPEM}}\right) = \int_{-\infty}^{+\infty} y^{\mathrm{WPEM}}(x) \, \mathrm{d}x = I_{\mathrm{area}},
\end{equation}
where $I_{\mathrm{area}}$ denotes the total integrated intensity, corresponding to the area enclosed by the background-free XRD profile. 
Consequently,
\begin{equation}
\int_{-\infty}^{+\infty} \left( \sum_{i=1}^{m} w_i \, y_i(x) \right) \mathrm{d}x
= \Omega\!\left(y^{\mathrm{WPEM}}\right),
\end{equation}
which yields the normalization condition $\sum_{i=1}^{m} w_i = I_{\mathrm{area}}$, with $w_i \in (0, I_{\mathrm{area}})$. 
For XRD patterns defined over the $\widetilde{2\theta}$ domain, thin-tailed p.d.f.s are recommended to model the PSF, ensuring both physical plausibility and numerical stability.

Each PSF, described by an individual p.d.f.,\ $y_i$, is further modeled using a PV function $p_{\mathrm{pv}}$, which itself is a weighted mixture of Lorentzian and Gaussian components. 
The Lorentzian component captures the intrinsic diffraction emission and the time-of-reflection probability associated with the interaction between a crystal plane and the incident X-ray beam, whereas the Gaussian component accounts for stochastic broadening arising from instrumental effects and random noise. 
Each PV function is a thin-tailed, continuous p.d.f.\ satisfies

\begin{equation}
\int_{-\infty}^{+\infty} p_{\mathrm{pv}}(x) \, \mathrm{d}x = 1.
\end{equation}
It is defined as
\begin{equation}
p_{\mathrm{pv}}(x) =
\Delta \cdot \frac{1}{\pi} \frac{\gamma}{(x-\mu)^2 + \gamma^2}
+ (1-\Delta) \cdot \frac{1}{\sqrt{2\pi}\sigma}
\exp\!\left( -\frac{(x-\mu)^2}{2\sigma^2} \right),
\end{equation}
where $\Delta \in (0,1)$ adjusts the angle-dependent noise level and represents the weight of the Lorentzian component. 
Substituting this p.d.f.\ into Eq.~\ref{m-mixture-eq} yields
\begin{equation}
y^{\mathrm{WPEM}}(x) =
\sum_{i=1}^{m} w_i
\left[
\frac{\Delta_i}{\pi} \frac{\gamma_i}{(x-\mu_i)^2 + \gamma_i^2}
+ \frac{1-\Delta_i}{\sqrt{2\pi}\sigma_i}
\exp\!\left( -\frac{(x-\mu_i)^2}{2\sigma_i^2} \right)
\right],
\label{whole_eq}
\end{equation}
where $\mu_i$ denotes the Bragg peak position of the $i$-th diffraction peak, $\gamma_i$ is the HWHM of the Lorentzian distribution, and $\sigma_i^2$ is the variance of the Gaussian distribution.

In the Le Bail method and related whole powder pattern decomposition approaches, the number of refinement parameters is reduced by enforcing an identical FWHM ($\Gamma$) for the Lorentzian and Gaussian components of the PV function. Specifically,
\begin{equation}
2\gamma = 2\sqrt{2\ln 2}\,\sigma = \Gamma,
\end{equation}
where $\Gamma$ is refined using a semi-empirical model, for example,
\begin{equation}
\Gamma^2 = U \tan^2 \widetilde{\theta} + V \tan \widetilde{\theta} + W.
\end{equation}
In our workflow, the Lorentzian and Gaussian width parameters, $\gamma$ and $\sigma^2$, are treated as independent variables and are directly optimized within the EM--Bragg process.

\textbf{Construction of the Q Function in the EM--Bragg Process.}
The EM--Bragg process is formulated to maximize the posterior likelihood of the latent variable $T_{ij}$, where $i$ indexes the PV components and $j$ denotes the sampling index corresponding to $\widetilde{2\theta}$. The model parameters are defined as
$
\{\Delta_i, w_i, \mu_i, \gamma_i, \sigma_i^2\} \triangleq \boldsymbol{\theta},
$
the observations as
$
(x_1, x_2, \ldots, x_j) \triangleq \mathbf{X},
$
and the corresponding measure function as
$
y^{\mathrm{WPEM}}(x) \triangleq P.
$
Under this notation, parameter estimation amounts to maximizing the log-likelihood
\begin{equation}
\arg\max_{\boldsymbol{\theta}} \log P(\mathbf{X} \mid \boldsymbol{\theta}).
\end{equation}

To facilitate the optimization, the latent variable $T_{ij}$ is introduced, yielding
\begin{equation}
\arg\max_{\boldsymbol{\theta}}
\log \left( \int P(\mathbf{X}, T_{ij} \mid \boldsymbol{\theta}) \, \mathrm{d}T_{ij} \right),
\end{equation}
since $P(\mathbf{X} \mid \boldsymbol{\theta})$ is the marginal distribution of the joint probability
$P(\mathbf{X}, T_{ij} \mid \boldsymbol{\theta})$. Assuming that the distribution of $T_{ij}$ is known as
$P(T_{ij})$, Jensen’s inequality leads to
\begin{align}
\arg\max_{\boldsymbol{\theta}}
\log \left( \int P(\mathbf{X}, T_{ij} \mid \boldsymbol{\theta}) \, \mathrm{d}T_{ij} \right)
&=
\arg\max_{\boldsymbol{\theta}}
\log \left( \int
\frac{P(\mathbf{X}, T_{ij} \mid \boldsymbol{\theta})}{P(T_{ij})}
P(T_{ij}) \, \mathrm{d}T_{ij} \right) \nonumber \\
&\ge
\arg\max_{\boldsymbol{\theta}}
\int
\log \left(
\frac{P(\mathbf{X}, T_{ij} \mid \boldsymbol{\theta})}{P(T_{ij})}
\right)
P(T_{ij}) \, \mathrm{d}T_{ij}.
\label{lower_bound}
\end{align}

Eq.~\ref{lower_bound} corresponds to evaluating the expectation of
$\log\!\left(P(\mathbf{X}, T_{ij} \mid \boldsymbol{\theta}) / P(T_{ij})\right)$
with respect to the latent variable $T_{ij}$, which constitutes the E-step of the EM-Bragg process.
Here, $P(T_{ij})$ depends only on the parameters obtained at the $t$-th iteration and is therefore written as
$P(T_{ij} \mid \boldsymbol{\theta}^{(t)})$. Substituting this relation into
Eq.~\ref{lower_bound} yields
{\footnotesize
\begin{align}
\arg\max_{\boldsymbol{\theta}}
\int
\log \left(
\frac{P(\mathbf{X}, T_{ij} \mid \boldsymbol{\theta})}
{P(T_{ij} \mid \boldsymbol{\theta}^{(t)})}
\right)
P(T_{ij} \mid \boldsymbol{\theta}^{(t)}) \, \mathrm{d}T_{ij}
=
\arg\max_{\boldsymbol{\theta}}
\Bigg\{
&\int
\log \left( P(\mathbf{X}, T_{ij} \mid \boldsymbol{\theta}) \right)
P(T_{ij} \mid \boldsymbol{\theta}^{(t)}) \, \mathrm{d}T_{ij}
\nonumber \\
&-
\int
\log \left( P(T_{ij} \mid \boldsymbol{\theta}^{(t)}) \right)
P(T_{ij} \mid \boldsymbol{\theta}^{(t)}) \, \mathrm{d}T_{ij}
\Bigg\}.
\label{e-step}
\end{align}
}

The second term in Eq.~\ref{e-step} is independent of $\boldsymbol{\theta}$ for a fixed
$\boldsymbol{\theta}^{(t)}$ and can therefore be omitted during maximization.
The remaining term defines the \emph{Q function} in the EM--Bragg process,
\begin{equation}
Q(\boldsymbol{\theta} \mid \boldsymbol{\theta}^{(t)})
=
\int
\log \left( P(\mathbf{X}, T_{ij} \mid \boldsymbol{\theta}) \right)
P(T_{ij} \mid \boldsymbol{\theta}^{(t)}) \, \mathrm{d}T_{ij},
\end{equation}
which constitutes a lower bound on the log-likelihood.
Accordingly, the M-step updates the parameters as
\begin{equation}
\boldsymbol{\theta}^{(t+1)} =
\arg\max_{\boldsymbol{\theta}}
Q(\boldsymbol{\theta} \mid \boldsymbol{\theta}^{(t)}).
\label{maximization-step}
\end{equation}
At each iteration, the Bragg likelihood is reconstructed using the updated peak positions
$\mu_i^{(t+1)}$, and the corresponding parameters are re-optimized, as detailed below.

\textbf{Initialization of Model.}
According to Bragg’s law, diffraction occurs when the scattering vector coincides with a reciprocal lattice vector, which can be expressed as
\begin{equation}
2 d_{(HKL)} \sin \widetilde{\theta}_{\mathrm{Brg}} = \lambda,
\end{equation}
where $d_{(HKL)}$ denotes the interplanar spacing and $(H,K,L)$ are the Miller indices of the diffracting planes.

The preliminary peak positions $\mu_i^{(0)}$ are computed directly from Bragg’s equation using lattice constants obtained from an initial structure identification model, such as XQueryer \cite{cao2025xqueryer}. Based on $\mu_i^{(0)}$, the initial values of the Lorentzian and Gaussian width parameters, $\gamma_i^{(0)}$ and $\sigma_i^{2(0)}$, are estimated from the FWHM, $\Gamma_i$, of the diffraction peaks,
\begin{equation}
\gamma_i^{(0)} = \Delta_i^{(0)} \frac{\Gamma_i}{2},
\qquad
\sigma_i^{2(0)} = \left(1 - \Delta_i^{(0)}\right) \frac{\Gamma_i}{2},
\end{equation}
where $\Delta_i^{(0)}$ denotes the Lorentzian mixing weight in the pseudo-Voigt function. Unless otherwise specified, $\Delta_i^{(0)}$ is set to 0.8, reflecting that the Lorentzian component dominates the PSFs.

To ensure numerical stability during the iterative optimization, both $\gamma_i$ and $\sigma_i^2$ are constrained to remain larger than a small positive constant $\varepsilon$ (default $\varepsilon = 5\times10^{-4}$), although the precise threshold may be adjusted depending on the application.

The initial peak weights $w_i^{(0)}$ are defined as
\begin{equation}
w_i^{(0)} =
\frac{y_i^{\mathrm{nb}}}{\sum_{i=1}^{m} y_i^{\mathrm{nb}}}
\, I_{\mathrm{area}},
\qquad i = 1, \ldots, m,
\end{equation}
where $y_i^{\mathrm{nb}}$ denotes the background-eliminated intensity and $I_{\mathrm{area}}$ is the total integrated intensity of the diffraction profile.

These initial values are used to initialize the EM–Bragg process, after which reliable results are obtained through posterior likelihood maximization.

\paragraph{EM--Bragg Process.}
In the $m$-component mixture model, each observation contributes to all components $y_i$, and the current parameter estimates are used to assign responsibilities of the data to individual diffraction peaks. Notably, there are $y_j^{\mathrm{nb}}$ samples at the $j$-th angular position $x_j$. During the maximization step, these responsibilities are incorporated into a weighted maximum posterior likelihood estimation together with constraints imposed by Bragg’s law.

Accordingly, the posterior probability $P(T_{ij})$ is defined to quantify the responsibility of observation $x_j$ for the $i$-th peak $y_i$, given by
\begin{align}
P(T_{ij})
&=
\frac{w_i \, p_{\mathrm{pv},i}(x_j)}
{\sum_{i=1}^{m} w_i \, p_{\mathrm{pv},i}(x_j)} \nonumber \\
&=
\frac{
w_i
\left[
\frac{\Delta_i}{\pi}
\frac{\gamma_i}{(x_j-\mu_i)^2+\gamma_i^2}
+
\frac{1-\Delta_i}{\sqrt{2\pi}\sigma_i}
\exp\!\left(-\frac{(x_j-\mu_i)^2}{2\sigma_i^2}\right)
\right]
}{
\sum_{i=1}^{m}
w_i
\left[
\frac{\Delta_i}{\pi}
\frac{\gamma_i}{(x_j-\mu_i)^2+\gamma_i^2}
+
\frac{1-\Delta_i}{\sqrt{2\pi}\sigma_i}
\exp\!\left(-\frac{(x_j-\mu_i)^2}{2\sigma_i^2}\right)
\right]
},
\quad
\begin{aligned}
j &= 1,\ldots,n,\\
i &= 1,\ldots,m .
\end{aligned}
\label{prob_of_tao}
\end{align}

Here, $m$ denotes the number of mixture components and $n$ the number of observed $\widetilde{2\theta}$ sampling points. The latent variable $T_{ij}$ represents the fractional contribution of observation $j$ to the $i$-th pseudo-Voigt component $p_{\mathrm{pv},i}$, commonly referred to as the responsibility of component $i$ for observation $j$. Under the soft-assignment assumption, the normalization condition $\sum_{i=1}^{m} T_{ij} = 1$ holds for each $j$.

The Q function \( Q(\boldsymbol{\theta} \mid \boldsymbol{\theta}^{(t)}) \) derived above can be explicitly written for the EM--Bragg process as
\begin{align}
Q(\boldsymbol{\theta} \mid \boldsymbol{\theta}^{(t)})
&=
\ln
\left[
\prod_{j=1}^{N}
\left(
\sum_{i=1}^{m} w_i \, p_{\mathrm{pv},i}(x_j)
\right)^{y_j^{\mathrm{nb}}}
\right] \nonumber \\
&=
\sum_{j=1}^{N}
y_j^{\mathrm{nb}}
\ln
\left(
\sum_{i=1}^{m} w_i \, p_{\mathrm{pv},i}(x_j)
\right) \nonumber \\
&=
\sum_{j=1}^{N}
y_j^{\mathrm{nb}}
\ln
\sum_{i=1}^{m}
w_i
\left[
\frac{\Delta_i}{\pi}
\frac{\gamma_i}{(x_j-\mu_i)^2+\gamma_i^2}
+
\frac{1-\Delta_i}{\sqrt{2\pi}\sigma_i}
\exp\!\left(-\frac{(x_j-\mu_i)^2}{2\sigma_i^2}\right)
\right].
\label{Q_function}
\end{align}

The sufficient statistics in this formulation are given by the pairs
$(x_j, T_{ij}^{(t)})$.
However, direct maximization of Eq.~\ref{Q_function} is not tractable in practice.
This difficulty arises from the Lorentzian component in each mixture term, whose expectation do not exist, as the Lorentzian profile corresponds to a Cauchy distribution.

To address this issue, we adopt a mathematical construction to derive a surrogate lower bound, which guarantees a monotonic increase of the log-likelihood lower bound at each iteration, i.e.,
$
Q(\boldsymbol{\theta}^{(t+1)}) > Q(\boldsymbol{\theta}^{(t)}),
$
thereby ensuring monotonic convergence of the EM--Bragg algorithm. The specific update rules are detailed below.

\paragraph{Mathematical Construction.}
Eq.~\ref{prob_of_tao} can be decomposed into Lorentzian and Gaussian contributions, yielding
\begin{align}
T_{ji}^{L}
&=
\frac{
w_i \dfrac{\Delta_i}{\pi}
\dfrac{\gamma_i}{(x_j-\mu_i)^2+\gamma_i^2}
}{
\sum_{i=1}^{m} w_i p_{\mathrm{pv},i}(x_j)
}, \\
T_{ji}^{R}
&=
\frac{
w_i \dfrac{1-\Delta_i}{\sqrt{2\pi}\sigma_i}
\exp\!\left(-\dfrac{(x_j-\mu_i)^2}{2\sigma_i^2}\right)
}{
\sum_{i=1}^{m} w_i p_{\mathrm{pv},i}(x_j)
}.
\end{align}

We further define the auxiliary quantity
\begin{align}
\beta_{ji}^{L}
\triangleq
T_{ji}^{L}
\frac{\gamma_i}{\pi\!\left[(x_j-\mu_i)^2+\gamma_i^2\right]},
\quad
j=1,\ldots,n,\; i=1,\ldots,m .
\end{align}

In the maximization step, all parameters are updated by maximizing the
$Q$ function given the latent variables
$\mathbb{P}(T_{ij}^{(t)})$.
This leads to the following constrained optimization problem:
\begin{equation}
\hat{\boldsymbol{\theta}}^{(t+1)}
=
\arg\max_{\boldsymbol{\theta}}
\left\{
\sum_{j=1}^{N}
y_j^{\mathrm{nb}}
\ln
\sum_{i=1}^{m}
w_i
\left[
\frac{\Delta_i}{\pi}
\frac{\gamma_i}{(x_j-\mu_i)^2+\gamma_i^2}
+
\frac{1-\Delta_i}{\sqrt{2\pi}\sigma_i}
\exp\!\left(
-\frac{(x_j-\mu_i)^2}{2\sigma_i^2}
\right)
\right]
\right\},
\label{eq:Mstep_obj}
\end{equation}
subject to the normalization constraint
\begin{equation}
\sum_{i=1}^{m} w_i = I_{\mathrm{area}} .
\end{equation}

The sufficient conditions for $\hat{\boldsymbol{\theta}}$ are obtained
using the Lagrange multiplier method.
We define the Lagrangian
\begin{align}
\mathcal{L}_1
&=
\sum_{j=1}^{N}
y_j^{\mathrm{nb}}
\ln
\sum_{i=1}^{m}
w_i
\left[
\frac{\Delta_i}{\pi}
\frac{\gamma_i}{(x_j-\mu_i)^2+\gamma_i^2}
+
\frac{1-\Delta_i}{\sqrt{2\pi}\sigma_i}
\exp\!\left(
-\frac{(x_j-\mu_i)^2}{2\sigma_i^2}
\right)
\right]
\nonumber\\
&\quad
+
\lambda
\left(
\sum_{i=1}^{m} w_i - I_{\mathrm{area}}
\right),
\label{6a}
\end{align}
where $\lambda$ is the Lagrange multiplier.

Taking first-order derivatives with respect to each parameter yields
\begin{align}
\frac{\partial \mathcal{L}_1}{\partial w_i}
&=
\sum_{j=1}^{N}
y_j^{\mathrm{nb}}
\frac{
\frac{\Delta_i}{\pi}
\frac{\gamma_i}{(x_j-\mu_i)^2+\gamma_i^2}
+
\frac{1-\Delta_i}{\sqrt{2\pi}\sigma_i}
\exp\!\left(
-\frac{(x_j-\mu_i)^2}{2\sigma_i^2}
\right)
}{
\sum_{i=1}^{m}
w_i p_{\mathrm{pv},k}(x_j)
}
+
\lambda
=0,
\label{6b}\\[6pt]
\frac{\partial \mathcal{L}_1}{\partial \mu_i}
&=
\sum_{j=1}^{N}
y_j^{\mathrm{nb}}
\frac{
w_i
\left[
\frac{\Delta_i}{\pi}
\frac{2\gamma_i}{\big[(x_j-\mu_i)^2+\gamma_i^2\big]^2}
+
\frac{1-\Delta_i}{\sqrt{2\pi}\sigma_i^3}
\exp\!\left(
-\frac{(x_j-\mu_i)^2}{2\sigma_i^2}
\right)
\right]
}{
\sum_{i=1}^{m}
w_i p_{\mathrm{pv},k}(x_j)
}
(x_j-\mu_i)
=0,
\label{6c}\\[6pt]
\frac{\partial \mathcal{L}_1}{\partial \gamma_i}
&=
\sum_{j=1}^{N}
y_j^{\mathrm{nb}}
\frac{
w_i \frac{\Delta_i}{\pi}
\frac{(x_j-\mu_i)^2-\gamma_i^2}
{\big[(x_j-\mu_i)^2+\gamma_i^2\big]^2}
}{
\sum_{i=1}^{m}
w_i p_{\mathrm{pv},k}(x_j)
}
=0,
\label{6d}\\[6pt]
\frac{\partial \mathcal{L}_1}{\partial \sigma_i}
&=
\sum_{j=1}^{N}
y_j^{\mathrm{nb}}
\frac{
w_i \frac{1-\Delta_i}{\sqrt{2\pi}\sigma_i}
\exp\!\left(
-\frac{(x_j-\mu_i)^2}{2\sigma_i^2}
\right)
}{
\sum_{i=1}^{m}
w_i p_{\mathrm{pv},k}(x_j)
}
\left(
\frac{(x_j-\mu_i)^2}{\sigma_i^3}
-
\frac{1}{\sigma_i}
\right)
=0,
\label{6e}\\[6pt]
\frac{\partial \mathcal{L}_1}{\partial \Delta_i}
&=
\sum_{j=1}^{N}
y_j^{\mathrm{nb}}
\frac{
w_i
\left[
\frac{1}{\pi}
\frac{\gamma_i}{(x_j-\mu_i)^2+\gamma_i^2}
-
\frac{1}{\sqrt{2\pi}\sigma_i}
\exp\!\left(
-\frac{(x_j-\mu_i)^2}{2\sigma_i^2}
\right)
\right]
}{
\sum_{i=1}^{m}
w_i p_{\mathrm{pv},k}(x_j)
}
=0.
\label{6f}
\end{align}

Solving Eq.~\ref{6b} yields
\begin{equation}
w_i^{(t+1)}
=
I_{\mathrm{area}}
\frac{
\sum_{j=1}^{N}
y_j^{\mathrm{nb}} T_{ji}^{(t)}
}{
\sum_{j=1}^{N}
y_j^{\mathrm{nb}}
},
\end{equation}
which has a clear physical interpretation as the mixing probability
of the $i$-th pseudo-Voigt component.
Interestingly, the remaining parameters
$\{\Delta_i,\mu_i,\gamma_i,\sigma_i^2\}$
can be updated directly in the M-step by incorporating additional
information beyond $T_{ji}^{(t)}$ to obtain a ``better'' solution by our mathematical construction.

Specifically, we substitute the parameters
$\boldsymbol{\theta}^{(t)}
=
\{\Delta_i^{(t)},\mu_i^{(t)},\gamma_i^{(t)},\sigma_i^{2(t)}\}$
into Eqs.~\ref{6c}--~\ref{6f}, which significantly simplifies the
optimization at the $(t+1)$-th iteration.
The resulting first-order conditions become
\begin{align}
\frac{\partial \mathcal{L}_1}{\partial \mu_i}
&=
\sum_{j=1}^{N}
y_j^{\mathrm{nb}}
\left[
\frac{2\pi}{\gamma_i}
\beta_{ji}^{L}
+
\frac{1}{\sigma_i^2}
T_{ji}^{R}
\right]
(x_j-\mu_i)
=0,
\label{7a}\\
\frac{\partial \mathcal{L}_1}{\partial \gamma_i}
&=
\frac{\pi}{\gamma_i^2}
\sum_{j=1}^{N}
y_j^{\mathrm{nb}}
\beta_{ji}^{L}
(x_j-\mu_i)^2
-
\pi
\sum_{j=1}^{N}
y_j^{\mathrm{nb}}
\beta_{ji}^{L}
=0,
\label{7b}\\
\frac{\partial \mathcal{L}_1}{\partial \sigma_i}
&=
\sum_{j=1}^{N}
y_j^{\mathrm{nb}}
T_{ji}^{R}
\left(
\frac{(x_j-\mu_i)^2}{\sigma_i^3}
-
\frac{1}{\sigma_i}
\right)
=0,
\label{7c}\\
\frac{\partial \mathcal{L}_1}{\partial \Delta_i}
&=
\sum_{j=1}^{N}
y_j^{\mathrm{nb}}
\left(
\frac{T_{ji}^{L}}{\Delta_i}
-
\frac{T_{ji}^{R}}{1-\Delta_i}
\right)
=0.
\label{7d}
\end{align}

Consequently, the closed-form update rules in the $(t+1)$-th
maximization step are given by
\begin{align}
\mu_i^{(t+1)}
&=
\frac{
\frac{2\pi}{\gamma_i^{(t)}}
\sum_{j=1}^{N}
y_j^{\mathrm{nb}}
\beta_{ji}^{L(t)} x_j
+
\frac{1}{\sigma_i^{2(t)}}
\sum_{j=1}^{N}
y_j^{\mathrm{nb}}
T_{ji}^{R(t)} x_j
}{
\frac{2\pi}{\gamma_i^{(t)}}
\sum_{j=1}^{N}
y_j^{\mathrm{nb}}
\beta_{ji}^{L(t)}
+
\frac{1}{\sigma_i^{2(t)}}
\sum_{j=1}^{N}
y_j^{\mathrm{nb}}
T_{ji}^{R(t)}
},
\label{8a}\\
\gamma_i^{(t+1)}
&=
\left[
\frac{
\sum_{j=1}^{N}
y_j^{\mathrm{nb}}
\beta_{ji}^{L(t)}
\big(x_j-\mu_i^{(t)}\big)^2
}{
\sum_{j=1}^{N}
y_j^{\mathrm{nb}}
\beta_{ji}^{L(t)}
}
\right]^{1/2},
\label{8b}\\
\sigma_i^{2(t+1)}
&=
\frac{
\sum_{j=1}^{N}
y_j^{\mathrm{nb}}
T_{ji}^{R(t)}
\big(x_j-\mu_i^{(t)}\big)^2
}{
\sum_{j=1}^{N}
y_j^{\mathrm{nb}}
T_{ji}^{R(t)}
},
\label{8c}\\
\Delta_i^{(t+1)}
&=
\frac{
\sum_{j=1}^{N}
y_j^{\mathrm{nb}}
T_{ji}^{L(t)}
}{
\sum_{j=1}^{N}
y_j^{\mathrm{nb}}
T_{ji}^{(t)}
}.
\label{8d}
\end{align}

In each EM--Bragg iteration, the derived peak locations are correlated
with a specific set of lattice constants by minimizing the following
objective function:
\begin{equation}
\mathcal{L}_2
=
\frac{1}{2m'}
\sum_{i\in m'}
\left(
\mu_i^{\mathrm{WPEM}}
-
2\arcsin\!\left(
\frac{\lambda}{2d_{H_iK_iL_i}}
\right)
\right)^2,
\label{9}
\end{equation}
where $m'$ denotes a subset of stable diffraction peaks selected for
lattice refinement, $\lambda$ is the X-ray wavelength, and
$d_{H_iK_iL_i}$ is the interplanar spacing associated with the
$(H_iK_iL_i)$ reflection.
After this correction, all peak positions are unified and strictly
calibrated according to Bragg’s law.

A simple gradient-descent scheme is employed to solve
Eq.~\ref{9} for lattice calibration, yielding
\begin{equation}
\label{lc_update}
\mathrm{lc}^*
=
\mathrm{lc}
-
r
\frac{1}{m'}
\sum_{i\in m'}
\left(
\mu_i^{\mathrm{WPEM}}
-
\mu_i^{\mathrm{Brg}}
\right)
\frac{\lambda}{
d_{H_iK_iL_i}^2
\sqrt{1-\lambda^2/(4d_{H_iK_iL_i}^2)}
}
\frac{\partial d_{H_iK_iL_i}}{\partial \mathrm{lc}},
\end{equation}
where $\mathrm{lc}\in\{a,b,c,\alpha,\beta,\gamma\}$ denotes the lattice
constants and lattice angles, and
$\mu_i^{\mathrm{Brg}} = 2\arcsin(\lambda / 2d_{H_iK_iL_i})$ is the Bragg
peak position computed from the current lattice parameters.
The learning rate $r$ is estimated using a finite-difference scheme,
\begin{equation}
r = \frac{2\tau}{
\displaystyle
\frac{\partial \mathcal{L}_2\!\left(\mu_i^{\mathrm{WPEM}}, \mu_i^{\mathrm{Brg}}; lc+\tau\right)}{\partial (lc+\tau)}
-
\frac{\partial \mathcal{L}_2\!\left(\mu_i^{\mathrm{WPEM}}, \mu_i^{\mathrm{Brg}}; lc-\tau\right)}{\partial (lc-\tau)}
},
\end{equation}
where the step size is set to $\tau = 0.05$.
The optimization terminates when
$|\mathrm{lc}^* - \mathrm{lc}| < 10^{-8}$.
The updated lattice constants are subsequently used to compute Bragg
peak positions for the next EM--Bragg iteration.

For multi-component (materials) mixtures, peaks from all components are first pooled and sorted by diffraction angle, while their component labels (peak identities) are stored for later retrieval. In the maximization step, all peaks are updated within the unified framework of Eqs.~\ref{8a}--~\ref{8d}. The updated peak positions are then mapped back to each component and adjusted using its own lattice geometry via Eq.~\ref{lc_update}. Peaks belonging to the same crystalline component are grouped into batches, within which lattice parameters are refined independently by enforcing Bragg’s law. The optimized peak positions are finally mapped back to their original stored indices, propagated to the next EM--Bragg iteration, and used for subsequent assignment. This mixed-phase polycrystalline powder-diffraction workflow is termed \emph{batch EM--Bragg optimization}.

\paragraph{Volume Fraction Determination.}
The scattering of X-rays or neutrons by matter originates from their
interaction with constituent atoms. For X-rays, scattering arises from
the interaction between incident photons and the electrons surrounding
atomic nuclei~\cite{dinnebier2015powder}. According to Thomson scattering
theory, the intensity scattered by an ideal electron at a distance $R$
is given by
\begin{equation}
I = I_0 \frac{e^4}{m^2 c^4 R^2} \sin^2 \varphi,
\label{eq:11a}
\end{equation}
where $I_0$ is the incident X-ray intensity, $e$, $m$, and $c$ denote the
electron charge, mass, and the speed of light, respectively, and
$\varphi$ is the angle between the scattered X-ray direction and the
electric field vector of the incident radiation.

For unpolarized incident X-rays, the electric field direction is
arbitrary within the plane perpendicular to the incident beam. The
corresponding scattered intensity of a single electron becomes
\begin{equation}
I = I_0 \frac{e^4}{m^2 c^4} \frac{1+\cos^2 2\theta}{2},
\label{eq:11b}
\end{equation}
Here, $2\theta$ denotes the scattering angle, and the factor $(1+\cos^2 2\theta)/2$ represents the polarization term. In deriving the EM--Bragg procedure, we introduce $\widetilde{\theta}$ to distinguish model-parameter notation. For clarity and consistency with standard diffraction conventions, we use $\theta$ to denote the diffraction angle throughout the remainder of this section.

The atomic arrangement of a crystalline material can be described as an
infinitely periodic repetition of a unit cell in three-dimensional
space. The structure factor is therefore introduced to characterize the
diffraction amplitude of a unit cell relative to a single electron and
is expressed as
\begin{equation}
F_{HKL} =
\sum_{j=1}^{N}
f_j
\left[
\cos\!\left(2\pi(Hx_j + Ky_j + Lz_j)\right)
+ i \sin\!\left(2\pi(Hx_j + Ky_j + Lz_j)\right)
\right],
\label{eq:11c}
\end{equation}
where $f_j$ is the atomic scattering factor of the $j$-th atom, and
$(x_j,y_j,z_j)$ are its fractional coordinates within the unit cell.

The observed diffraction peak intensity is a combined result of electron
scattering, systematic extinction, diffraction geometry, absorption,
multiple scattering, and thermal vibration effects. It can be written as
\begin{equation}
I =
\frac{1}{32\pi D}
I_0
\frac{e^4}{m^2 c^4}
\frac{\lambda^3}{V_0^2}
V
F_{HKL}^2
P
\frac{1+\cos^2 2\theta}{\sin^2\theta \cos\theta}
e^{-2M}
A(\theta),
\label{eq:12}
\end{equation}
where $D$ is the specimen--detector distance, $\lambda$ is the X-ray
wavelength, $V_0$ is the unit-cell volume, $V$ is the irradiated powder
volume, $P$ is the reflection multiplicity, $e^{-2M}$ is the
Debye--Waller factor accounting for thermal vibration, and $A(\theta)$
denotes the absorption factor.

Based on diffraction intensity theory, the volume fractions of mixture
components can be determined quantitatively. Eq.~\ref{eq:12} can
be decomposed into an angle-independent constant and an
angle-dependent term. For powder components measured under identical
experimental conditions, the Debye--Waller factor $e^{-2M}$ can be
approximated as constant. The absorption factor is further simplified as
$A(\theta) \approx c/\mu$, where $c$ is a constant and $\mu$ is the
linear absorption coefficient, which depends only on the intrinsic
material properties.

Under these approximations, the diffraction intensity of component $i$
can be expressed as
\begin{equation}
I_i = C Q_i V_i,
\end{equation}
where $C$ is an angle-independent constant and
\begin{equation}
Q_i =
F_{HKL,i}^2
P_i
\frac{1+\cos^2 2\theta}{\sin^2\theta \cos\theta}
\frac{1}{\mu_i},
\end{equation}
is a component-dependent factor determined by the crystal structure,
atomic positions, diffraction angle, and physical properties of
component $i$.

The volume fraction ratio between components $i$ and $j$ is therefore
given by
\begin{equation}
\frac{V_i}{V_j}
=
\frac{I_i / Q_i}{I_j / Q_j}.
\label{eq:13}
\end{equation}

Since the structure factor $F_{HKL}^2$ depends on detailed atomic
coordinates and scattering factors that are not always readily
available, we provide three alternative schemes
for volume fraction estimation:
\begin{enumerate}
\item \emph{Strict intensity-based method}: the volume fraction is
computed directly using Eq.~\ref{eq:13}.
\item \emph{Approximate intensity-based method}: only the angular factor
$(1+\cos^2 2\theta)/(\sin^2\theta \cos\theta)$ is retained, yielding
$Q_i \approx (1+\cos^2 2\theta)/(\sin^2\theta \cos\theta)$.
\item \emph{Integral intensity method}: the volume fraction is estimated
from the integrated diffraction intensities,
\begin{equation}
\frac{V_i}{V_j}
=
\frac{\int I_i \,\mathrm{d}\theta}
{\int I_j \,\mathrm{d}\theta},
\label{eq:14}
\end{equation}
where $\int I_i$ and $\int I_j$ are the integrated areas of the
decomposed diffraction components.
\end{enumerate}

\subsection{XRDinspector: a screening layer for PXRD phase identification and refinement assessment}
\label{sec:XRDinspector}

\begin{figure}[H]
\centering
\includegraphics[width=1\textwidth]{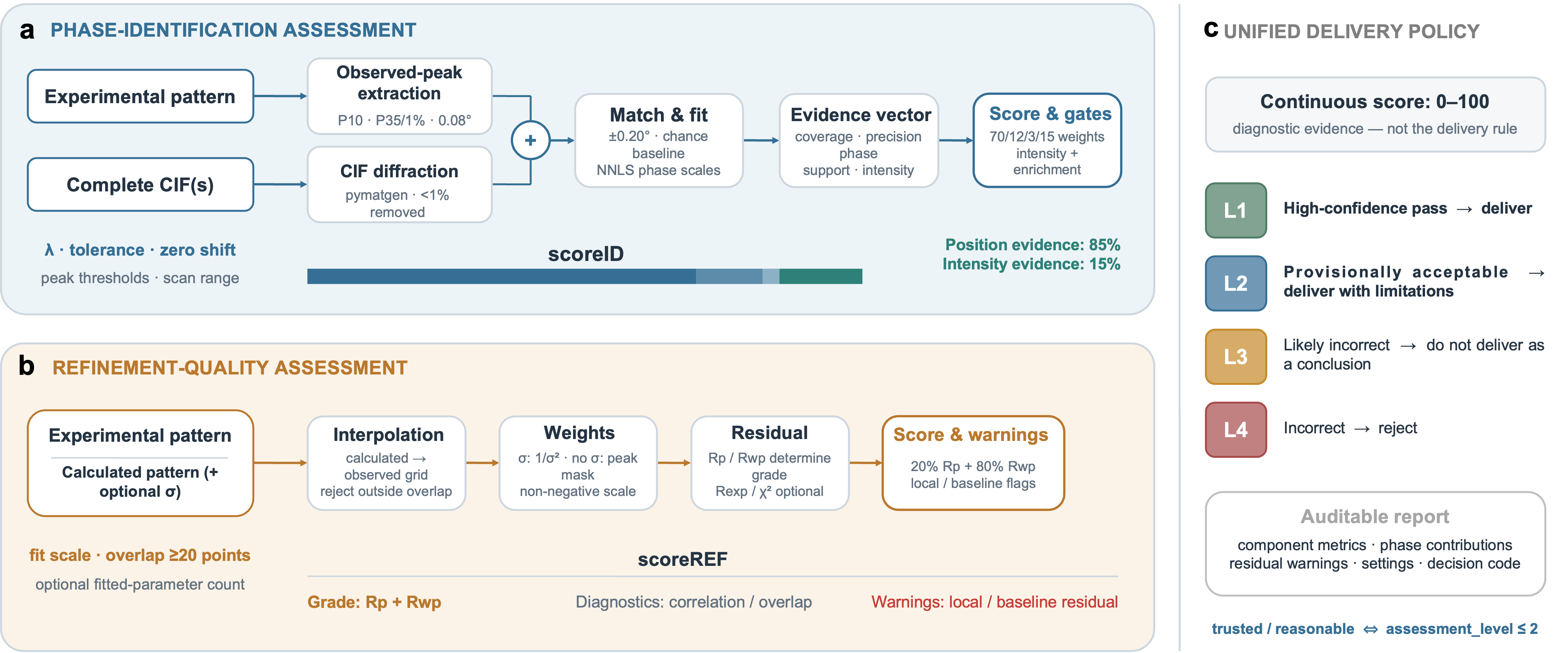}
\caption{\textbf{XRDinspector provides transparent quality assessment for PXRD results.} \textbf{a}, Candidate CIF(s) and an experimental pattern are aligned to quantify peak coverage, precision, phase support and intensity agreement. \textbf{b}, Observed and calculated profiles are evaluated using $R_p$, $R_{wp}$, correlation, overlap and local-residual diagnostics. \textbf{c}, The unified policy converts these diagnostic scores into four delivery levels, from high-confidence delivery (L1) to rejection (L4), and returns an auditable report with metrics, warnings and settings.}
\label{fig:xinspector-overview}
\end{figure}

PXRD is frequently used as a rapid decision tool in materials discovery, process control and characterization. Yet the path from a diffractogram to a conclusion combines several distinct tasks: candidate generation, structural calculation, profile modelling and expert judgement. These tasks should not be collapsed into a single opaque confidence label. In particular, a candidate may align with prominent features while leaving meaningful measured intensity unexplained; conversely, a physically plausible minor phase can remain below the practical detection limit. A useful quality layer must therefore expose evidence and limits rather than overstate certainty. XRDinspector accepts either (i) an observed pattern and one or more complete CIF structures, or (ii) observed and calculated patterns exported by external refinement software (Fig.~\ref{fig:xinspector-overview} \textbf{a}). 

\paragraph*{Position-resolved evidence for phase hypotheses}
For each supplied CIF, theoretical reflections are calculated with pymatgen's \texttt{XRDCalculator}, including systematic absences and calculated relative intensities\cite{ong2013python}. Reflections below a configurable relative-intensity threshold are removed. Local maxima in the observed pattern are baseline-corrected and separated by a minimum angular distance. A configurable $2\theta$ tolerance, with an optional zero shift, then relates observed and predicted positions.

The identification score aggregates complementary quantities. \textbf{Observed-peak coverage} measures the fraction of extracted experimental peak prominence accounted for by at least one predicted position. \textbf{Predicted-peak precision} measures the fraction of predicted positions supported by the experiment. \textbf{Phase support} is averaged across candidate phases so that an unsupported minor candidate cannot be concealed by a strongly supported major phase. In multiphase inputs, a non-negative least-squares fit of phase-relative intensities supplies a separate intensity-agreement component. These components, matching counts and phase-wise evidence states are retained in the returned report (Fig.~\ref{fig:xinspector-overview} \textbf{b}).

The score is continuous, but communication demands actionable boundaries. XRDinspector maps evidence to four delivery levels: high-confidence pass, provisionally acceptable, uncertain/likely incorrect and incorrect. The mapping is deliberately conservative: a level 1 or 2 conclusion can be delivered, with limitations for level 2; levels 3 and 4 are not deliverable as scientific conclusions. In a multiphase model, a zero fitted contribution is described as ``not detected or below detection limit'', not as proof that a candidate is incorrect. 

\paragraph*{Whole-pattern refinement checks}
For refinement outputs, the calculated trace is interpolated onto the experimental $2\theta$ grid. Evaluation is blocked when the common range is too short, preventing an artificially favourable partial-range comparison. A non-negative weighted global scale factor may be fitted before computing $R_p$, $R_{wp}$ and whole-pattern correlation. When uncertainty values are supplied through the Python API, $R_\mathrm{exp}$, $\chi^2$ and degrees of freedom are also reported.

The refinement score combines a residual component dominated by $R_{wp}$ with a correlation component. A maximum relative residual flags local disagreement even where a global score seems reasonable (Fig.~\ref{fig:xinspector-overview} \textbf{c}). The report therefore supports a compact first pass through large batches of refinement exports while preserving the diagnostics needed to inspect the difference pattern. 
\paragraph*{Identification scoring}
Single-phase candidates are scored by combining observed-peak coverage,
predicted-position precision, and positional support:
\begin{equation}
S_{\mathrm{id}}^{(1)}
= \kappa_{\mathrm{id}}
\left(
\alpha_1 C_{\mathrm{obs}}
+\beta_1 P_{\mathrm{pred}}
+\gamma_1 S_{\mathrm{phase}}
\right).
\end{equation}
For multiphase candidates, the score additionally accounts for
relative-intensity agreement:
\begin{equation}
S_{\mathrm{id}}^{(m)}
= \kappa_{\mathrm{id}}
\left(
\alpha_m C_{\mathrm{obs}}
+\beta_m P_{\mathrm{pred}}
+\gamma_m S_{\mathrm{phase}}
+\delta_m A_{\mathrm{int}}
\right).
\end{equation}
Here, $C_{\mathrm{obs}}$ denotes intensity-weighted observed-peak coverage,
$P_{\mathrm{pred}}$ predicted-position precision, $S_{\mathrm{phase}}$
mean per-phase positional support, and $A_{\mathrm{int}}$ non-negative
relative-intensity agreement. The coefficients are non-negative and
normalized within each scoring scheme. The default X-ray wavelength and
positional tolerance are denoted by $\lambda_0$ and $\Delta_{2\theta}$,
respectively.

\paragraph*{Refinement scoring}
Refinement quality is reported using
\begin{equation}
R_p
=\frac{\sum_i |y_{\mathrm{obs},i}-y_{\mathrm{calc},i}|}
       {\sum_i |y_{\mathrm{obs},i}|},
\qquad
R_{wp}
=\left[
\frac{\sum_i w_i(y_{\mathrm{obs},i}-y_{\mathrm{calc},i})^2}
     {\sum_i w_i y_{\mathrm{obs},i}^2}
\right]^{1/2}.
\end{equation}
When measurement uncertainties are unavailable, the weights $w_i$ are
estimated under an approximate Poisson noise model. The refinement score
combines a residual-based term with the Pearson correlation $r$ after
interpolation and, where applicable, non-negative scaling:
\begin{equation}
S_{\mathrm{ref}}
=\kappa_{\mathrm{ref}}
\left[
\eta\exp\!\left(-\frac{R_{wp}}{\tau}\right)
+(1-\eta)\frac{r+1}{2}
\right],
\end{equation}
where $\eta$ and $\tau$ are scoring parameters and
$\kappa_{\mathrm{ref}}$ sets the score scale.

\subsection{Gan Jiang enables evidence-governed refinement across FullProf and GSAS-II}
\label{sec:fullprof-gsas}
We developed an auditable workflow in which Gan Jiang coordinates refinement while the crystallographic engines perform the numerical optimization. For each task, Gan Jiang examines the diffraction data, instrument metadata, candidate structures, and scientific objective; formulates a focused, testable refinement hypothesis; selects an appropriate engine; and interprets the resulting metrics, residual profiles, and refined parameters. FullProf and GSAS-II complement our WPEM (PyXplore) engine, with each selected engine retaining responsibility for profile calculation and least-squares optimization. The integration of FullProf and GSAS-II into this workflow is described below.

\paragraph{Controlled hypothesis testing with FullProf}
This branch targets powder-diffraction data for which the candidate phase set
is known or supported by prior identification evidence. The inputs comprise a
diffraction pattern (for example, \texttt{.xy}), a structural model
(\texttt{.cif} or \texttt{.pcr}), instrument metadata and a stated refinement
objective. Gan Jiang audits these inputs and constructs a narrowly scoped PCR
branch for the current hypothesis. The automation layer then creates an
isolated working directory, copies the immutable inputs, invokes the FullProf
\texttt{fp2k} backend, preserves the execution logs and parses the refinement
metrics. Gan Jiang reviews these outputs before deciding whether to accept the
change, revert it or formulate the next refinement step.

\begin{table}[htbp]
\centering
\caption{Roles in the Gan Jiang--FullProf refinement workflow.}
\begin{tabularx}{\linewidth}{@{}>{\centering\arraybackslash}p{3.0cm}>{\centering\arraybackslash}X@{}}
\toprule
\textbf{Component} & \textbf{Role} \\
\midrule
Experiment & Provides diffraction data, structural candidates and instrument metadata. \\
Gan Jiang & Audits inputs, defines hypotheses, evaluates outputs and reports conclusions. \\
Automation & Isolates each run, executes the backend and records logs. \\
FullProf & Calculates diffraction profiles and performs least-squares refinement. \\
\bottomrule
\end{tabularx}
\end{table}

\begin{figure}[h]
\centering\includegraphics[width=\textwidth]{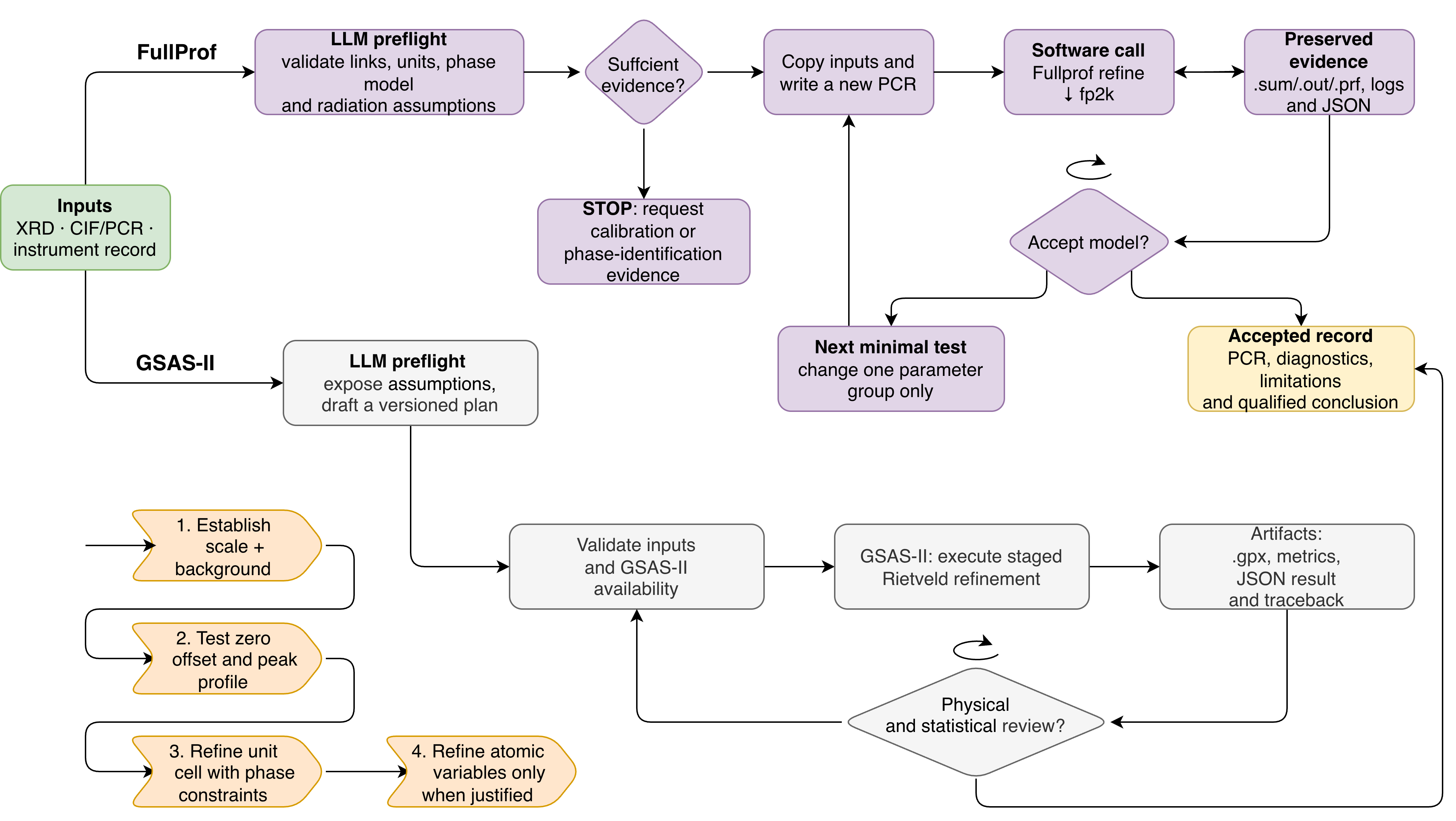}
\caption{\textbf{Evidence-governed refinement workflow for the FullProf and GSAS-II branches.} Experimental XRD data, structural CIFs or PCR files, and instrument records first undergo language-model preflight checks of links, units, phase model and radiation assumptions. Insufficient evidence triggers a request for calibration or phase-identification information. The FullProf branch (top) copies inputs into a new PCR file, performs refinement and either preserves the resulting profiles, logs and JSON summary or proceeds to the next minimal parameter-group test. The GSAS-II branch (bottom) validates its staged inputs, executes Rietveld refinement, and subjects the result to physical and statistical review before recording project files, metrics and tracebacks. Yellow chevrons denote the ordered refinement stages; diamonds denote decision gates. Accepted analyses retain diagnostics, limitations and a qualified conclusion.}
\label{fig:workflow}
\end{figure}

Figure~\ref{fig:workflow} summarizes the operational loop, which separates data
stewardship, numerical execution and scientific acceptance. Each iteration
tests one explicit hypothesis, such as releasing the background coefficients
or introducing a single preferred-orientation direction. Gan Jiang avoids
initially releasing strongly correlated quantities together; for example, it
stabilizes the global peak-position correction before refining the unit-cell
parameters. It then proceeds through the following reversible stages and
inspects the observed, calculated and difference profiles after each stage.

\begin{enumerate}
  \item \textbf{Preflight.} Verify the data link, scan range, wavelength, space group, atom types and phase list. An unknown sample first requires peak analysis, indexing or phase-search evidence.
  \item \textbf{Scale and global position.} With the structural, cell and profile model fixed, fit scale. If the residual supports a common position shift, test exactly one global correction and hold competing corrections fixed.
  \item \textbf{Background, cell and profile.} Begin with the smallest smooth background model; refine cell parameters only after the global position is stable; then add only the peak-width, shape or asymmetry terms supported by the residual.
  \item \textbf{Structure and specimen effects.} Release coordinates, constrained occupancies, preferred orientation or size/strain only when resolution, counting statistics and chemistry support their identifiability. Test one effect at a time.
  \item \textbf{Restricted final cycle.} Jointly release only parameters that were stable in earlier stages. Revert and simplify if values hit bounds, depend on the starting point or violate chemistry.
\end{enumerate}

\paragraph{Reproducible refinement planning with GSAS-II}
The GSAS-II branch connects the same reasoning loop to the scriptable GSAS-II
API \cite{toby2013gsas}. Its inputs, including the diffraction data, instrument parameter file, and one or more structural CIFs, remain immutable (Fig.~\ref{fig:workflow}). For every attempt, Gan
Jiang creates a new isolated working directory and records the intended model
changes in \texttt{refinement\_plan.json}. The automation layer returns both a
GSAS-II project and \texttt{refinement\_result.json}; the latter records the
execution state, plan, file paths, fit metrics and, when a run fails, the error
traceback. Consequently, each action can be reconstructed independently of the
dialogue that initiated it.

Across both branches, Gan Jiang serves as the planner between the scientist and
the deterministic refinement engine. It maintains a ledger of what changed,
why the change was introduced and which diagnostic was expected to improve.
Fit statistics such as $R_{wp}$ are treated as diagnostics rather than final
verdicts. Acceptance additionally requires inspection of the residual profile,
phase identity, space group, composition, parameter uncertainties and the
physical plausibility of lattice parameters, displacement parameters and site
occupancies. Gan Jiang also flags parameters that are strongly correlated,
sensitive to initialization or unable to account for systematic misfit. This
evidence-governed sequencing prevents reductions in $R_{wp}$ obtained through
non-physical compensation while producing a complete, reviewable record of
the refinement trajectory.

\clearpage
\section{Supplementary results}
 \subsection{Gan Jiang API}
\label{sec:GJapi}

\input{api}

\subsection{Gan Jiang structural dataset}
\label{sec:mp500}

A crystal structure database is essential to PXRD-based workflows. We maintain the Gan Jiang database, which combines structures sourced from the Materials Project with experimental structures from RRUFF, opXRD, and our previously collected crystals. The current system searches these two sources for candidate structures. It also supports external databases, such as ICSD and CCDC, through plugins that parse and manage their records dynamically; these external sources are not used in the present system.

\paragraph{MP500 dataset}
 We assembled MP500, a compact, ASE-compatible SQLite database of periodic crystal structures indexed by Materials Project-style identifiers, to serve as a structural backbone for Gan Jiang. The present release contains 100,315 structures spanning 84 chemical elements and comprising 3,143,789 atoms in total. Each entry stores atomic numbers, Cartesian coordinates, a unit cell, periodic-boundary metadata, atom count, and unit-cell volume, along with a unique internal identifier, a sequential numeric label, and an \texttt{mpid} field. Structures contain 1 to 360 atoms per stored cell, with a median of 20 atoms; the median volume per atom is 15.09~\AA$^3$.

\begin{figure}[t]
  \centering
  \includegraphics[width=\linewidth]{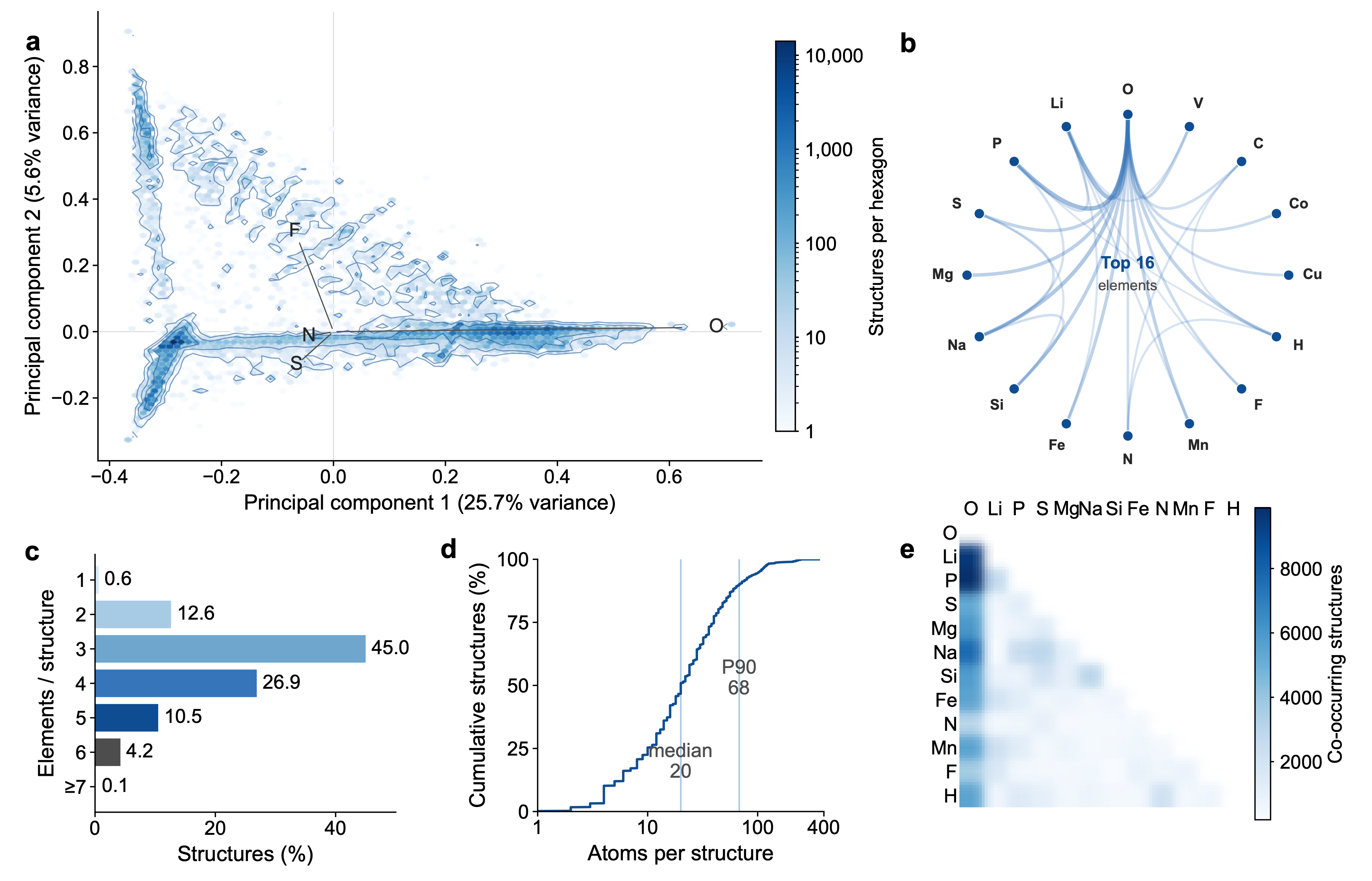}
  \caption{\textbf{Compositional and structural landscape of the MP500 dataset.}
  \textbf{a}, Principal-component map of elemental composition for all 100,315
  stored structures; colour denotes the number of structures per hexagonal bin.
  \textbf{b}, Co-occurrence network of the 16 most prevalent elements; links
  represent the 28 largest pairwise co-occurrence counts. \textbf{c},
  Distribution of chemical complexity, expressed as the number of distinct
  elements per structure. \textbf{d}, Empirical cumulative distribution of
  atoms per stored unit cell; vertical lines mark the median and 90th
  percentile. \textbf{e}, Pairwise co-occurrence matrix of prevalent elements.}
  \label{fig:mp500}
\end{figure}

The collection spans broad elemental and compositional space
(Fig.~\ref{fig:mp500}). Tri-element structures form the largest chemical-order
class (45,132 structures, 45.0\%), followed by four-element structures (26,974
structures, 26.9\%) and binary structures (12,677 structures, 12.6\%). This
composition is useful for benchmarking methods that must remain informative
beyond simple unary and binary materials. We report structure-level elemental
prevalence rather than raw atom counts, so that elements occurring in large
conventional cells do not dominate the dataset summary.
Figure~\ref{fig:mp500} is a descriptive analysis of the complete database. The figure is intended to
answer five complementary questions: which compositional regions are represented
(\textbf{a}), which elements recur together (\textbf{b} and \textbf{e}), how chemically complex the stored
structures are (\textbf{c}), and how large their crystallographic cells are (\textbf{d}).

For structure $i$, we constructed an 84-dimensional composition vector
$\mathbf{x}_i=(x_{i1},\ldots,x_{i84})$, where $x_{ij}=n_{ij}/N_i$ is the
fraction of atoms of element $j$, $n_{ij}$ is its stored atomic count and
$N_i=\sum_j n_{ij}$ is the total number of atoms in the stored cell. Principal
component analysis (PCA) was fitted after mean-centering these vectors across
all 100,315 structures. The first two components explain 25.7\% and 5.6\% of
the total compositional variance, respectively. Each hexagon aggregates
structures with similar two-dimensional PCA coordinates, and its colour encodes
the number of structures in that bin on a logarithmic scale. Thus, dense regions
identify frequently represented composition families, whereas isolated hexagons
identify sparsely represented compositions. The labelled arrows correspond to
selected high-loading elements and indicate the directions in which their
atomic fractions contribute most strongly to the displayed PCA axes (Fig.~\ref{fig:mp500}~\textbf{a}).

For each element pair $(p,q)$, we calculated the structure-level co-occurrence
count $C_{pq}=\sum_i I_{ip}I_{iq}$, where $I_{ip}=1$ when structure $i$
contains element $p$ and zero otherwise. Nodes are the 16 elements with the
highest structure prevalence; node area increases with the number of structures
containing that element. To keep the graphic readable, only the 28 pairs with
the largest raw values of $C_{pq}$ are drawn. Link width and opacity both
increase with $C_{pq}$. The panel therefore reveals recurring compositional
associations in the stored database, not chemical bond strengths, interaction
energies or conditional probabilities. In particular, high prevalence of an
element can increase its raw co-occurrence counts; panel~\textbf{e} provides the full
matrix representation for the leading elements (Fig.~\ref{fig:mp500}~\textbf{b}).

Chemical complexity was defined as $K_i=\sum_j I_{ij}$, the number of distinct
elements in structure $i$. We calculated the percentage
$100\,N^{-1}\sum_i I(K_i=k)$ for each value of $k$, grouping the sparse tail as
$K_i\geq7$. The resulting distribution is centred on ternary materials:
45,132 structures (45.0\%) contain three elements. Four-element and binary
structures account for 26.9\% and 12.6\%, respectively, whereas unary
structures account for 0.6\%. The dataset is consequently weighted towards
ternary and quaternary compositions, a point that should be considered when
constructing train--test splits or reporting performance by chemical order (Fig.~\ref{fig:mp500}~\textbf{c}).

For each entry, cell size is the stored atom count $N_i$, not the number of
symmetry-inequivalent sites and not a primitive-cell atom count. We plot the
empirical cumulative distribution
$F(n)=N^{-1}\sum_i I(N_i\leq n)$ on a logarithmic horizontal axis, which
preserves both small and large cells. The median is 20 atoms per stored cell and
the 90th percentile is 68 atoms; the complete range is 1--360 atoms (Fig.~\ref{fig:mp500}~\textbf{d}).

The panel~\textbf{e}  displays the same count $C_{pq}$ as Fig.~\ref{fig:mp500}~\textbf{b} for the
12 elements with the highest structure prevalence. The diagonal is suppressed
and one triangular half is shown because $C_{pq}=C_{qp}$. Darker cells denote
more structures containing both indicated elements. Unlike the network, this
matrix retains every pair within the selected 12-element set, including weak
associations omitted from the network for visual clarity. Together, panels \textbf{b}
and \textbf{e} distinguish the database's dominant co-occurrence backbone from its
lower-frequency composition pairs (Fig.~\ref{fig:mp500}~\textbf{e}).

\subsubsection{Experimental dataset}
\label{sec:experimental-structures}

To complement the structures in MP500, we assembled an
experiment-linked structure collection from RRUFF and a second branch that
combines opXRD with our private data. RRUFF provides
crystallographic models for experimentally characterized minerals together
with diffraction, spectroscopic, and chemical records
\cite{lafuente2015rruff}. opXRD is a multi-institutional collection of
experimental powder X-ray diffraction data; among its labeled records, a
subset is accompanied by a complete crystal structure
\cite{hollarek2026opxrd}. The present Gan Jiang snapshot uses 1,159
RRUFF structures and 848 structures from the combined opXRD + private-data
branch, giving 2,007 source-qualified records in total. Only records containing
a complete, parseable three-dimensional periodic structure are included in the
candidate library.

\begin{figure*}[t]
    \centering
    \includegraphics[width=0.9\textwidth]{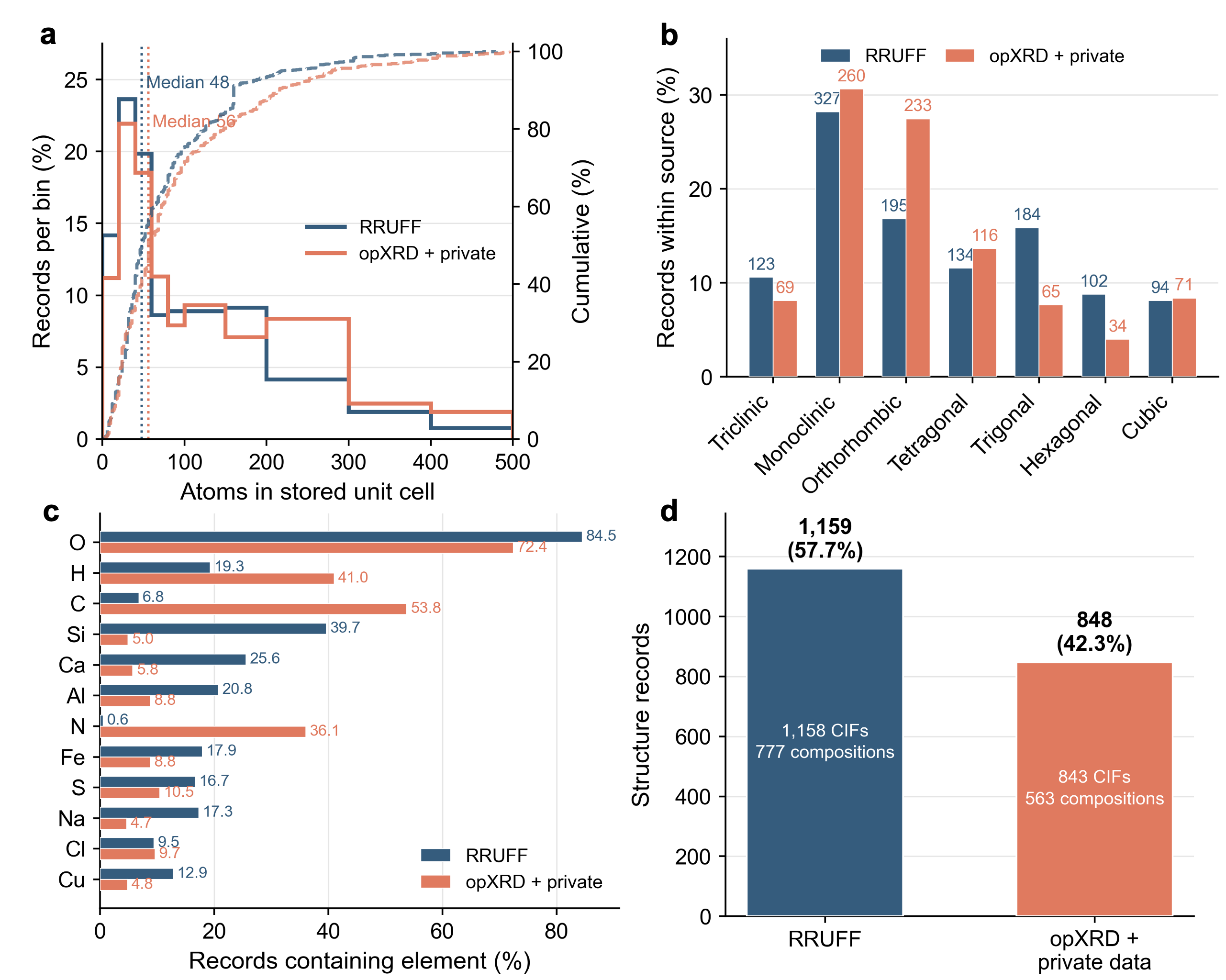}
    \caption{Overview of the experimental candidate library. \textbf{a}, Distribution
    of the number of atoms in the stored unit cells; solid outlines show binned
    fractions, dashed curves show empirical cumulative distributions, and
    dotted vertical lines mark the medians. \textbf{b}, Crystal-system distributions
    within RRUFF and the combined opXRD + private-data branch; symmetries were
    identified with \textsc{spglib} using
    $\mathrm{symprec}=10^{-2}\,\text{\AA}$, and numbers above the bars give
    record counts. \textbf{c}, Occurrence of the 12 most prevalent elements, with each
    element counted at most once per structure record. \textbf{d}, Source composition
    of the collection, with distinct CIF counts determined from the stored
    SHA-256 checksums.}
    \label{fig:experimental-structure-statistics}
\end{figure*}

During ingestion, each structure is stored with its source and source-specific
identifier, sample identifier, original CIF filename, and a SHA-256 checksum.
These fields preserve provenance and permit exact duplicate detection without
conflating structures that merely share a composition. The current snapshot
contains 2,001 distinct CIF checksums and 1,339 distinct reduced compositions,
spans 80 chemical elements, and contains cells ranging from 2 to 496 atoms
(median 54). For the descriptive statistics in
Fig.~\ref{fig:experimental-structure-statistics}, space groups were reassigned
from the stored cells using \textsc{spglib} with a symmetry tolerance of
$10^{-2}\,\text{\AA}$ \cite{togo2024spglib}. The resulting structures cover
141 space groups across all seven crystal systems; monoclinic structures are
the largest class (587 records), followed by orthorhombic structures (428
records). The two sources are chemically complementary: RRUFF is enriched in
oxygen-bearing mineral systems containing Si, Ca, and Al, whereas the combined
opXRD + private-data branch contains a larger fraction of C-, H-, and N-bearing
structures. This diversity broadens the experimental candidate space beyond
the computed inorganic structures in MP500.

\subsection{Indexing of Gan Jiang crystal dataset }

Efficient structure search and match requires a searchable link between
crystal structures and their expected powder-diffraction patterns. To
establish this link, theoretical PXRD patterns are generated from the
structures in the Gan Jiang database and associated with their source
records. Each indexed record retains the structure identifier, provenance,
and a diffraction signature comprising predicted peak positions and
relative intensities. This allows a measured pattern to be compared with
database candidates without recalculating every theoretical pattern during
each search.
During identification, the observed diffraction peaks are used to retrieve
and rank plausible structures. Candidate matches are then examined using
the full pattern and, where appropriate, passed to subsequent refinement.
Keeping the diffraction index linked to the underlying structural records
allows search results to be traced back to their source structures and
updated when database entries change. 

\begin{figure}[h]
\centering\includegraphics[width=0.8\textwidth]{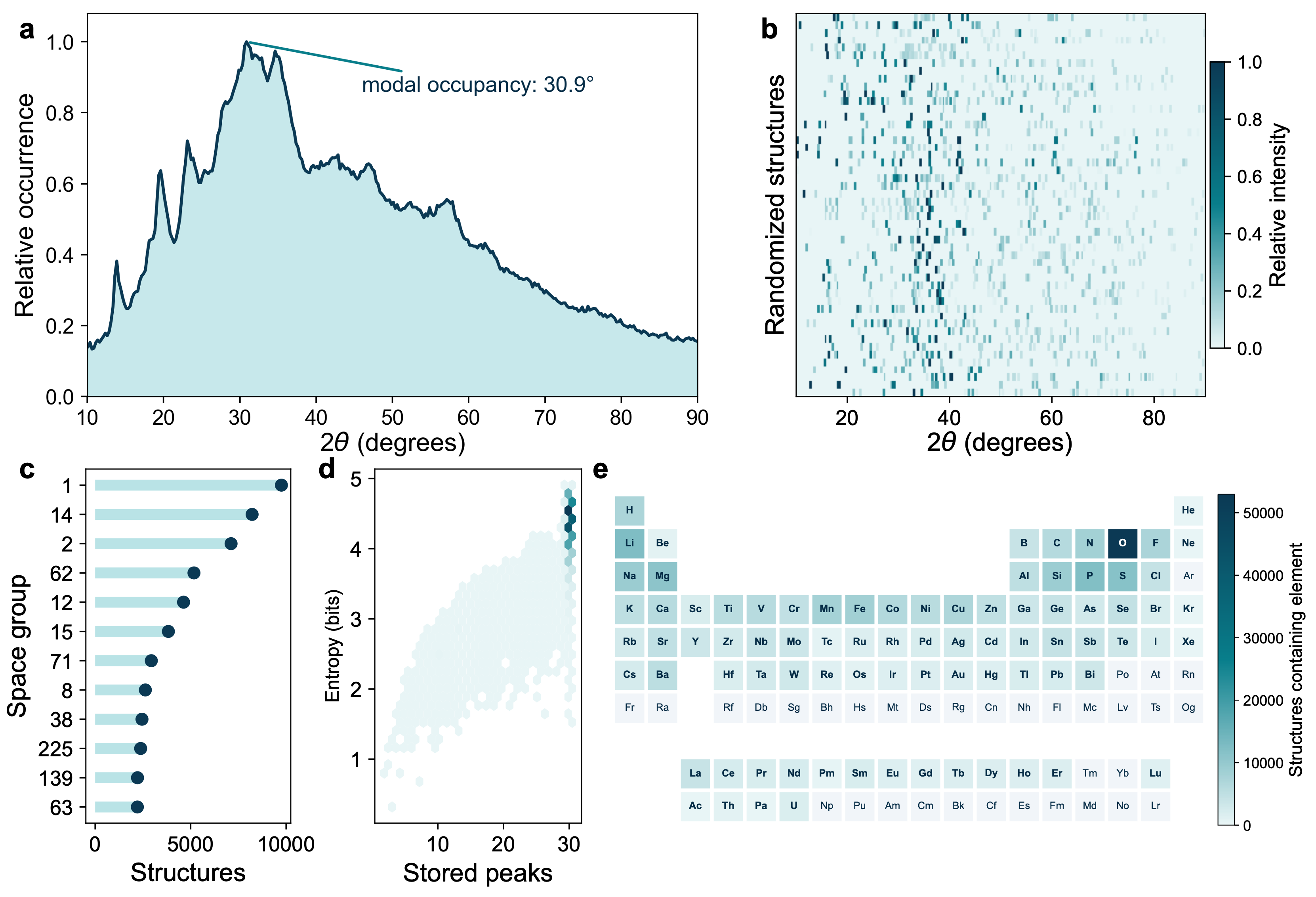}
\caption{\textbf{Five-dimensional index of the MP500 XRD archive.} \textbf{a}, Archive-wide occupancy of 2,934,906 stored theoretical reflection positions, with a modal value at 30.9$^\circ$ 2$\theta$. \textbf{b}, Structure-resolved diffraction barcode for 50 fixed-seed records; colour denotes normalized reflection intensity. \textbf{c}, Rank-frequency distribution of leading space groups. \textbf{d}, Hexagonally binned relation between stored-peak count and Shannon entropy of normalized peak intensity. \textbf{e}, Elemental support of the inverted index; colour denotes the number of structures containing an element.}
\label{fig:indexing}
\end{figure}

We index archive coverage along five coupled dimensions (Fig.~\ref{fig:indexing}). In Fig.~\ref{fig:indexing}\textbf{a}, every stored reflection contributes to a 0.25$^\circ$ 2$\theta$ histogram, normalized to its maximum. The resulting curve is structured rather than spectrally uniform: occupancy rises sharply in the low-angle region, reaches a modal value at 30.9$^\circ$, and decays steadily above approximately 60$^\circ$. Thus, high-angle agreement is intrinsically supported by fewer reference reflections and should not be interpreted as carrying the same prior weight as agreement near the modal region. Figure~\ref{fig:indexing}\textbf{b} converts the 50 fixed-seed example patterns into a structure-by-angle barcode: each 0.1$^\circ$ bin is coloured by the maximum normalized intensity in a five-bin local window. The barcode shows repeated low-angle bands across multiple records, but also shows that the dark, high-intensity cells are discontinuous from one structure to the next. In other words, common individual peak positions coexist with highly specific \emph{co-occurrence patterns}. This is the retrieval-relevant result: a single peak near a frequently occupied angle is weak evidence, whereas agreement among several peaks in the same barcode pattern is much more discriminative for phase identification.

Figure~\ref{fig:indexing}\textbf{c} ranks the 12 most frequent space groups by record count. It exposes strong crystallographic concentration: the leading group alone contributes nearly 10,000 structures, followed by a progressive decline across the remaining groups. This distribution tells readers that a chemically filtered retrieval will still confront an uneven structural prior. Figure~\ref{fig:indexing}\textbf{e} maps the inverted elemental index onto the periodic table; each tile is coloured by the number of structures containing that element. Oxygen is especially prevalent, whereas pale tiles identify elements absent from the archive. These are explicit no-coverage regions for an element-constrained query, rather than an unexplained retrieval failure.

Finally, Fig.~\ref{fig:indexing}\textbf{d} measures the information compactness of every record. The horizontal axis is the number of retained reflections and the vertical axis is Shannon entropy, $H=-\sum_i p_i\log_2 p_i$, calculated from the normalized peak intensities $p_i$ of one structure. Each hexagon represents records falling within the same peak-count--entropy interval; darker hexagons contain more records. The rising envelope means that patterns with more retained peaks can distribute intensity across more reflections and therefore reach higher entropy. Conversely, the low-entropy region at small peak counts corresponds to fingerprints dominated by one or a few reflections. The dense vertical accumulation at 30 peaks is expected: 30 is the archive's storage cap, so all patterns with at least 30 usable reflections terminate at the same horizontal coordinate. The median entropy is 4.39 bits. For retrieval, this panel distinguishes a low-entropy candidate, whose score may be driven by only a few dominant peaks, from a high-entropy candidate whose agreement is distributed over a richer fingerprint.

\subsection{Skill learning through refinement case studies}
\label{sec:skill}

We compare revisions to the FullProf, GSAS-II, and WPEM refinement skills. Paired blocks reproduce selected skill excerpts before and after revision. FullProf represents a historical workflow redesign, whereas the GSAS-II and WPEM excerpts below come from skill-only optimization. 

\paragraph{FullProf: from a general refinement sequence to a bounded profile search.}

\begin{skilltext}{skillbefore}{Before}{staged parameter refinement}
5. \textbf{Sequential Logic}: Follow the mandatory refinement loop strictly: \texttt{scale} $\rightarrow$ \texttt{zero} $\rightarrow$ \texttt{cell} $\rightarrow$ \texttt{profile\_w}. Only introduce \texttt{background} when there is clear evidence of a non-zero baseline in the residual audit, and handle failures gracefully.
\end{skilltext}

\begin{skilltext}{skillafter}{After}{task-specific search policy}
Use the fixed forward-model tools to refine Gaussian width and angular zero while retaining the supplied structure. Experimental XRD, neutron, TOF and quantitative phase analysis require their own radiation, background and uncertainty models.

Start from the exact native fit proposed by the tool's approximate initialization. Test a neighborhood with width radius 0.008 degrees and zero radius 0.004 degrees. Then reduce the radii to 0.004/0.002 and 0.002/0.001 degrees. Inspect whether the best point moves, the weighted residual falls and the relative L2 error improves. When the best point keeps moving, repeat a neighborhood at that resolution before making it smaller.
\end{skilltext}

The revised text makes the search variables, step sizes, and progression rule explicit. 

\paragraph{GSAS-II: from general rollback advice to explicit execution safeguards.}
The original playbook already required rejecting unhelpful or unstable parameter additions. The skill-only revision retained that principle and added a specific warning about simultaneous refinement and saved-project consistency.

\begin{skilltext}{skillbefore}{Before}{general acceptance and rollback rule}
Disable and roll back a new parameter class when it produces only a trivial Rwp reduction, hits a bound, increases esds, has high correlation, or does not improve the corresponding residual shape.
\end{skilltext}

\begin{skilltext}{skillafter}{After}{added staged-refinement safeguard}
Prefer a staged, individually verified sequence of small refinements over one large simultaneous refinement: a fully simultaneous pass can leave the saved project inconsistent with its own final calculated curve even when it converges cleanly.
\end{skilltext}

The revision also added the following response contract. There was no corresponding block in the frozen skill; the values shown are the original candidate's example settings.

\begin{skillcode}{skillafter}{After}{newly added parameter-response contract}
{
  "parameters": {
    "background_terms": 4, "cycles": 12,
    "profile_order": "after_cell", "profile": "W",
    "zero": false, "joint_final": false,
    "initial_W": 43.5
  },
  "reason": "One-sentence physical justification per override."
}
\end{skillcode}

The new contract requires overrides inside \texttt{parameters}, preventing a documented failure in which top-level keys were silently ignored. The skill further requires confirmation from run output before claiming that a suggestion was applied. Its added rules set \texttt{joint\_final=false} by default following saved-curve verification failures, impose a practical \texttt{initial\_W} floor of 2 above the schema minimum of 0.25 to avoid invalid starting widths with the default instrument file, and limit large patterns to six cycles, at most eight, with no joint pass under the historical 600~s budget. 

\paragraph{WPEM: reconciling the known-CIF workflow.}
The main workflow initially required structure solving for every supplied phase. The revised workflow and examples consistently bypassed that step for known structures.

\begin{skillcode}{skillbefore}{Before}{structure solving in the default call sequence}
# Run once per phase, before CIFpreprocess.
WPEM.StructureSolve(
    no_bac_intensity_file="./ConvertedDocuments/no_bac_intensity.csv",
    cif_file="phase-0.cif",
)
\end{skillcode}

\begin{skillcode}{skillafter}{After}{known-CIF default in the call example}
# Known-CIF default: skip StructureSolve unless explicitly requested.
# WPEM.StructureSolve(...)  # Only when pre_solve=true
\end{skillcode}

The revision supplemented generic tuning advice with an explicit response to Bragg-step divergence and timeouts.

\begin{skilltext}{skillbefore}{Before}{general convergence adjustments}
- For inadequate convergence, first verify CIF identity/order, wavelength, range, and peak lists. Then adjust \texttt{InitializationEpoch}, \texttt{iter\_max}, \texttt{subset\_number}, \texttt{low\_bound}, or \texttt{up\_bound} conservatively.
\end{skilltext}

\begin{skilltext}{skillafter}{After}{budget-aware tuning and targeted recovery}
- For inadequate convergence, first verify CIF identity/order, wavelength, range, and peak lists. Then adjust \texttt{InitializationEpoch}, \texttt{iter\_max}, or \texttt{subset\_number} conservatively, respecting the runtime rules above.

- For Bragg-step divergence or timeouts on low-symmetry, reflection-rich cells, prefer \texttt{fixed\_cell=true} over more iterations.
\end{skilltext}

The updated skill recommends three to five EM iterations for dense patterns or reflection-rich low-symmetry cells under the historical 600~s, one-CPU budget. The interface default remains \texttt{fixed\_cell=\allowbreak false}; the example's \texttt{true} value is a suggested override. The text also strengthens default-first, one-group-at-a-time tuning, citing a training example in which a multi-group intervention gave relative $L_2\approx0.81$, compared with approximately 0.23 for defaults.

\clearpage
\section{Extended results}

Table \ref{tab:egyptian-lattice} summarizes the refined lattice parameters of the Egyptian cosmetic powder.

\begin{table}[h]
\centering
\caption{Refined lattice parameters of the five mineral phases identified in
the ancient Egyptian cosmetic powder. Values in parentheses are the estimated
uncertainties in the final reported digit.}
\label{tab:egyptian-lattice}
\small
\setlength{\tabcolsep}{4pt}
\begin{tabular}{@{}lcccccc@{}}
\toprule
Phase & $a$ (\AA) & $b$ (\AA) & $c$ (\AA) & $\alpha$ ($^\circ$) & $\beta$ ($^\circ$) & $\gamma$ ($^\circ$) \\
\midrule
Gypsum     & 5.6800(9) & 15.2144(0) & 6.5303(7) & 90 & 118.4841(5) & 90 \\
Phosgenite & 8.1600(0) & 8.1600(0)  & 8.8834(3) & 90 & 90          & 90 \\
Cerussite  & 5.1794(7) & 8.4922(9)  & 6.1418(1) & 90 & 90          & 90 \\
Galena     & 5.9388(0) & 5.9388(0)  & 5.9388(0) & 90 & 90          & 90 \\
Laurionite & 9.7002(6) & 4.0200(3)  & 7.1108(1) & 90 & 90          & 90 \\
\bottomrule
\end{tabular}
\end{table}

\begin{figure}[H]
  \centering
  \includegraphics[width=\linewidth]{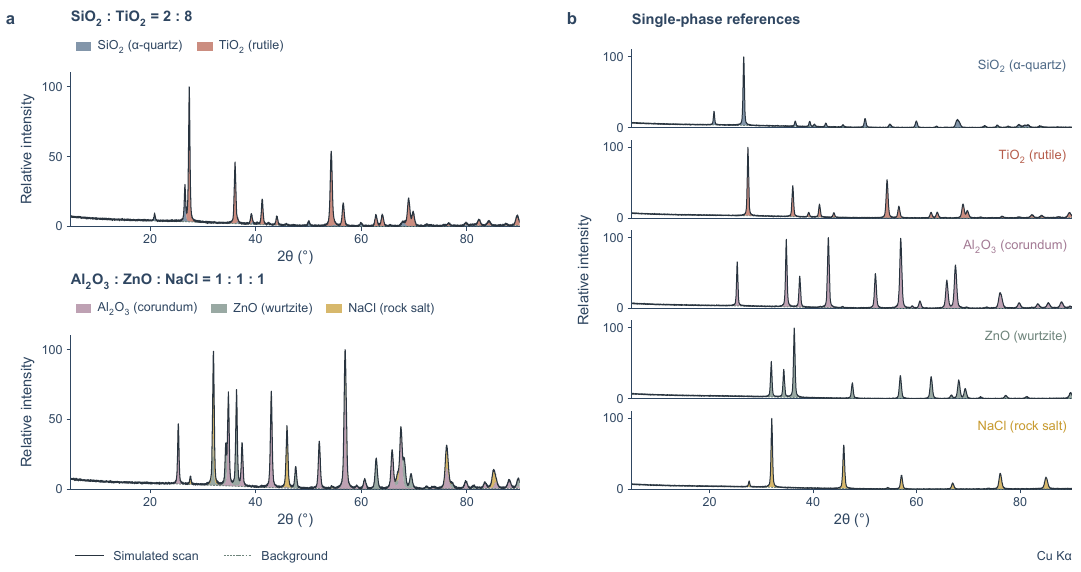}
  \caption{\textbf{Whole-pattern decomposition of binary and ternary powder mixtures.} \textbf{a}, Fits to a nominal 2:8 \ce{SiO2} ($\alpha$-quartz)--\ce{TiO2} (rutile) mixture (top) and a nominal 1:1:1 \ce{Al2O3} (corundum)--\ce{ZnO} (wurtzite)--\ce{NaCl} (rock salt) mixture (bottom). Black curves show the measured scans, grey dashed curves show the fitted background and coloured fills show the phase-resolved contributions. \textbf{b}, Component diffraction profiles recovered for each phase in the two mixtures, using the same colour code.}
  \label{fig:multi-phase}
\end{figure}

Figure~\ref{fig:multi-phase} illustrates whole-pattern decomposition of controlled binary and ternary mixtures. For the nominal 2:8 \ce{SiO2}--\ce{TiO2} mixture and the nominal 1:1:1 \ce{Al2O3}--\ce{ZnO}--\ce{NaCl} mixture, Gan Jiang jointly fits the measured scan, smooth background and phase-resolved diffraction contributions. The recovered component profiles retain the characteristic reflections of their assigned phases, showing that the system can separate overlapping signals while preserving a direct, inspectable connection to the measured pattern.

\newpage
\renewcommand{\refname}{References}
\putbib
\end{bibunit}

\end{document}

%% file: api.tex
\paragraph{Agent mode and direct CLI access.}

Gan Jiang is primarily designed as an autonomous
diffraction-analysis agent that coordinates specialised
scientific tools through executable skills and
evidence-guided decision policies. In the standard agent
mode, a user provides diffraction data together with a
scientific objective, and Gan Jiang determines which
operations to perform, how the intermediate results should
be interpreted, and which subsequent analysis steps are
required.

In addition to this agent-based interaction, we expose a
subset of the computational capabilities integrated in
Gan Jiang for direct programmatic use. These capabilities
can be accessed through a command-line interface (CLI),
providing an alternative mode in which the caller explicitly
selects individual operations. The underlying diffraction
tools are shared between the two modes; the distinction is
whether their execution is automatically orchestrated by
Gan Jiang or specified directly by the user or an external
program.

For example, in agent mode, a request to analyse an unknown
powder-XRD pattern may cause Gan Jiang to perform peak
detection, retrieve candidate structures, inspect the
resulting evidence and subsequently launch refinement.
In CLI mode, the same operations can instead be invoked
individually, for example by directly requesting phase
matching, calculating a diffraction pattern from a CIF,
or submitting a specified structural model for refinement.

\paragraph{CLI invocation.}

Programmatic access is provided through the
\texttt{delta-cli} client on the Delta infrastructure
platform \cite{delta_infra_dashboard}. After installing and authorizing via simple command:

\begin{skillcode}{skillafter}{CLI}{Installation and authorization}
# install
npx @delta-infra/cli@latest install

# authorize
delta-cli auth login
\end{skillcode}

Scientific operations follow the general form

\begin{skillcode}{skillafter}{CLI}{Scientific tool invocation}
# caling science tools
delta-cli science invoke --tool xrd \
  --endpoint <ENDPOINT> \
  [--data <JSON> | --file <FILE> --params <JSON>]
\end{skillcode}

Here, \texttt{--tool xrd} selects the diffraction
toolchain, while \texttt{--endpoint} specifies the
operation to execute. Structured requests are supplied
through \texttt{--data}. File-based inputs are uploaded
through \texttt{--file}, together with metadata provided
through \texttt{--params}.

The interface exposes operations for service inspection,
artifact management, phase identification, multiphase
analysis, structure-based diffraction calculation,
refinement and asynchronous job management.
Table~\ref{tab:ganjiang-api-endpoints} summarises the
major functional groups. The listed names are endpoint
identifiers supplied to the CLI rather than literal HTTP
paths.

\begin{table}[htbp]
    \centering
    \caption{
        Programmatic access to selected capabilities of the
        Gan Jiang diffraction toolchain. Endpoint identifiers
        are supplied to the \texttt{--endpoint} argument of
        \texttt{delta-cli science invoke --tool xrd}.
    }
    \label{tab:ganjiang-api-endpoints}

    \begingroup
    \small
    \setlength{\tabcolsep}{5pt}
    \renewcommand{\arraystretch}{1.2}

    \begin{tabularx}{\linewidth}{
        @{}
        >{\centering\arraybackslash}p{0.28\linewidth}
        >{\centering\arraybackslash}X
        @{}
    }
        \toprule
        \textbf{Function} &
        \textbf{Endpoint identifiers} \\
        \midrule

        Service status and capabilities &
        \nolinkurl{healthz},
        \nolinkurl{readyz},
        \nolinkurl{capabilities},
        \nolinkurl{xmatcher-status}
        \\

        \addlinespace[3pt]
        Artifact management &
        \nolinkurl{xmatcher-artifacts},
        \nolinkurl{xmatcher-artifact},
        \nolinkurl{xmatcher-artifact-content}
        \\

        \addlinespace[3pt]
        Identification and decomposition &
        \nolinkurl{xmatcher-detect-peaks},
        \nolinkurl{xmatcher-match},
        \nolinkurl{xmatcher-multiphase-match},
        \nolinkurl{xqueryer-predictions},
        \nolinkurl{xdecomposer-predictions}
        \\

        \addlinespace[3pt]
        Structure export and pattern calculation &
        \nolinkurl{mp500-cif-exports},
        \nolinkurl{xmatcher-cif-xrd},
        \nolinkurl{xmatcher-pdf-peaks-xlsx},
        \nolinkurl{xmatcher-refinement-patterns}
        \\

        \addlinespace[3pt]
        Refinement &
        \nolinkurl{refinement-gsasii},
        \nolinkurl{refinement-fullprof},
        \nolinkurl{refinement-wpem}
        \\

        \addlinespace[3pt]
        Job management &
        \nolinkurl{refinement-job},
        \nolinkurl{refinement-job-events},
        \nolinkurl{refinement-job-artifacts},
        \nolinkurl{refinement-job-cancel}
        \\

        \bottomrule
    \end{tabularx}

    \endgroup
\end{table}

\paragraph{Artifact-based data exchange.}

Files exchanged through the CLI are represented as
artifacts. Diffraction patterns, structural CIFs and
derived computational inputs are first uploaded through
\nolinkurl{xmatcher-artifacts}. The upload specifies the
filename, artifact kind and media type, and returns an
\texttt{artifact\_id} together with a SHA-256 checksum.

Subsequent operations reference these identifiers rather
than a client-local file path. The artifact kind records
the intended role of the file, for example
\texttt{observed\_pattern},
\texttt{model\_input\_pattern} or \texttt{cif}.
Derived representations, such as refinement-ready
patterns and provenance manifests, are stored as separate
artifacts rather than overwriting the original measurement.

This design allows the same input to be reused across
different operations while retaining its identity and
provenance. For example, a candidate CIF obtained through
structure retrieval can subsequently be passed to
diffraction-pattern calculation and to a refinement
backend. Artifact metadata and file content can be
retrieved through \nolinkurl{xmatcher-artifact} and
\nolinkurl{xmatcher-artifact-content}, respectively.

The same artifact representation is used internally when
the operations are selected by the Gan Jiang agent.
Consequently, an analysis can move between agent-controlled
orchestration and direct CLI invocation without requiring
a separate data representation.

\paragraph{Phase identification and diffraction calculations.}

Several identification capabilities of Gan Jiang are
available as individual CLI operations.
\nolinkurl{xmatcher-detect-peaks} extracts peaks from an
observed diffraction pattern, while
\nolinkurl{xmatcher-match} performs single-phase
search--match against the structural database.
\nolinkurl{xmatcher-multiphase-match} provides the
corresponding multiphase search capability.

The learned retrieval and decomposition components are
exposed through \nolinkurl{xqueryer-predictions} and
\nolinkurl{xdecomposer-predictions}, respectively.
Their outputs can be consumed directly by an external
workflow or used as intermediate evidence within Gan Jiang.

The CLI also exposes structure-based calculations.
Candidate structures from MP500 can be exported through
\nolinkurl{mp500-cif-exports}. Given one or more structural
CIFs, \nolinkurl{xmatcher-cif-xrd} calculates their
diffraction patterns, whereas
\nolinkurl{xmatcher-pdf-peaks-xlsx} exports a table of
calculated reflections. User-supplied CIFs can be passed
to these operations directly, without requiring a
preceding phase-identification step.

\paragraph{Example: structure-based diffraction calculation.}

The following example illustrates a direct tool-level
workflow. A structural CIF is first uploaded as an
artifact and is then reused to request its calculated
diffraction-peak table.

\begin{skillcode}{skillafter}{CLI}{CIF upload and diffraction-peak export}
# Inspect the available diffraction services.
delta-cli science invoke --tool xrd \
  --endpoint capabilities

# Upload a structural CIF.
delta-cli science invoke --tool xrd \
  --endpoint xmatcher-artifacts \
  --file "Mn2O3.cif" \
  --content-type application/octet-stream \
  --params '{
    "filename": "Mn2O3.cif",
    "kind": "cif",
    "media_type": "chemical/x-cif"
  }' > cif-upload-response.json

# Reuse the returned artifact reference.
delta-cli science invoke --tool xrd \
  --endpoint xmatcher-pdf-peaks-xlsx \
  --data '{
    "schema_version": 1,
    "cifs": [
      {
        "artifact": {
          "artifact_id": "<CIF_ARTIFACT_ID>",
          "sha256": "<CIF_SHA256>"
        },
        "name": "Mn2O3"
      }
    ]
  }' > peak-table-response.json
\end{skillcode}

The \texttt{artifact\_id} and \texttt{sha256} values in
the second request are obtained from the upload response.
The resulting table contains the Miller indices,
diffraction angles, interplanar spacings and relative
intensities of the calculated reflections. Binary output
is represented in the CLI response using a base64-encoded
payload and can be decoded to recover the corresponding
XLSX file.

This example illustrates the difference between the two
access modes. When the same operation is used within
Gan Jiang, the agent determines why the calculation is
needed and how the resulting reflections contribute to
the current structural hypothesis. Through direct CLI
access, the caller explicitly requests the calculation
and receives its computational output.

\paragraph{Structural refinement.}

Refinement capabilities are exposed in the same way.
An observed diffraction pattern is first prepared for
refinement through
\nolinkurl{xmatcher-refinement-patterns}, which produces
a refinement-input artifact and an associated provenance
manifest. The prepared pattern is then combined with one
or more structural CIF artifacts and submitted to the
desired backend.

The available refinement operations include
\nolinkurl{refinement-gsasii},
\nolinkurl{refinement-fullprof} and
\nolinkurl{refinement-wpem}, corresponding to GSAS-II,
FullProf and PyWPEM/PyXplore. Multiple backends may also
be configured for the same input, allowing their outputs
to be retained independently.

Refinement is executed asynchronously. A submission
returns a job identifier, which can subsequently be used
to inspect the job state through
\nolinkurl{refinement-job}, retrieve its event history
through \nolinkurl{refinement-job-events}, and obtain
the produced artifacts through
\nolinkurl{refinement-job-artifacts}. Jobs can also be
cancelled through \nolinkurl{refinement-job-cancel}.

Output artifacts can include refined CIFs, calculated
diffraction profiles, fit diagnostics and backend-specific
summaries. Separating submission from result retrieval
allows longer refinement calculations to be executed
without keeping the original CLI process active.

\paragraph{Example: direct multiphase refinement.}

Consider a diffraction pattern for which
\(\mathrm{Li_2CO_3}\) and \(\mathrm{NaCl}\) have already
been selected as candidate phases. The observed pattern
and the two corresponding CIFs are uploaded as separate
artifacts. The observation is converted into a
refinement-ready representation using
\nolinkurl{xmatcher-refinement-patterns}, after which the
prepared pattern and phase structures are submitted to,
for example, \nolinkurl{refinement-gsasii}.

The returned job identifier provides access to the
subsequent execution state and output artifacts.
A completed refinement may produce updated CIFs for the
two phases, the calculated whole-pattern profile, and
backend diagnostics. These outputs can then be consumed
directly by an external program or supplied to subsequent
Gan Jiang analysis components.

In agent mode, the corresponding sequence can instead
be generated automatically: Gan Jiang selects the
refinement operation after evaluating the preceding
phase evidence, retrieves the computational outputs and
passes them to the refinement Inspector before deciding
whether the current structural hypothesis should be
accepted, revised or rejected. Direct CLI invocation
exposes the same computational operation while leaving
these workflow decisions to the caller.

\paragraph{Execution results and scientific interpretation.}

The direct interface deliberately distinguishes execution
of a computation from interpretation of its scientific
meaning. A refinement job reaching
\texttt{completed} with
\texttt{pipeline\_status=success} indicates that the
requested computational workflow has completed. It does
not, by itself, establish that the supplied phase model
is scientifically correct.

The returned outputs therefore retain fit diagnostics,
convergence information, calculated profiles and
structural artifacts required for subsequent inspection.
When invoked through Gan Jiang, these outputs are
evaluated by the corresponding evidence and acceptance
policies described in Fig.~2. When the same operation is
invoked directly through the CLI, the caller can perform
an equivalent downstream assessment or incorporate the
results into an external workflow.

In this way, the CLI mode provides direct programmatic
access to selected scientific capabilities already
integrated in Gan Jiang, while the agent mode provides
the higher-level orchestration, evidence interpretation
and iterative decision-making required for an autonomous
diffraction-analysis workflow.